\documentclass[11pt]{article}

\usepackage{amsmath}
\usepackage{amsfonts}
\usepackage{amssymb}
\usepackage{graphicx}
\usepackage{multirow}
\usepackage{color}
\usepackage{rotating}
\usepackage{tabularx}

\usepackage{caption}
\usepackage{subcaption}

\newcommand{\specialcell}[2][c]{%
  \begin{tabular}[#1]{@{}c@{}}#2\end{tabular}}

\newcommand{\bm}[1]{\boldsymbol{#1}}
\newcommand{\der}{\mathrm{d}}

\def\spacingset#1{\renewcommand{\baselinestretch}%
{#1}\small\normalsize}
\numberwithin{equation}{section}
\numberwithin{table}{section}
\numberwithin{figure}{section}

\newcommand{\logit}{\text{logit}}
\newcommand{\N}{\mathcal{N}}
\newcommand{\C}{\; | \;}
\newcommand{\p}{\bm{p}}
\newcommand{\Y}{\bm{Y}}
\newcommand{\y}{\bm{y}}

\newcommand{\x}{\bm{x}}
\newcommand{\z}{\bm{z}}
\newcommand{\g}{\bm{\gamma}}
\newcommand{\al}{\bm{\alpha}}
\newcommand{\be}{\bm{\beta}}
\newcommand{\ps}{\bm{\psi}}
\newcommand{\om}{\bm{\omega}}
\newcommand{\vol}{\texttt{vol}}
\newcommand{\age}{\texttt{age}}
\newcommand{\beds}{\texttt{beds2008}}
\newcommand{\cd}{\bm{\cdot}}
\newcommand{\ha}{\hat\alpha}
\newcommand{\hP}{\hat P}

\newcommand{\indsim}{\stackrel{\normalfont\mbox{ind}}{\sim}}
\newcommand{\iidsim}{\stackrel{\normalfont\mbox{iid}}{\sim}}

\begin{document}
\spacingset{1.3}
%\vspace*{.5in}
\begin{center}
{\Large {\bf Mortality Rate Estimation and Standardization \\
for Public Reporting: Medicare's Hospital Compare}}\\
\vspace{.1in}
%{\large {\bf by}}\\
%\vspace{.1in}
{\large {By E.I. George, V. Ro\v{c}kov\'a, P.R. Rosenbaum, V.A. Satop\"a\"a and J.H. Silber
%\spacingset{1}
\footnote{Edward I. George is Professor of Statistics,  Department of Statistics, University of Pennsylvania, Philadelphia, PA, 19104, edgeorge@wharton.upenn.edu; Veronika Ro\v{c}kov\'a is {Assistant Professor of Econometrics and Statistics at the Booth School of Business of the University of Chicago, Chicago, IL, 60637, Veronika.Rockova@chicagobooth.edu; Paul R. Rosenbaum is Professor of Statistics,  Department of Statistics, University of Pennsylvania, Philadelphia, PA, 19104, rosenbap@wharton.upenn.edu; Ville Satop\"a\"a is Assistant Professor of Technology and Operations Management at INSEAD, 77305 Fontainebleau, France, ville.satopaa@insead.edu; and Jeffrey H. Silber is Professor of Pediatrics and Anesthesiology \& Critical Care, The University of Pennsylvania School of Medicine and Professor of Health Care Management, The Wharton School, Philadelphia, PA, 19104, silberj@wharton.upenn.edu.   This work was supported by the Agency for Healthcare Research and Quality grant No.~R21-HS021854; and grants SBS-1260782 and DMS-1406563 from the National Science Foundation. The authors are especially grateful to Nabanita Mukherjee, an associate editor and anonymous referees for their many constructive suggestions.}
} }}\\
\vspace{.1in}
{\large {\it
University of Pennsylvania, University of Chicago and INSEAD}}\\
\vspace{.1in}
{March 2018\\ (typo corrected version of August 2017)}
\end{center}
\spacingset{1.0}
{\small Bayesian models are increasingly fit to large
administrative data sets and then used to make individualized recommendations.
In particular, Medicare's Hospital Compare webpage provides information to
patients about specific hospital mortality rates for a heart attack or
Acute Myocardial Infarction (AMI). \ Hospital Compare's current recommendations are based on a
random-effects logit model with a random hospital indicator and patient risk factors. \
%The recommendations can describe a few hospitals near where you happen to live.
\ Except for the
largest hospitals, these individual recommendations or predictions are not
checkable against data, because data from smaller hospitals are too limited
to provide a meaningful check.
%What is the most relevant way to check the process that produces these individualized recommendations?
Before individualized Bayesian recommendations, people derived general advice from
empirical studies of many hospitals; e.g., prefer hospitals of type 1 to type
2 because the risk is lower at type 1  hospitals.
%The Bayesian model makes not only individualized predictions, but also aggregated predictions for how such studies of general advice will turn out. \
Here we calibrate these Bayesian recommendation systems by checking, out of sample, whether their predictions
aggregate to give correct general advice derived from another sample. 
%A Bayesian model that accurately predicts such general advice has the virtue that its predictions agree in aggregate with the general advice one would have had to use if the Bayesian model were unavailable. \ 
This process of calibrating
individualized predictions against general empirical advice leads to
substantial revisions in the Hospital Compare model for AMI mortality.
In order to make appropriately calibrated predictions, our revised models incorporate information about hospital volume, nursing staff, medical residents, and the hospital's ability to
perform cardiovascular procedures.   For the ultimate purpose of comparisons, hospital mortality rates must be standardized to adjust for patient mix variation across hospitals.  We find that indirect standardization, as currently used by Hospital Compare, fails to adequately control for differences in patient risk factors and systematically underestimates mortality rates at the low volume hospitals.  To provide good control and correctly calibrated rates, we propose direct standardization instead.
\medskip
}%

\noindent {\small {\it Key Words and phrases:} Bayesian inference; calibration by matching; hierarchical random effects modeling; individualized prediction; predictive Bayes factors; standardized mortality rates.}
\newpage
\spacingset{1.3}
\section{Are Mortality Rates For AMI Lower at Some Hospitals?}
\subsection{Individualized Bayes Predictions Should Calibrate with Sound,
Empirically-Based General Advice}

With a view to providing the public with information about the quality of
hospitals, Medicare runs a website called \textquotedblleft Hospital
Compare\textquotedblright\ (http://www.medicare.gov/hospitalcompare/). \ Among
other things, for each hospital, Hospital Compare provides information about
the mortality rate of patients treated for a heart attack, or ``acute myocardial
infarction'' (AMI). \ If you enter your zip code at the website, Hospital
Compare will tell you about hospitals near where you live. \ Nationally,
for a person who arrives at the hospital alive, the 30 day mortality rate
following AMI is in the vicinity of 15\%. \ The website's reported
hospital-specific mortality rates are based on Medicare claims data and a
random effects logit model in which hospitals enter as a random intercept and
adjustments are made for risk factors describing individual patients, for
instance age and prior heart attacks. \ The number reported by Hospital
Compare is essentially an indirectly standardized mortality rate for each
hospital, adjusting for measured risk factors describing the patient. \ An
indirectly standardized rate is a constant multiple of a ratio of two
predictions for the mortality of the patients actually treated at that
hospital, namely, in the numerator, the model's predicted mortality rate if
these patients were treated at thiÄs hospital, and in the denominator, the
model's predicted mortality for the same patients if treated at { 
what Hospital Compare considers to be a ``typical''
hospital.} \ A ratio substantially above one is interpreted as
\textquotedblleft worse than average risk\textquotedblright\ and a ratio
substantially below one is interpreted as \textquotedblleft better than
average risk\textquotedblright. \ The website describes most hospitals as
\textquotedblleft no different than the national average.\textquotedblright

Some small hospitals treat a few AMIs per year, whereas there is a hospital in
New York that treats on average about two AMIs per day. \ Mortality rates from
small hospitals are quite unstable, and the random intercepts model used by
Hospital Compare shrinks these rates to resemble the National average. \ Their
model says: \textquotedblleft if there is not much data about your hospital,
then we predict it to be average.\textquotedblright\ \ For any one small
hospital, there is not much data to contradict that prediction. \ So their model
claims that the mortality rate at each small hospital is close to the National
average. \ Is this a discovery or an assumption?

If it is a discovery, then it is a surprising discovery. \ A fairly consistent
finding in health services research is that, adjusting for patient risk
factors, mortality rates are typically higher at low volume hospitals
(Gandjour, Bannenberg, and Lauterbach 2003; Halm, Lee, and Chassin 2002; Luft,
Hunt, and Maerki 1987; Shahian and Normand 2003). \ Indeed, this pattern is
unambiguously evident in the data used to fit the Hospital Compare model.
Therefore, sound general advice would be to avoid low volume hospitals for
treatment of AMI.

So, is the finding of average risk at small hospitals a discovery or an
assumption? \ Actually, it is neither: it is a mistake. The model is not
properly calibrated; see Dawid (1982) for discussion of calibration.
\ Although there is very little data about any one small hospital, hence very
little data to check a statement about one small hospital, there is plenty of
data about small hospitals as a group. \ When Hospital Compare's
predictions for all small hospitals are added up, 
it is unambiguously clear that the risk
at small hospitals as a group is well above the national average; see Silber
et al. (2010).

There is, here, a general principle. \ A Bayesian model can use all of the
data to make an individualized prediction that is difficult to check as a
single prediction. \ It is possible that this individualized prediction is
better than relying upon general advice, because it is possible that this
individualized prediction is tapping into distinctions evident in the data but
not reflected in general advice. \ But if the general advice is correct as
general advice, the individualized predictions should not aggregate to
contradict it.
%correct general advice. 
%For instance, if the risk at low-volume hospitals as a group is well above the national average, so that general advice would direct you away from low-volume hospitals if you had a nearby
%alternative, then individualized predictions should not aggregate to
%contradict that general advice. \ If the data warrants, individualized predictions may identify particular low volume hospitals with average or better than average mortality, but checkable aggregates should agree.
As a check on whether a Bayesian model is calibrated, checking 
individualized predictions against general advice has two virtues. \ First, what it checks
can fail to hold, so it can reject some models as inadequate. \ Second, what
it checks is relevant: the check is against the advice you would fall back
upon if individualized predictions were unavailable. \ A model may be
detectably false in an irrelevant way --- it may use a double exponential
distribution where a logistic distribution would have been better --- but that
model failure may have negligible consequences for its recommendations.
\ However, if the model contradicts correct general advice, then there is
reason to worry about its individualized predictions. 
 These general considerations are illustrated in Section \ref{sec:calibrations}.

\subsection{Outline: Modeling, Calibrating and Reporting Hospital Mortality Rates}

{ In the current paper, we show how the Hospital Compare model can be elaborated to yield improved predictions that no longer contradict general advice. \ We confirm such improvements by fitting the model in one sample and making predictions for another: in particular, we predict the   outcome of the general advice that would be obtained from the second sample by an empirical study that made no use of the model. For the public reporting of these improved predictions, we propose a direct standardization approach that is effective at adjusting hospital mortality rate comparisons for patient mix differences between hospitals.}

In Section 2, we apply {a Bayesian implementation of} the Hospital Compare model to recent Medicare data for AMI, obtaining results similar to those reported on the Hospital Compare web-page.  Observing how it treats the various sources of mortality rate variation, we
then consider, in Section 3, whether the Hospital Compare model adequately describes the
data.  Specifically, in Section 3, we describe a sequence of hierarchical random effects logit models
predicting AMI mortality from attributes of patients prior to admission, such
as age, prior MI, and diabetes, and from the identity and attributes of individual
hospitals, such as a hospital's volume, its capabilities in interventional
cardiology and cardiac surgery, and the adequacy of its support staff in terms of nurses and
residents.  We also consider an interaction between a patient attribute and a
hospital attribute, so the model becomes able to say that the best hospital for
one patient may differ from the best hospital for another patient.

The models are evaluated in Section \ref{sec:predcomp} on the basis of { predictive Bayes factors which gauge} their ability to make out-of-sample predictions.   Then Section \ref{sec:calibrations} checks whether the models are calibrated by (i) performing matched studies of general advice using out-of-sample data, studies that make no
use of the model, and (ii) using the model's out-of-sample predictions to predict the
results of those matched studies.  In other words, the model's individualized predictions
are aggregated to predict the results of a study that might have been used to generate
general advice without individualized predictions.  Some models are much better calibrated
than others.

We turn to standardization for public reporting in Section 5.  The parameters of a Bayesian model would be difficult
for the public to understand.  Hospital Compare reports indirectly standardized rates.  We
contrast and evaluate directly and indirectly standardized rates.  We conclude that indirectly
standardized rates should not be used for public reporting, but directly standardized rates
work well.  In Section 6, we describe what can be learned from our recommended approach to hospital mortality rate estimation.  This includes mortality rate uncertainty intervals, classification of hospitals by Low, Average and High mortality rates, and the influence of hospitals attributes on mortality.  Section 7 concludes with a discussion.  In supplemental appendices, we provide technical aspects of our Bayesian computational approach, as well as extensions and details of the various analyses considered throughout.

\section{Hierarchical Bayesian Models for Adjusted Mortality Rates}
\subsection{The Data}
Our data were obtained from Medicare records on $N = 377,615$ AMI cases for patients admitted to $H = 4,289$ hospitals from July 1, 2009 to December 31, 2011.   The first two years of data, up to June 30, 2011, were used to fit the models under consideration here and in Section \ref{sec:hrm}.    The remaining six months of data were then used for the model validations in Section \ref{sec:meval}.  We will refer to these two datasets as the training data and the validation data respectively.

Each case in our data contains an indicator of patient death within 30 days of admission, patient-specific demographics and risk factors (gender, age, history of diabetes, etc.), and hospital-specific attributes (volume, number of beds, etc.)   We denote these variables as follows.  For patient $j$ in hospital $h$, $j = 1, 2, \dots, n_h$ and $h = 1, 2, \dots, H$, let $y_{hj} \in \{0, 1\}$ be the binary outcome for whether the patient died $\left(y_{hj} = 1 \right)$ or did not die $\left(y_{hj} = 0 \right)$ within 30 days of admission.  Let $\boldsymbol{x}_{hj}$ and $\bm{z}_h$ be the accompanying vectors of patient attributes and hospital attributes, respectively.  The number of patients per hospital $n_h$ varied a great deal over our data, ranging from 1 to 2782 patients, with a median value of 79.  The three year volume of all Medicare AMI admissions at hospital $h$, which we denote $\vol_h$, is a particular characteristic that will turn out to figure prominently in our modeling of mortality rates throughout.

\subsection{The Hospital Compare Random Effects Model} \label{sec:HC model}

\begin{figure}[h!]
\centering
\begin{subfigure}{0.30\textwidth}
\includegraphics[width=\textwidth, height = \textwidth]{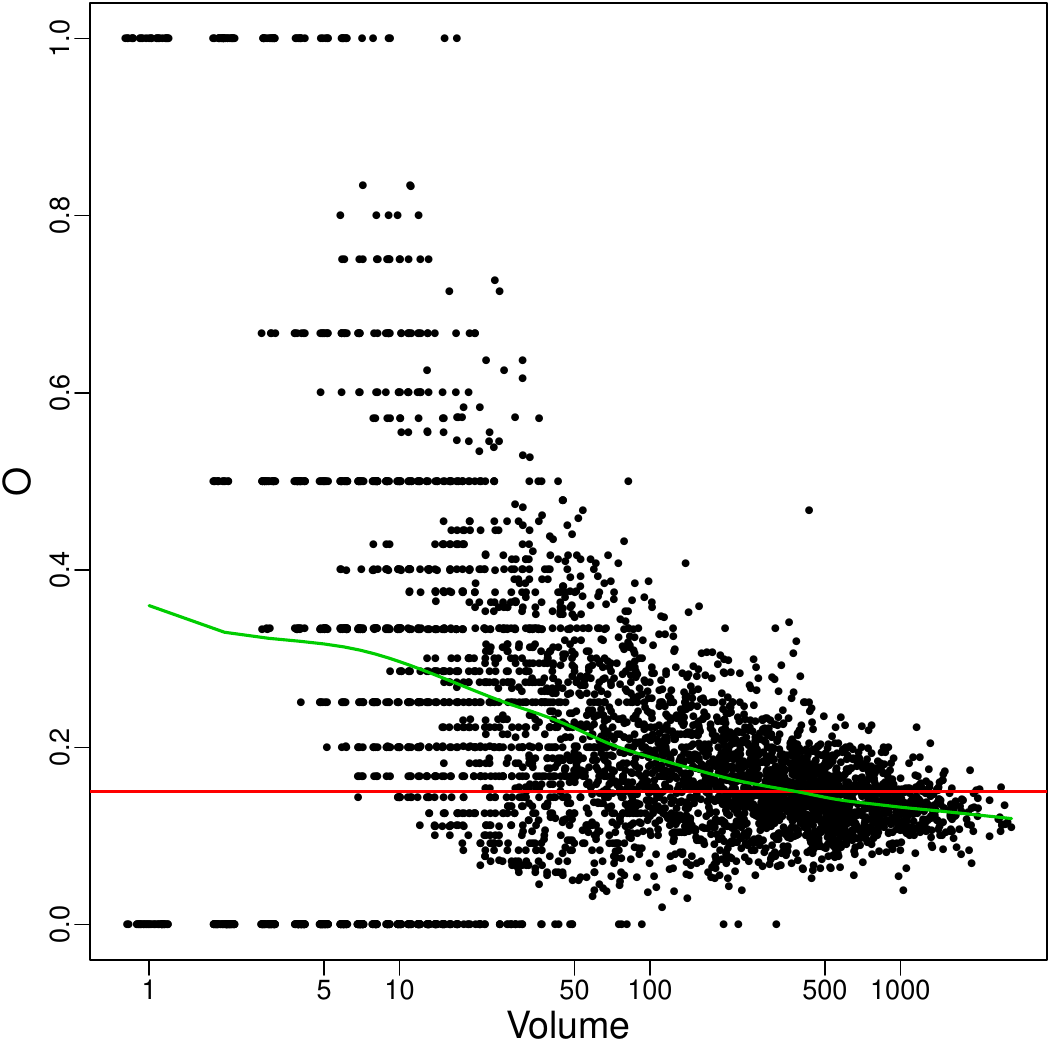}
\end{subfigure}
\centering
\caption{\small Raw observed hospital mortality rates $O_h$ by $\vol_h$. Overall average rate indicated by the red horizontal line.  Average rate by $ \vol_h$  summarized by the green superimposed smoothing spline. }
\label{Fig1}
\end{figure}

To motivate the modeling of mortality rates for our Medicare data, let us begin with Figure \ref{Fig1}, a plot of the raw observed hospital mortality rates $O_h = \frac{1}{n_h} \sum_j y_{hj}$ by volume $\vol_h$.
As would be expected, $O_h$ variation is largest at low volume hospitals where $n_h$ is small, and then steadily decreases as hospital volume increases.  Also evident in the plot is a steadily decreasing average mortality rate, summarized by a smoothing spline, which is highest at low volume hospitals.  This spline crosses the overall average patient mortality rate line $\bar y =0.1498$ at a hospital volume of about 450. An issue of central interest is the extent to which this average patient mortality rate curve can be explained by patient attributes and/or hospital characteristics.

Recognizing patient and hospital effects as potential sources of mortality rate variation, Medicare's Hospital Compare  (Yale New Haven Health Services Corporation 2014, Appendix 7A) uses a random-effects logit model to estimate underlying hospital mortality rates.   Proposed by Krumholz et al. (2006) for this context, this model is of the form
\begin{eqnarray} \label{eq:HC1 model}
Y_{hj} \C \alpha_h, \be,\boldsymbol{x}_{hj} & \indsim & \text{Bernoulli}(p_{hj}) \mbox{ where }
\logit (p_{hj})  = \alpha_h +\bm{x}_{hj}'\be  \\ \label{eq:HC2 model}
\alpha_h \C  \mu_\alpha, \sigma^2_{\alpha} & \iidsim & \mathcal{N}(\mu_{\alpha}, \sigma^2_{\alpha}).
\end{eqnarray}
Here, $P(Y_{hj} = 1) = p_{hj}  = \logit^{-1}(\alpha_h +\x_{hj}'\be)$
is the $hj$th patient's underlying 30-day mortality rate, which is determined by a hospital effect $\alpha_h$ and a patient effect $\bm{x}_{hj}'\bm{\beta}$.  The hospital effects $\alpha_h$  are modeled as independent normal random effects drawn from a single normal distribution which does not depend on any hospital attributes.  On the other hand, the patient effects $\bm{x}_{hj}'\bm{\beta}$, which explicitly depend on patient attributes $\bm{x}_{hj}$, are transmitted through a common fixed effect vector $\boldsymbol{\beta}$.  Under this model, the underlying average 30-day mortality rate for patients treated at hospital $h$ is given by
\begin{equation}\label{P_h}
P_h = \frac{1}{n_h} \sum_{j=1}^{n_h} p_{hj}.
\end{equation}
To mesh with our labeling of models proposed later in Section \ref{sec:genmodel}, we will refer to the Hospital Compare model \eqref{eq:HC1 model}-\eqref{eq:HC2 model} as the (C,C) model because it constrains both the mean and the variance of the $\alpha_h$ distribution to be constant.

For the implementation of this (C,C) model, we propose a fully Bayesian approach with the relatively noninfluential, neutral  conjugate priors
\begin{equation}\label{bprior}
\be \C \sigma_\beta^2 \sim \mathcal{N}_d\left( \boldsymbol{0}, \sigma_\beta^2 \boldsymbol{I}_{d}\right),   \quad \sigma_\beta^2 \sim \mathcal{IG}(1, 1)
\end{equation}
for the fixed effects parameters in \eqref{eq:HC1 model}, and
\begin{equation}\label{alphapriorhc}
\mu_{\alpha} \sim \N(0, g \sigma_\alpha^2), \quad g \sim \mathcal{IG}(1, 1),  \quad \sigma_\alpha^2 \sim \mathcal{IG}(1, 1)
\end{equation}
for the hyperparameters of the random effect distribution in \eqref{eq:HC2 model}. For compatibility with the model elaborations proposed in Section \ref{sec:hrm}, we have used a heavy tailed conjugate hyper $g$-prior for $\mu_{\alpha}$.

With such a fully Bayes model, all inferences about mortality rates, hospital effects, patient effects, and functions of these, can be obtained from the posterior distributions $\pi(\p \C \y)$, $\pi(\al \C \y)$ and $\pi(\be \C \y)$, where $\p$, $\al$, $\be$ and $\y$ are the complete vectors of mortality rates, hospital effects, fixed effects and observed mortality indicators, respectively. In particular, we use posterior means $\hat\p$, $\hat\al$ and $\hat\be$ as estimates throughout,
%as estimates of $\p$, $\al$ and $\be$ throughout,
with 95\% posterior density intervals to describe their uncertainty.   As described in Appendix \ref{sec:MCMC}, these can all be efficiently computed by  Markov Chain Monte Carlo (MCMC) posterior simulation using a P\'olya-Gamma latent variable augmentation, (Polson, Scott and Windle 2013).  As would be expected under heavy tailed priors such as \eqref{bprior} and \eqref{alphapriorhc}, our posterior mean estimates from the (C,C) model are very similar to the constrained likelihood estimates used by Hospital Compare with the SAS 9.3 GLIMMIX software. Indeed, the $\ha_h$ GLIMMIX estimates and the $\ha_h$ Bayes estimates here had a correlation of 0.9994.

%For example, Figure \ref{Gvsalpha} shows how the GLIMMIX and the Bayes recentered estimates of $\alpha_h$ here are nearly identical with the Bayes estimates slightly more spread out. The correlation is 0.9994.

%\begin{figure}[h!]
%\centering
%\begin{minipage}[t]{0.32\textwidth}
%\includegraphics[width=\textwidth, height = \textwidth]{GLIMMIX}
%\caption{GLIMMIX vs Bayes Estimates of $\alpha$}
%\label{Gvsalpha}
%\end{minipage}
%\centering
%\end{figure}

\begin{figure}[h!]
\centering
\begin{subfigure}{0.24\textwidth}
\includegraphics[width=\textwidth, height = \textwidth]{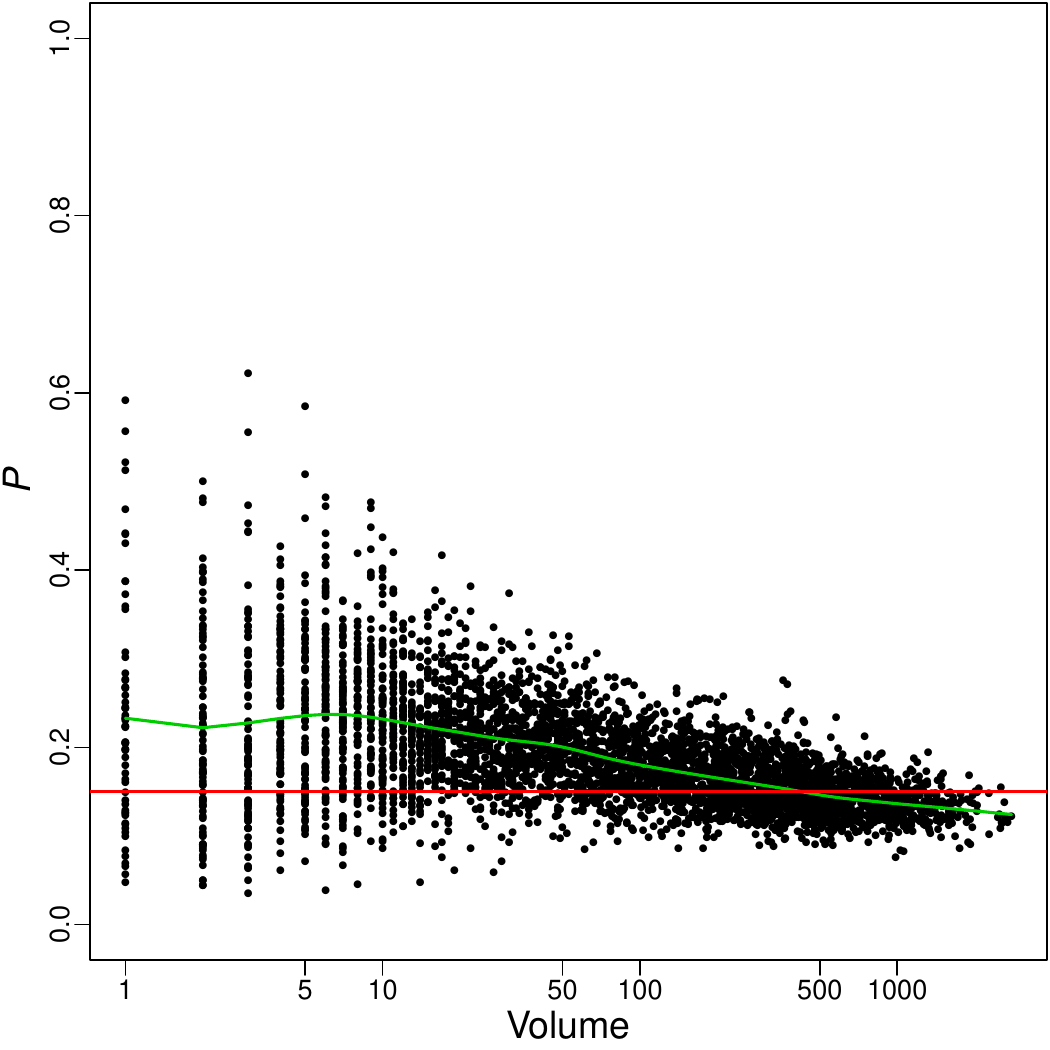}
\caption{\scriptsize $\hP_h$ under (C,C)}
\end{subfigure}
\begin{subfigure}{0.24\textwidth}
\includegraphics[width=\textwidth, height = \textwidth]{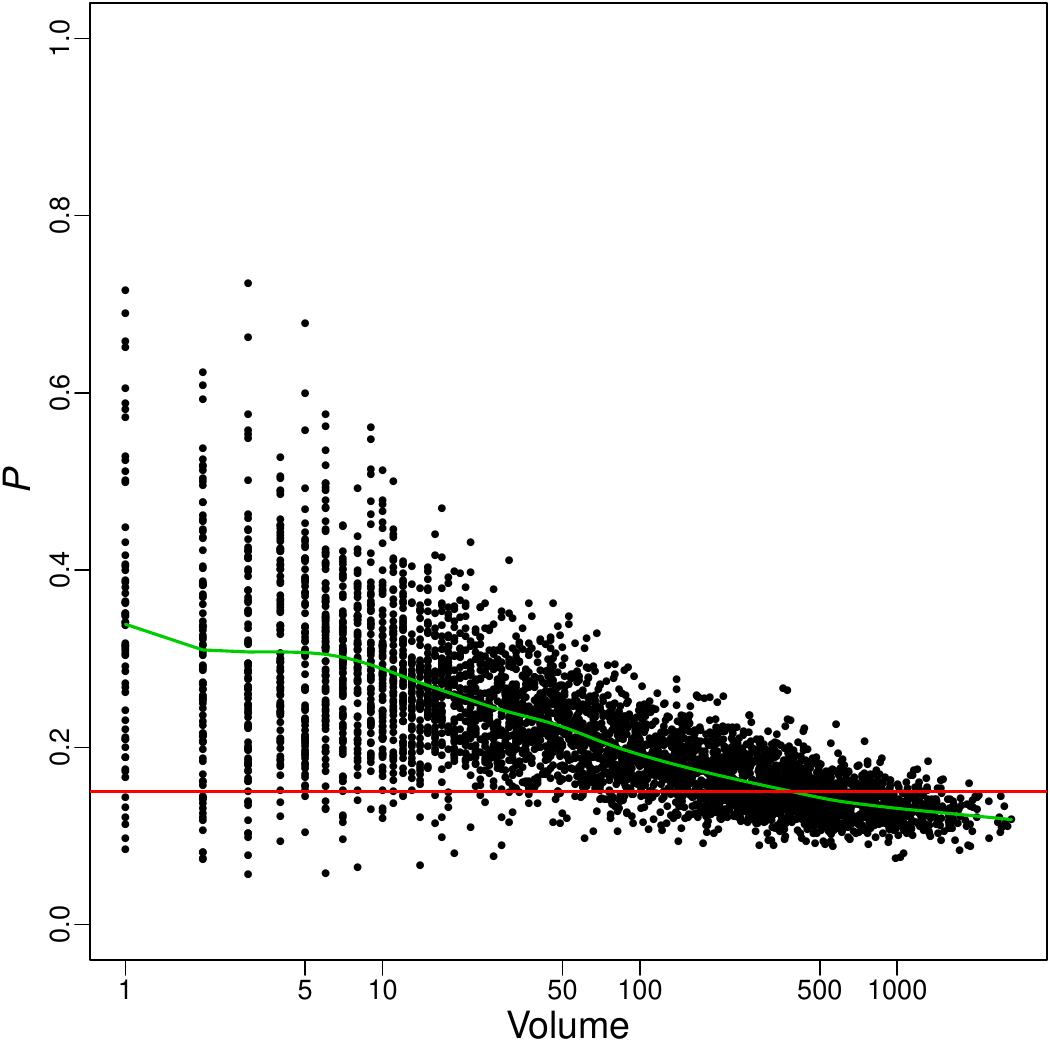}
\caption{\scriptsize $\hP_h$ under (L,C)}
\end{subfigure}
\begin{subfigure}{0.24\textwidth}
\includegraphics[width=\textwidth, height = \textwidth]{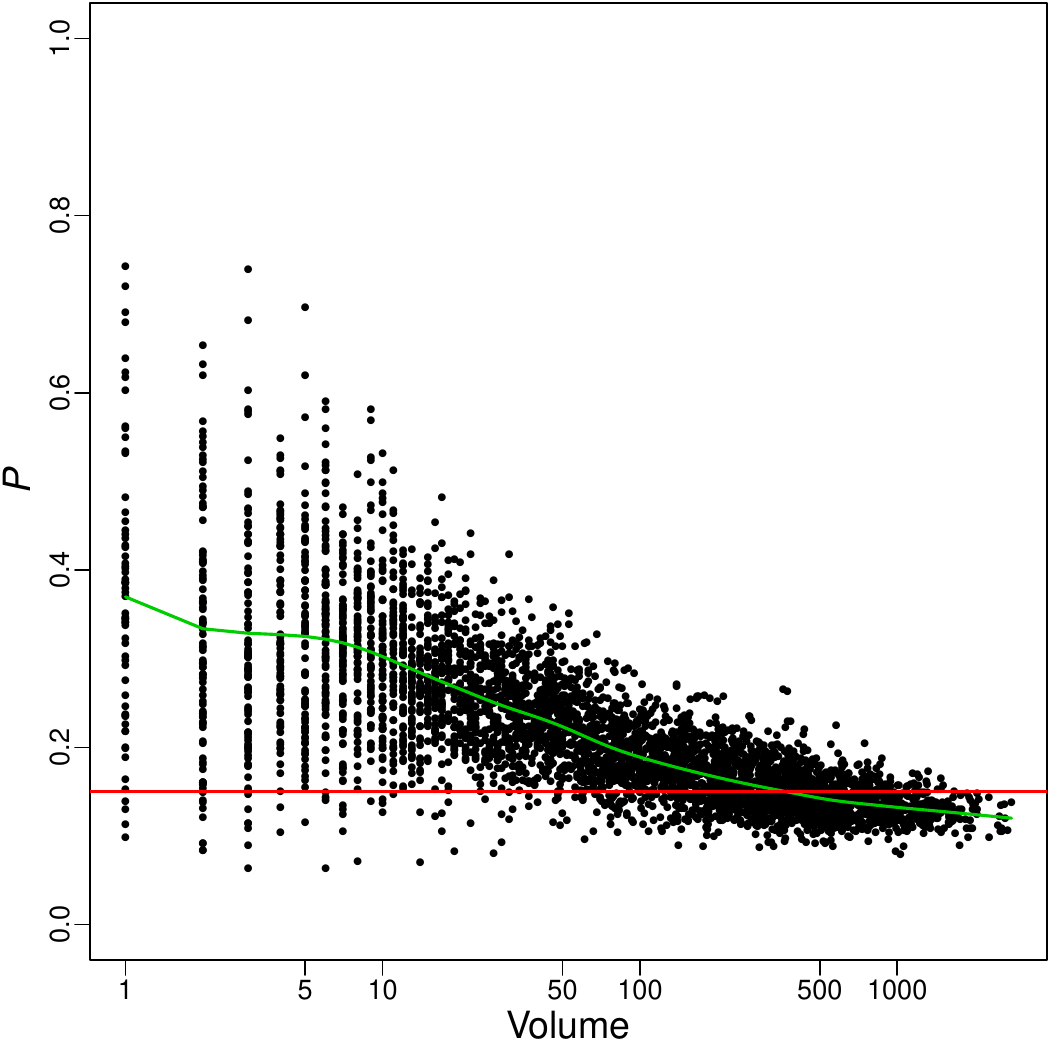}
\caption{\scriptsize $\hP_h$ under (S,L)}
\end{subfigure}
\begin{subfigure}{0.24\textwidth}
\includegraphics[width=\textwidth, height = \textwidth]{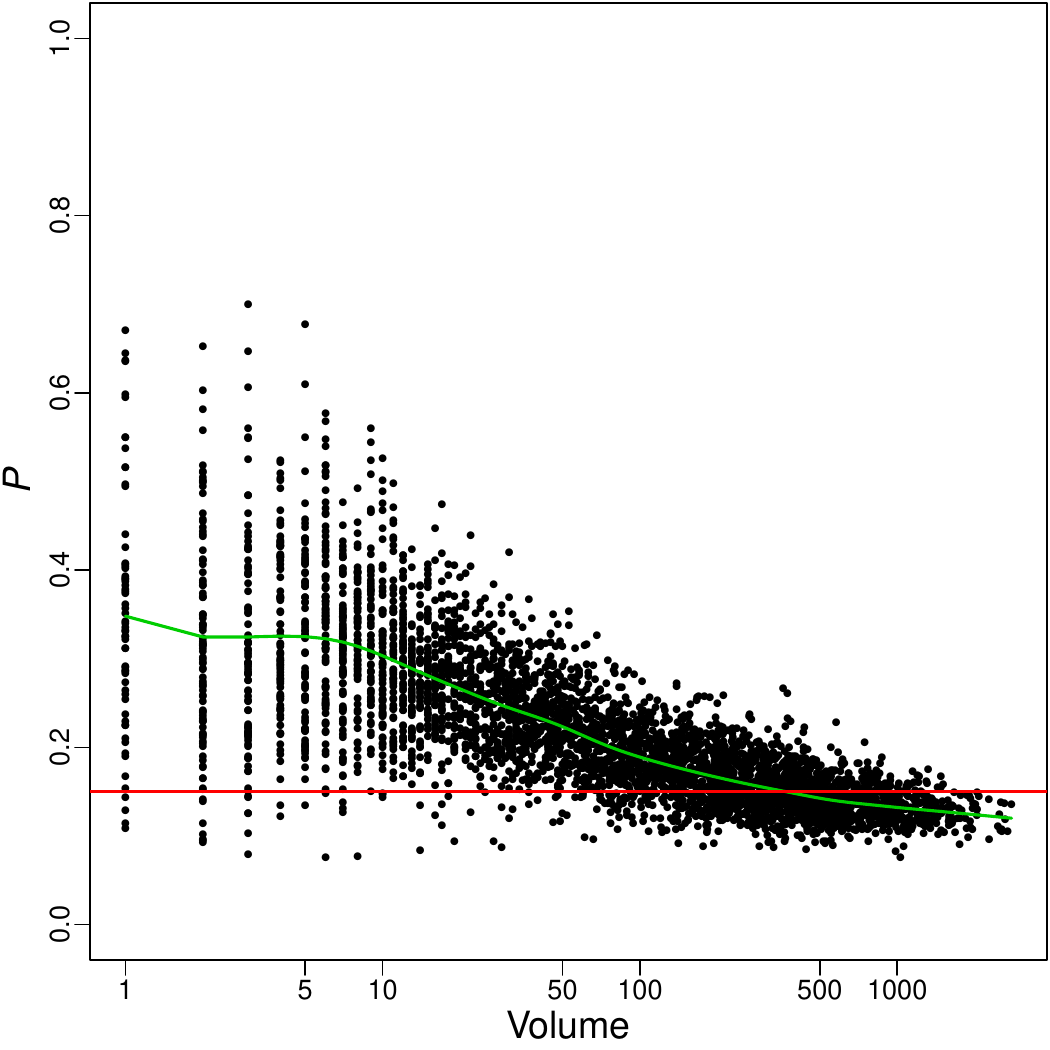}
\caption{\scriptsize $\hP_h$ under (SLI,L) }
\end{subfigure}
\centering
\vspace{-.1cm}
\caption{$\hP_h$ vs $\vol_h$. }
\label{Fig2}
\end{figure}

\begin{figure}[h!]
\centering
\begin{subfigure}{0.24\textwidth}
\includegraphics[width=\textwidth, height = \textwidth]{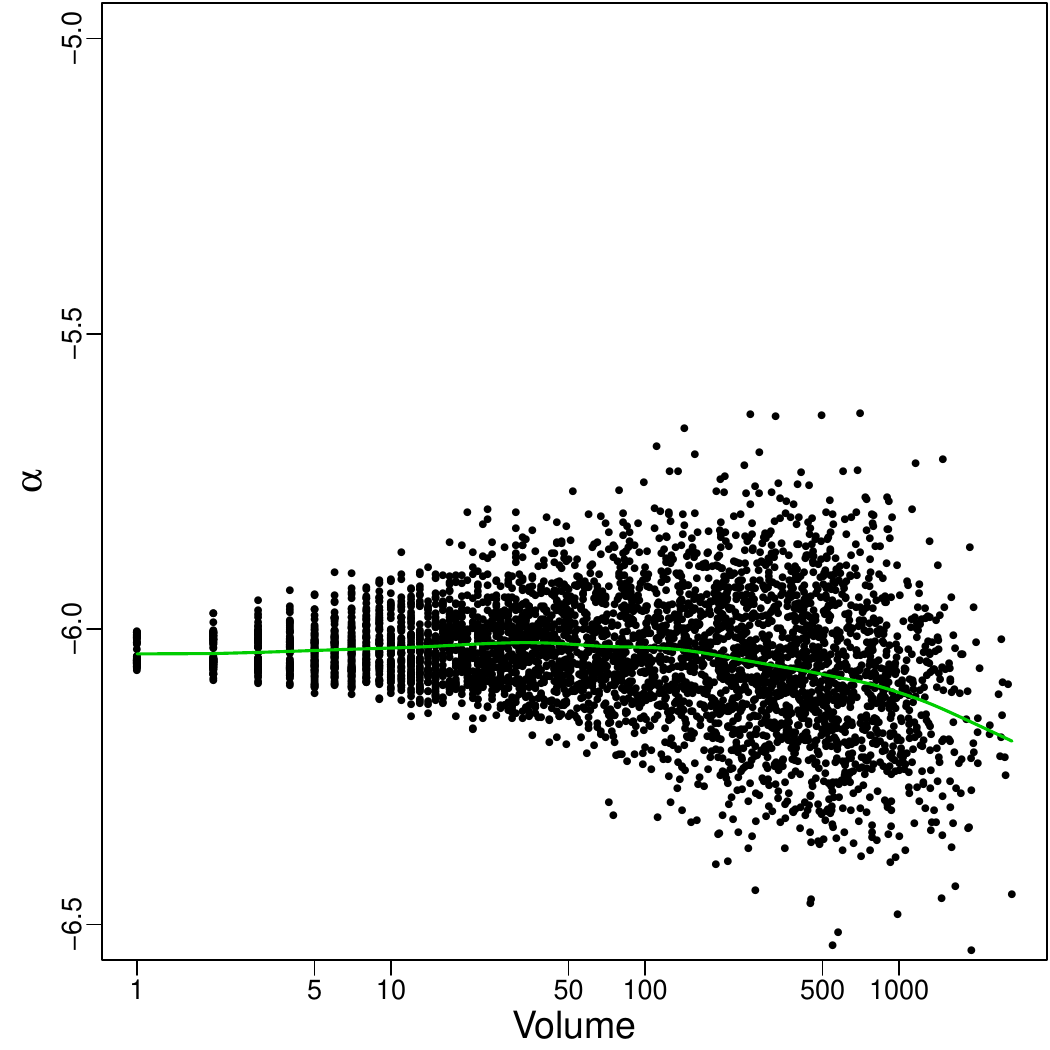}
\caption{\scriptsize $\ha_h$ under (C,C)}
\end{subfigure}
\begin{subfigure}{0.24\textwidth}
\includegraphics[width=\textwidth, height = \textwidth]{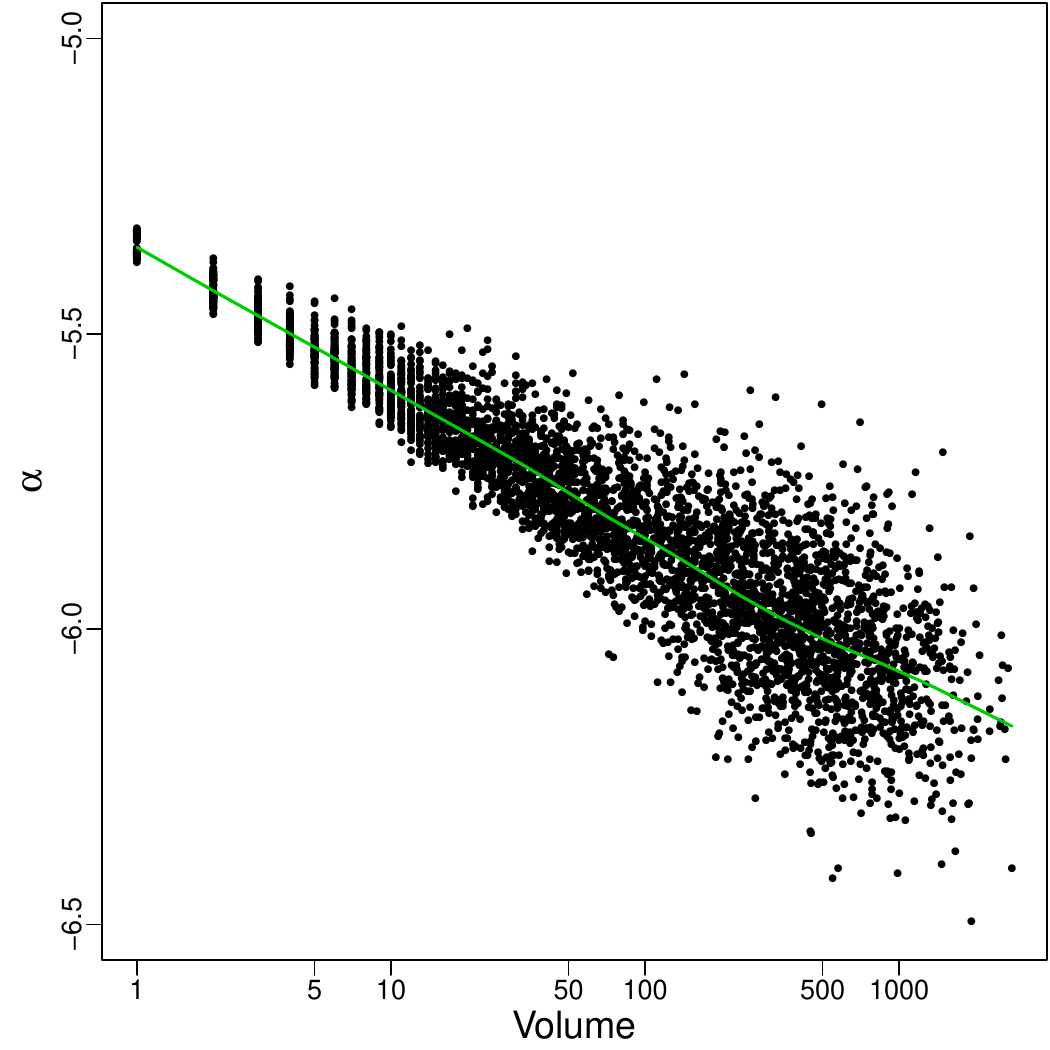}
\caption{\scriptsize $\ha_h$ under (L,C)}
\end{subfigure}
\begin{subfigure}{0.24\textwidth}
\includegraphics[width=\textwidth, height = \textwidth]{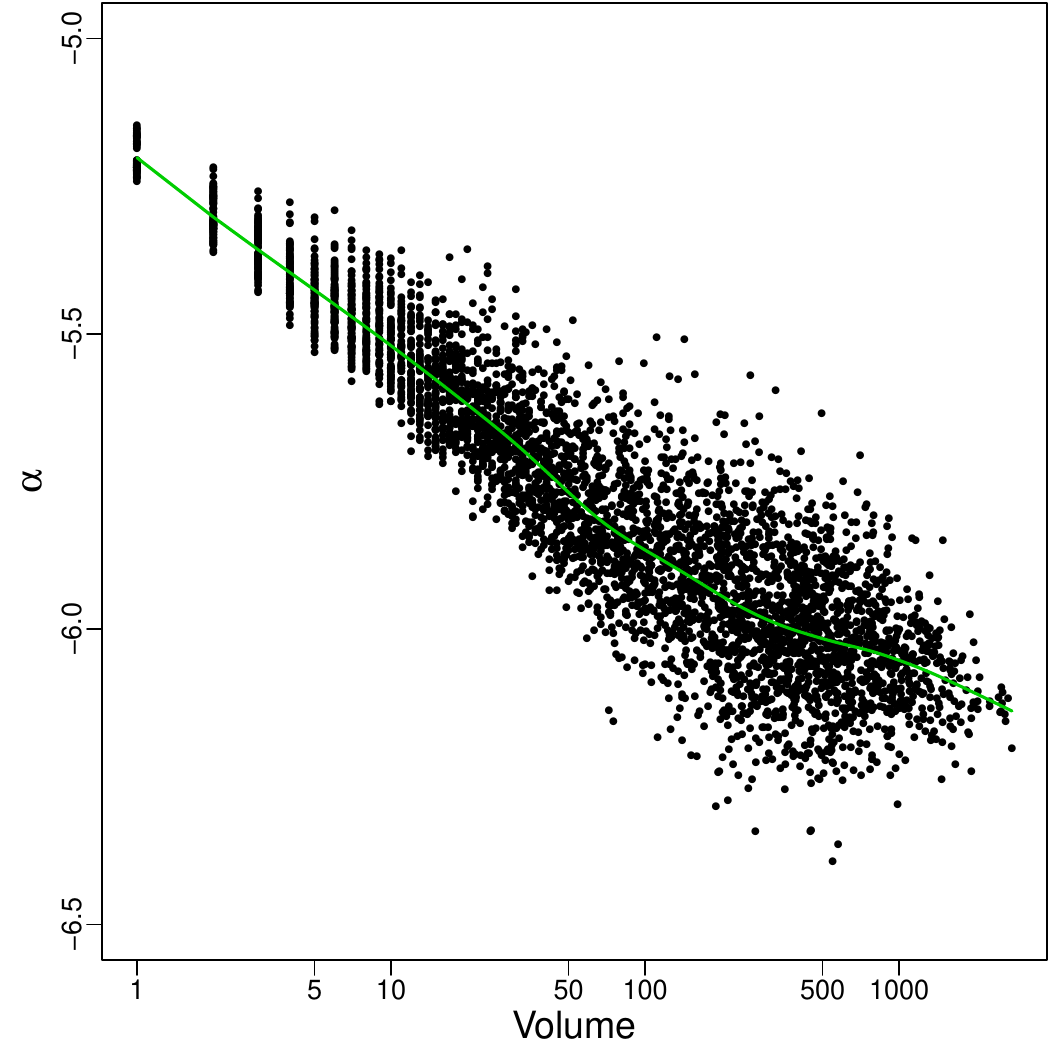}
\caption{\scriptsize $\ha_h$ under (S,L)}
\end{subfigure}
\begin{subfigure}{0.24\textwidth}
\includegraphics[width=\textwidth, height = \textwidth]{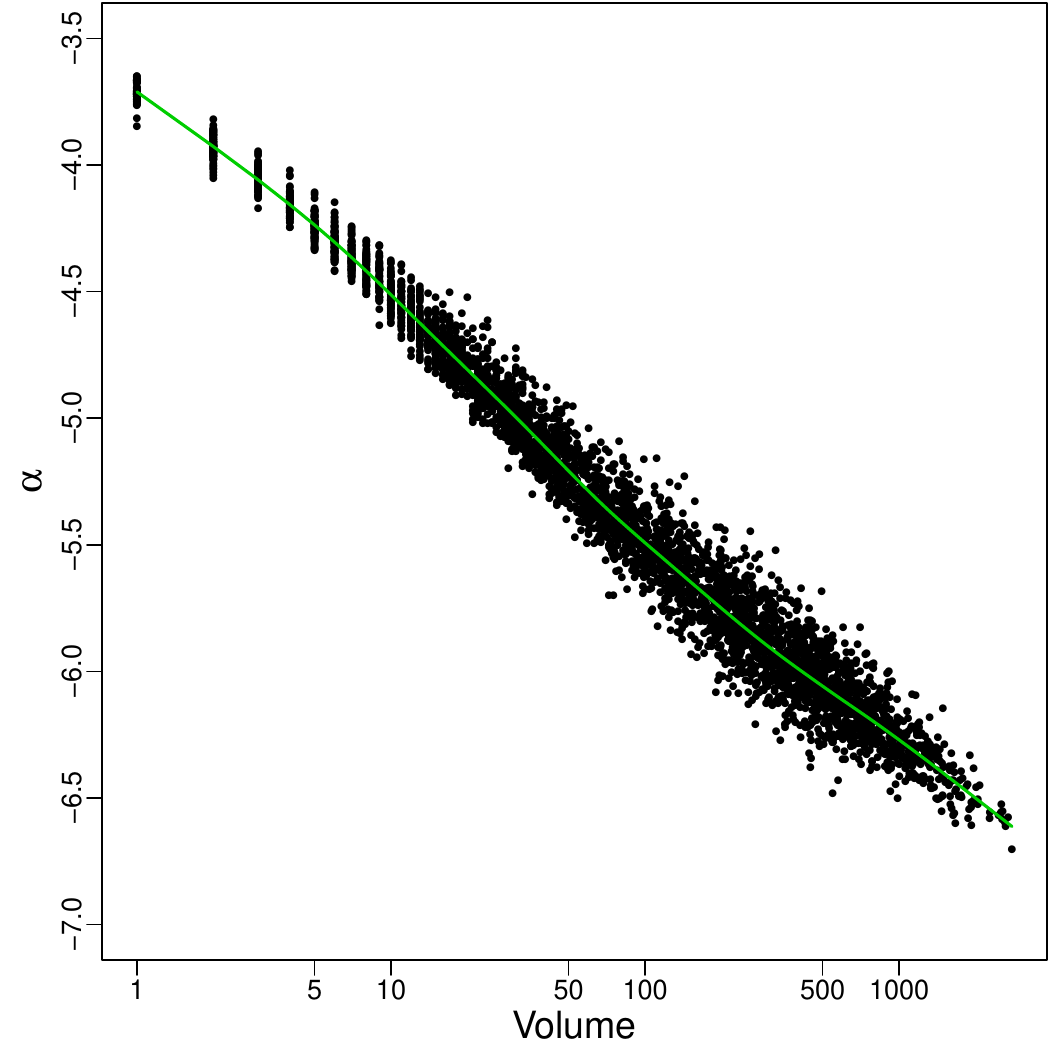}
\caption{\scriptsize  $\ha_h$ under (SLI,L) }
\end{subfigure}
\centering
\vspace{-.1cm}
\caption{$\ha_h$ vs $\vol_h$. }
\label{Fig3}
\end{figure}

Figure \ref{Fig2}a plots the $\hP_h$ posterior mean estimates by $\vol_h$ for the (C,C) model.  We see immediately that both the $O_h$ values and the decreasing average mortality rate spline from Figure \ref{Fig1} have been shrunk towards the overall mean mortality rate line $\bar y =0.1498$.  As we would expect, the $O_h$ realizations have simply added extra variability to their underlying $P_h$ values.  Nevertheless, it appears that substantial $\hP_h$ variability of  remains, especially at the small volume hospitals where $n_h$ in \eqref{P_h} is small.

Insight into the source of the $\hP_h$ variability under the (C,C) model is obtained from Figure \ref{Fig3}a, a plot of the $\ha_h$ posterior mean estimates by $\vol_h$.   Whereas the $\hP_h$ manifest larger variation at the low volume hospitals, as well as an elevated and decreasing average mortality rate, the $\ha_h$ manifest exactly the opposite behavior, (though on a different scale than the $\hP_h$).  The variation of the $\ha_h$ is smallest at the low volume hospitals, where the  average $\ha_h$ smoothing spline is flat and unrelated to hospital volume.
%gradually increasing with increasing volume. Furthermore, the decreasing average mortality trend of the $P_h$ estimates at the lower volume hospitals is not at all present in these $\alpha_h$ estimates, the $\alpha_h$ average smoothing spline is flat and unrelated to hospital volume there.
Since $\alpha_h$ and $\x_{hj}'\be$  are the only components of $p_{hj}$ in \eqref{eq:HC1 model}, the variation of the (C,C) model $\hP_h$  at the low volume hospitals is being driven almost entirely by the variation of patient effects $\x_{hj}'\be, j = 1,\ldots, n_h$ across hospitals.  Thus, under the (C,C) model, the elevated average mortality rates at the lower volume hospitals is coming from a { riskier} patient case-mix distribution at those hospitals rather than from hospital effect differences.
%among the hospitals themselves.
A primary purpose of the Hospital Compare analysis is to adjust for patient case-mix variation with an indirect standardization of the hospital mortality estimates.   This standardization effectively eliminates all mortality rate differences between the low volume hospitals as will be seen in Figure \ref{PISplots}a of Section \ref{sec:standardization}.  As discussed in Section 1, such a conclusion is at odds with the general finding in the literature that patient risk-adjusted mortality rates are typically higher at low volume hospitals (Gandjour, Bannenberg, and Lauterbach 2003; Halm, Lee, and Chassin 2002; Luft, Hunt, and Maerki 1987; Shahian and Normand 2003).

%Note also that the decreasing average mortality trend in the $O_h$ and the $P_h$ at the lower volume hospitals is not at all present in these $\alpha_h$ estimates.  This (C,C) model would lead us to conclude that hospital effects at the lower volume hospitals are unrelated to hospital volume and play a very small role in differentiating between hospital mortality rates.

%The absence of a relationship between hospital effects and low volume hospitals in Figure \ref{Fig1abc}c is at odds with the general finding in the literature that mortality rates are typically higher at low volume hospitals  even after patient case-mix risk adjustment (Gandjour, Bannenberg, and Lauterbach 2003; Halm, Lee, and Chassin 2002; Luft,Hunt, and Maerki 1987; Shahian and Normand 2003).

\section{Hierarchical Modeling of the Random Hospital Effects}\label{sec:hrm}
\subsection{The Illusion of Safe Shrinkage Estimation}

The absence of an elevated level of low volume hospital effects in Figure \ref{Fig3}a turns out to be an artifact of the (C,C) model, as will be seen by comparison with alternative models.  In leaving hospital characteristics out of the model and treating the $\alpha_h$'s as independent of volume $\vol_h$, the (C,C) model has not allowed the data to speak to this possibility.  Indeed, the pattern in Figure \ref{Fig3}a is consistent with the strong random effects assumption \eqref{eq:HC2 model} of normally distributed $\alpha_h$'s with a common mean $\mu_\alpha$  and variance $\sigma^2_{\alpha}$.  Under such a normal  prior, we would expect all the $\ha_h$ estimates to be shrunk towards a single common mean.  Such shrinkage would be especially pronounced for those $\ha_h$ for which there is less sample information, namely the $\ha_h$'s for the low volume hospitals.  This is exactly what we see.

Although shrinkage estimation has the potential to improve noisy estimates, such as the raw small hospital mortality rates in our setting, it can only do so if the shrinkage targets, here the means of the $\alpha_h$'s, are appropriately specified.  Contrary to the commonly held belief that shrinkage estimation can do no harm, which can be the case in certain stylized contexts, shrinkage estimation with a model that is at odds with the data can be very detrimental.  With an unforgiving, nonrobust normal prior that imposes strong shrinkage,  the resulting estimates will be poor and misleading if shrinkage targets are incorrectly specified; see Berger (1985).

The evident and plausible relationship between mortality rates and volume suggests that it would be more reasonable to shrink mortality rates towards the mean rates of hospitals with similar volumes.  Unfortunately this does not happen with the random effects distribution (2.2) which shrinks all rates towards a single overall rate.
%Indeed, that the posterior under this model can make no accommodation for volume or any other hospital attributes is reflected by the fact that posterior estimates of $\alpha_h$ will also be normal with a common mean and variance.

\subsection{Modeling the Random Hospital Effects} \label{sec:genmodel}

The key to the development of a better hierarchical Bayes model for our data is to elaborate the random effects distribution \eqref{eq:HC2 model} in a way that will allow the data to inform us of any potential relationship between hospital mortality rates and volume and hospital attributes such as volume.  For this purpose, we propose hierarchical logit model formulations of the form
 \begin{eqnarray} \label{eq:H1 model}
Y_{hj} \C \alpha_h, \bm\beta,\bm{x}_{hj} & \indsim & \text{Bernoulli}(p_{hj}) \mbox{ where }
\mbox{logit} (p_{hj})  = \alpha_h +\x_{hj}'\bm{\beta}  \\ \label{eq:H2 model}
\alpha_h \C\mu_h(\bm{z}),\sigma^2_h(\bm{z})  & \indsim & \mathcal{N}(\mu_h(\bm{z}),\sigma^2_h(\bm{z})).
\end{eqnarray}
where $p_{hj} = P(Y_{hj} = 1)$ is the $hj$th patient's underlying 30-day mortality rate.  As in \eqref{eq:HC1 model}, $\mbox{logit} (p_{hj})$ in \eqref{eq:H1 model} is still the sum of a hospital effect $\alpha_h$ plus a fixed effect $\x_{hj}'\bm{\beta}$.   However, \eqref{eq:H2 model} now allows the mean $\mu_h(\bm{z})$ and the variance $\sigma^2_h(\bm{z})$ of the hospital effects distribution to be functions of the hospital attributes $\bm{z}$.  As before, the fixed effects $\x_{hj}'\bm{\beta}$  in \eqref{eq:H1 model} still account for patient risk variation via the patient attributes $\x_{hj}$, but the random effects distribution $\mathcal{N}(\mu_h(\bm{z}),\sigma^2_h(\bm{z}))$ in \eqref{eq:H2 model} now allows the $\alpha_h$'s to more fully account for hospital-to-hospital variation via the hospital attributes $\bm{z}$.  Note that { as a matter of convenience, we have separated the roles of $\x_{hj}$ in \eqref{eq:H1 model} and  $\bm{z}$ in \eqref{eq:H2 model} to be consistent with their variation at the patient and hospital levels respectively.  Because hospital attributes $\bm{z}$ do not vary within hospitals, this partition avoids} their inclusion within the patient level linear component \eqref{eq:H1 model}, where modeling their effects would be complicated by the collinearity of the intercept estimates with hospital attributes deployed at the patient level.

We now proceed to consider specific formulations for $\mu_h(\bm{z})$ and $\sigma^2_h(\bm{z})$ that yield better calibrated predictions for our data.  As noted earlier, we refer to the Hospital Compare  \eqref{eq:HC1 model}-\eqref{eq:HC2 model}  specification of \eqref{eq:H1 model}-\eqref{eq:H2 model}, namely the one for which $\mu_h(\bm{z}) \equiv \mu_{\alpha}$ and $\sigma^2_h(\bm{z}) \equiv \sigma^2_{\alpha}$, as the Constant-Constant (C,C) model.  Each of the formulations proposed below will be a relaxation which nests the (C,C) model as a special case, thereby allowing the data to ignore the hospital attributes if they do not lead to better predictions.  Thus, these formulations will let the data speak rather than override what the data have to say.

\subsection{Modeling $\alpha_h$ as a Function of Volume} \label{sec:volmodels}

To shed light on the issue of whether hospital mortality rates are related to volume after accounting for patient mix effects, we begin by considering formulations for the mean and variance of the $\alpha_h$ hospital effects in \eqref{eq:H2 model} as functions of the hospital attribute volume $\vol_h$ only.  We begin with a simple linear specification, and then proceed to consider a more flexible model which allows for a more refined description of the underlying relationship.   This flexible formulation will then serve as a foundation for the further addition of hospital covariates and interactions to our final model in Section \ref{sec:SLII}. 

{ 
Before proceeding, we should emphasize that, although the application of our models will be seen to unambiguously reveal a strong association between high hospital mortality rates and low volume hospitals,  we are not addressing the issue of whether this relationship is causal. Our goal is mainly to confirm the aforementioned finding of such an association in the literature,  and to show that by including hospital volume in our models, we get better and more informative predictions with the Medicare data.  Such predictions will help guide patients towards safer hospitals.  

To provide further insight into the relationship between mortality rates and hospital size, we also examined the relationship between mortality rates and $\beds$, the number of beds in 2008, a hospital attribute that is indisputably exogenous to our observed mortality rates. As shown in Appendix A.3 of the Supplemental material, a strong association between these two variables persists.
}

\subsubsection{A Simple Linear Emancipation of the Means}\label{sec:(L,C)}

We begin with perhaps the simplest elaboration of the mean and variance functions for \eqref{eq:H2 model},
\begin{equation}\label{LCv}
\mu_h(\bm{z}) = \gamma_0 + \gamma_1 \log (\vol_h +1), \quad\quad \sigma^2_h(\bm{z}) \equiv \sigma^2_{\alpha},
\end{equation}
a linear relaxation of the mean which keeps the variance constant.   For the full model \eqref{eq:H1 model}-\eqref{eq:H2 model} with this specification, we add the conjugate fixed effect prior \eqref{bprior} for $\be$, and add the conjugate hyperparameter priors
 \begin{equation}
(\gamma_0, \gamma_1)' \C g, \sigma_\alpha^2 \sim \mathcal{N}_2 (\bm{0}, g \sigma_\alpha^2 \bm{I}_2), \quad g \sim \mathcal{IG}(1, 1), \quad \sigma_\alpha^2 \sim \mathcal{IG}(1, 1),
\end{equation}
where $g$ and $\sigma_\alpha^2$ are apriori independent.  We will refer to the hierarchical logit model formulation with this Linear-Constant specification as the (L,C) model.  Note that the (L,C) model nests the (C,C) model as the special case for which $\gamma_1 = 0$.

%Perhaps the simplest such specification is with a linear form for the mean, yielding the Linear-Constant (L,C) model
%\begin{equation}\label{LC}
%\mu_h(\bm{z}) = [1, \bm{z}_h']\bm{\gamma}, \quad\quad\quad\sigma^2_h(\bm{z}) \equiv \sigma^2_{\alpha},
%\end{equation}
%where $\bm{\gamma}$ is an $r$-dimensional vector of regression coefficients.

%Of particular interest to us will be the special case of \eqref{LC} where $\bm{z}_h$ consists of the single hospital characteristic volume $\vol_h$, \begin{equation}\label{LCv}
%\mu_h(\bm{z}) = \gamma_0 + \gamma_1 \log (\vol_h), \quad\quad\quad\sigma^2_h(\bm{z}) \equiv \sigma^2_{\alpha}.
%\end{equation}
%The nested (C,C) model is here obtained when $\gamma_1 = 0$.

Application of the (L,C) model to our data produced the hospital mortality rate and hospital effect estimates $\hP_h$ and $\ha_h$ displayed in Figures \ref{Fig2}b and \ref{Fig3}b.  We see immediately that compared to their (C,C) counterparts in Figure \ref{Fig2}a, the (L,C) $\hP_h$'s are generally higher at the lower volume hospitals where the average rate smoothing spline has been raised substantially.  This is evidently a consequence of their component $\ha_h$'s in Figure \ref{Fig3}b, which are dramatically different from their (C,C) counterparts in Figure \ref{Fig3}a.  The (L,C) $\ha_h$'s are now substantially higher at the low volume hospitals, with a clear downward sloping linear trend in $\log (\vol_h +1)$ summarized by the superimposed smoothing spline.   With a posterior mean estimate $\hat\gamma_1 = -0.106$ and a 95\% credible interval $(-0.116, -0.096)$, the data have expressed an unambiguous preference for a downward sloping (L,C) mean specification \eqref{LCv} over the (C,C) constant specification \eqref{eq:HC2 model} for which $\gamma_1 = 0$.  As will be seen with the more formal predictive model comparisons in Section \ref{sec:predcomp}, the (L,C) model $\hP_h$ improve very substantially over their (C,C) counterparts.

\subsubsection{A Spline-Log-Linear Emancipation of the Means and Variances}

With only a simple linear relaxation of the mean function, the (L,C) model released volume to reveal higher mortality rates explained by dramatically higher hospital effects at the low volume hospitals.  To further release the explanatory power of volume, we consider a more flexible relaxation of both the mean and variance functions.  In particular, we consider a spline specification for $\mu_h(\bm{z})$ coupled with a log linear specification for $\sigma^2_h(\bm{z})$.

%As such a simple elaboration of the (C,C) model, it is remarkable that the (L,C) estimates would be so substantially changed We thus proceeded to consider a much more flexible elaboration with a spline specification for the mean $\mu_h(\bm{z})$ and a log linear specification of the variant $\sigma^2_h(\bm{z})$.

%Although the data have a demonstrated preference for $\alpha_h$ estimates with a downward sloping trend in $\log (\vol_h)$, the (L,C) model constrains this trend to be linear.   Observing how the (L,C) model served to uncover sizable systematic variation of the means of the hospital effects, it is natural to consider the possibility that systematic variation of their dispersion will be revealed with a more flexible choice of $\sigma^2_h(\bm{z})$.

For the spline mean specification, let $\bm{v} = (\log (\vol_1 +1),\ldots, \log(\vol_H+1)'$ be the vector of all hospital specific volumes.  We construct a $B$-spline basis of degree $d$ and number of knots $\kappa$, represented by the columns of $\bm{B}_{d,\kappa}(\bm{v})$, an $H \times k$,  $(k = (d+1)+\kappa)$  matrix.  Letting $\bm{b}_h(\bm{v})$ be the $h$th row of $\bm{B}_{d,\kappa}(\bm{v})$, our spline specification is obtained as
\begin{equation}\label{SC}
\mu_h(\bm{z}) = \bm{b}_h(\bm{v}) \bm{\gamma}_S,
\end{equation}
where $\g_S$ is a $k \times 1$ vector of spline regression coefficients.  To this we add a prior on $\g_S$ of the form
\begin{equation}\label{penprior}
\g_S \C g_S, \sigma_\alpha^2 \sim \mathcal{N}_k\left(\bm{0}, g_S \sigma_\alpha^2 \bm{P}^{-1}\right),  \quad\quad g_S \sim \mathcal{IG}(1, 1),
\end{equation}
where $\bm{P}$ is a banded matrix that penalizes second-order differences between adjacent spline coefficients, and $g_S$ serves as a roughness penalty that determines the wiggliness of the resulting curve. (The usual spline roughness penalty is $\lambda = 1/g_S$). With this penalization, a nested linear parametric form is obtained as $g_S\rightarrow0$. Note that the conjugate inverse-gamma prior on $g_S$ is often considered in the context of hyper-$g$-priors for variable selection.    This $P$-spline (penalized B-spline) formulation allows us to begin with a rich $B$-spline basis $\bm{B}_{d,\kappa}(\bm{\bm{v}})$ with many knots, and use regularization to circumvent the  difficulties of optimizing the number and position of knots.

%Here we opt for the conjugate inverse-gamma $g_S \sim \mathcal{IG}(1, 1),$ a form also considered in the context of hyper-$g$-priors for variable selection.
For the log-linear variance specification. (Box and Meyer 1986; Gu, Fiebig, Cripps and Kohn 2007), we set
\begin{equation}\label{Lvol}
\sigma^2_h(\bm{z}) =   \exp\{\delta\, \vol_h\} \, \sigma^2_{\alpha},
\end{equation}
which nests the previous case $\sigma^2_h(\bm{z}) \equiv \sigma^2_{\alpha}$ when {$\delta = 0$.}    To this we add the prior
\begin{equation}\label{deltapr}
\delta \C \sigma_\alpha^2 \sim \mathcal{N} (0, g_\delta \sigma_\alpha^2), \quad\quad g_\delta \sim \mathcal{IG}(1, 1),\end{equation}
and then complete the entire Bayesian specification of \eqref{SC}-\eqref{penprior}-\eqref{Lvol}-\eqref{deltapr} with
\begin{equation}
\sigma_\alpha^2 \sim \mathcal{IG}(1, 1),
\end{equation}
where $g_S$, $g_\delta$ and $\sigma_\alpha^2$ are all assumed apriori independent. We will refer to the hierarchical logit model formulation with this Spline-LogLinear specification as the (S,L) model.

Application of this (S,L) model with a B-spline basis of degree $d = 3$ and $\kappa = 17$ knots, yields the hospital rate and effect estimates $\hP_h$ and $\ha_h$ in Figures \ref{Fig2}c  and \ref{Fig3}c.  These estimates again support the story revealed by the (L,C) model, that low volume hospitals have generally higher mortality rates driven by substantially higher hospital effects.  Having been freed from the constraints of linearity and constant variance, the low volume $\ha_h$ estimates are now even higher and more dispersed than their (L,C) counterparts in Figure \ref{Fig3}b, manifesting a decreasing nonlinear trend.  And with a posterior mean estimate of { $\hat\delta = -0.00112$ and a 95\% credible interval of $(-0.00160, -0.00066)$}, the data have also expressed a preference for the log linear variance specification $\eqref{Lvol}$ as well.  Although the (L,C) and (S,L) $\hP_h$ plots in Figures \ref{Fig2}b and \ref{Fig2}c seem very similar, manifesting considerable patient-mix variability, the formal predictive comparisons which we will see in Section \ref{sec:predcomp} confirm the (S,L) model estimates as a further improvement.

%Corroborating what was revealed by the (L,C) model, these estimates are generally similar to their (L,C) counterparts in Figures \ref{Fig2}b  and \ref{Fig3}b, all being higher at the low volume hospitals than their (C,C) counterparts  in Figures \ref{Fig2}a and \ref{Fig3}a.
%The mortality rate $\hP_h$ estimates and hospital effect estimates $\alpha_h$ are again substantially higher for the low volume hospitals, exhibiting a clear decreasing trend as $\vol_h$ increases.  However, a few interesting differences have appeared.   The $\alpha_h$ levels for the low volume hospitals are have increased and the trend is now nonlinear, initially decreasing more rapidly (in the log scale).  It appears that the (S,L) model has allowed the data to express a preference for some nonlinearity.  That this nonlinearity in fact yields a further improvement to $P_h$ will be borne out by the formal predictive comparisons in Section \ref{sec:predcomp}.

\subsection{Adding Further Hospital Attributes and Interactions to the Model}\label{sec:SLII}

With our (S,L) model as a foundation, we now consider enlarging the model to incorporate further hospital attributes.  This is most simply done by adding them as linear terms to  $\mu_h(\bm{z})$ in \eqref{SC}.   This yields the spline-linear mean specification
\begin{equation}\label{SCr}
\mu_h(\z) =   \bm{b}_h(\bm{\bm{v}}) \g_S + \z_h'\g_L, \end{equation}
where $\z_h$ is an $r\times 1$ vector of hospital $h$ attributes and $\g_L$ is an $r\times1$ vector of linear regression coefficients.  This form is then completed with the priors
\begin{equation} \label{Spr}
\g_S \C g_S, \sigma_\alpha^2  \sim  \mathcal{N}_k\left(\bm{0}, g_S \sigma_\alpha^2 \bm{P}^{-1}\right), \quad\quad g_S \sim \mathcal{IG}(1, 1),
\end{equation}
\begin{equation} \label{Lpr}
\g_L \C g_L, \sigma_\alpha^2 \sim \mathcal{N}_r (\bm{0}, g_L \sigma_\alpha^2 \bm{I}_r), \quad\quad\quad g_L \sim \mathcal{IG}(1, 1),  \end{equation}
where $g_S$ and $g_L$ are priori independent of each other and of $\sigma_\alpha^2 \sim \mathcal{IG}(1, 1)$.

Going further, hospital attributes can also be incorporated as patient-hospital interactions of the form $x_{hj}*z_h$,  products of particular attributes in $\x_{hj}$ and $\z_h$, respectively.  Because the values of such interaction terms vary at the patient level, these would be added as covariates to the linear fixed effects part of the model.  Without  such interactions, the model would say that one hospital, $h$, is either better or worse than another, $h'$, for every patient.  Of course, there is no reason to restrict attention to models with this feature and no reason to expect the world to be well described by such a model.  Patient-hospital interactions remove this limitation.

Keeping the log-linear variance formulation \eqref{Lvol}-\eqref{deltapr}, we shall refer to the hierarchical logit formulation with this Spline-Linear-Interaction specification as the (SLI,L) model.   Applying an instance of this model to our data, we added three hospital attributes named NTBR, RTBR and PCI as linear terms in (\ref{SCr}). The Nurse-to-Bed-Ratio (NTBR) and the Resident-to-Bed-Ratio (RTBR) are continuous hospital variables that describe the density of support staff at a hospital. 
The binary hospital variable PCI is a catch-all for the ability of a hospital to perform any of the following procedures: percutaneous coronary interventions such as percutaneous transluminal coronary angioplasty (PTCA), a stent, or a coronary artery bypass graft (CABG) surgery.  The ability to perform these procedures is common in large volume hospitals and much less common in small volume hospitals.  

We also added a patient-hospital interaction, $\age_{hj}* \log(\vol_h +1)$,  appending it to $\x_h$ as an additional covariate for the fixed effects term $\x_{hj}'\bm{\beta}$.  With this interaction, the model may provide mortality rate estimates which favor one hospital for a younger Medicare patient, say aged 68, and a different hospital for another older Medicare patient, say aged 90.

Application of this particular  (SLI,L) model to our data produced the hospital rate and effect estimates $\hP_h$ and $\ha_h$ displayed in Figures \ref{Fig2}d and \ref{Fig3}d.  These estimates again convey the same message as the (L,C) and (S,L) estimates in Figures \ref{Fig2}bc and \ref{Fig3}bc, namely that low volume hospital mortality rates are generally higher, driven by low volume hospital $\ha_h$'s which exhibit a clear decreasing average trend as $\vol_h$ increases. However, the $\ha_h$ in  Figure \ref{Fig3}d now vary more from high to low (note the changed vertical scale), a consequence of adding the  $\age_{hj}* \log(\vol_h +1)$ interaction, which has served to model a portion of the hospital effect variation at the patient level.  With a patient-hospital interaction in the model, the $\ha_h$'s no longer capture the entirety of the hospital effects.  And once again, although the $\hP_h$'s for the three models in Figures \eqref{Fig2}bcd look very similar,  model comparisons in Section \ref{sec:predcomp} will show that the (SLI,L) model leads to still further predictive improvements.

\subsection{Further Potential Elaborations}

As will be confirmed in Section \ref{sec:meval}, our (L,C), (S,L) and (SLI,L) models have served to reveal the inadequacies of the (C,C) model.   Paving the way for further improvements, these hierarchical elaborations of the random effects model are hardly the end of the story.  Indeed, it is clear that many further elaborations may be promising, for example, by adding more hospital attributes as linear terms, spline terms or patient-hospital interactions.  One might also consider elaborations of the log-linear variance specification \eqref{Lvol} that include more hospital attributes, for example
%\begin{equation}\label{Lgen}
$\sigma^2_h(\bm{z}) =  \exp\{\bm{z}_h'\bm{\delta}\}\, \sigma^2_{\alpha}$,
%\end{equation}
where $\bm{z}_h$ is an $q\times 1$ vector of hospital $h$ characteristics (possibly different from $\bm{z}_h$ above) and $\bm{\delta}$ is a $q\times 1$ regression vector.

Going further, one could also consider different families for random effects distributions.  More robust parametric distributions such as the Cauchy or t-distributions would serve to downweight the influence of extremes.  Even more flexibility could be obtained with nonparametric prior distributions.   Indeed, Guglielmi et.~al.~(2014)
proposed modeling hospital coronary mortality rates with a Bayesian hierarchical logit model analogous to our (L,C) model but with a Dependent Dirichlet Process for the random effects.  Such an elaboration opened the door for clustering hospitals into groups with identical mortality rates.   Other interesting nonparametric random effect logit models that also incorporate hospital process indicators for modeling and clustering hospital coronary mortality rates are proposed by Grieco et.~al.~(2014).  Such nonparametric elaborations provide promising new routes for improved mortality rate modeling.

A model for hospital mortality rates can be used for a variety of purposes, not just public reporting.  
Spiegelhalter et al. (2012) discuss the issues that arise in different applications of such models.

%capturing the effect of hospital attributes on the variance, see Box and Meyer (1986) and Gu, Fiebig, Cripps and Kohn (2007).  For the Bayesian specification we add the priors
%\begin{equation}\label{Lgenpr}
%\delta \C \sigma_\alpha^2 \sim \mathcal{N} (0, g_\delta \sigma_\alpha^2), \quad\quad g_\delta \sim \mathcal{IG}(1, 1).\end{equation}
%The prior specification for \eqref{Spr}-\eqref{Lpr}-\eqref{Lgen}-\eqref{Lgenpr} is finally completed with $\sigma_\alpha^2 \sim \mathcal{IG}(1, 1)$.

\section{Model Evaluation}\label{sec:meval}

\subsection{Predictive Bayes Factor Model Comparisons}\label{sec:predcomp}

Following the traditional Bayesian model choice formalism, we use Bayes factors to compare the performance of the proposed models (L,C), (S,L), and (SLI,L) with the performance of the Hospital Compare (C,C) model.  For this purpose we turn to out-of-sample predictive Bayes factors rather than in-sample Bayes factors.  As is well recognized,  in-sample Bayes factors based on diffuse parameter priors, such as those we have used with our training data, are unreliable criteria for model comparisons (Cox 1961, Berger 2006).   Furthermore, because prediction is the intended use of these models, comparisons based on out-of-sample performance are of fundamental relevance here.  Thus, we use predictive Bayes factors evaluated on the validation data using posterior rather than prior predictive likelihoods (Gelfand and Dey 1994).

Posterior predictive likelihoods are obtained by averaging the probability of the validation data with respect to a data-updated ``prior''
distribution using the training data. Here, the predictive likelihood for model $\mathcal{M}$ is obtained as
\begin{equation}\label{pplik}
\pi(\bm{y}_{val}\C\bm{y}_{tr},\mathcal{M})=\int_{\al,\be}\pi(\bm{y}_{val}\C \al,\be,\mathcal{M})\pi(\al,\be \C\bm{y}_{tr},\mathcal{M})\,\der\al \der\be.
\end{equation}
where $\bm{y}_{val}$ and $\bm{y}_{tr}$ are the validation data and training data $\bm{y}$ values, respectively, and $\al = (\alpha_1,\dots,\alpha_H)'$.  Note that the training data posterior $\pi(\al,\be\C\bm{y}_{tr},\mathcal{M})$ now serves as a stable and non diffuse prior for the validation data.   The predictive Bayes factor for comparison of model $\mathcal{M}_1$ versus $\mathcal{M}_2$ is then naturally defined as the ratio of the two posterior predictive likelihoods (Gelfand and Dey 1994, Kass and Raftery 1995),
$$
BF_{\mathcal{M}_1,\mathcal{M}_2}=\frac{\pi(\bm{y}_{val}\C\bm{y}_{tr},\mathcal{M}_1)}{\pi(\bm{y}_{val}\C\bm{y}_{tr},\mathcal{M}_2)}.
$$

Evaluation of the predictive Bayes factor is obtained by Monte Carlo integration of the posterior predictive Bayes likelihoods using posterior parameter samples from the MCMC output, as described in Section \ref{sec:MCMC}, based on the training data.  Using the simulated values $\al^{(s)},\be^{(s)}\sim\pi(\al,\be\C \bm{y}_{tr},\mathcal{M}_i)$ from \eqref{MCMCsample}, these approximations to the posterior predictive likelihoods are obtained by the empirical averages
$$\widehat{\pi}(\bm{y}_{new}\C\bm{y}_{tr},\mathcal{M}_i)=\frac{1}{S}\sum_{s=1}^S\pi(\bm{y}_{val}\C\al^{(s)},\be^{(s)},\mathcal{M}_i), i = 1,2$$

%and $\pi(\bm{y}_{val}\C\al^{(s)},\be^{(s)},\mathcal{M}_h)$ is the likelihood evaluated at the coefficient vectors $\al^{(s)},\be^{(s)}$.

The log posterior predictive Bayes factors comparing each of the (L,C), (S,L), (SLI,L) models with the (C,C) model are reported in Table \ref{predictiveBF}.  The predictive improvement over the (C,C) model by every one of our models is very large.  Beginning with the (L,C) model, which simply allowed hospital effect means to be linear in volume rather than constant, the predictive likelihood increased by a huge factor of $e^{27.54}$.  Each subsequent elaboration led to a further increase - moving from linear to spline-log-liner in volume (S,L), adding three hospital covariates and a patient-volume interaction - culminating in a predictive likelihood increase of $e^{37.96}$ for our (SLI,L) model, which was the very best.

\begin{table}[ht]
\centering
\begin{tabular}{rrrrrrr}
  \hline \hline
Model & (L,C) & (S,L)  & (SLI,L) \\
%Model & (L,C) & (S,C)  & (S,L)  & (S,L) & (SLI,L) \\
  \hline
& 27.54 & 32.13 & 37.96 \\
%& 27.54 & 31.93 & 32.13 & 35.46 & 37.96 \\
   \hline
\end{tabular}
\caption{Out-of-sample log posterior predictive Bayes factor comparisons to the (C,C) model.}
%obtained by MCMC integration) across all hospitals and within each volume-quintile of hospitals. The log-likelihood of (C,C) has been subtracted from each respective  column to make comparisons easier. The first column is the sum of the last five columns.
%Calculations based on the last $5\,000$ samples from the posterior.}
\label{predictiveBF}
\end{table}

\subsection{Out of Sample Calibration of Aggregated Individual\\
Predictions Against Empirical Studies of General Advice}\label{sec:calibrations}

The Bayesian model predicts future { mortality} rates at individual
hospitals. \ For many hospitals, there are too few AMI patients to permit a
serious test of the model's predictions at that hospital.\ Here, we
calibrate the model by comparing its predictions to the general advice
people would otherwise fall back on if individualized predictions were not
available. \ Specifically, we conduct an out-of-sample observational study
checking the general advice; then, we determine which { models 
predict the results of that observational study with reasonable accuracy.  
%If, instead of conducting an observational study of general advice, one used a model to predict the results of that study for the same out-of-sample patients, which model would predict the results of that observational study?

To illustrate, we consider the advice that  one should avoid hospitals that rarely treat AMI}. 
\ As noted earlier, the literature strongly suggests this
is good general advice, although it is difficult to know whether it is good
advice for any single hospital that treats few AMIs --- after all, such a
hospital provides few patients upon which to base a mortality rate.

Using the validation sample that was not used to build the model,
we look at the 20\% of hospitals with the
lowest volume. \ This consisted of 747 low volume (LV) hospitals each with $%
\vol_{h}\leq 23$ AMI in Medicare patients over 3 years, that is, on average, at most 1 AMI patient in
Medicare every 1.57 months. \ In the 6-month Medicare validation sample, there were a
total of 1302 AMI patients at
such hospitals. \ Each such patient was matched to 5 patients at the 20\%
hospitals with the highest Medicare volume, defined to be the 753 hospitals with $%
\vol_{h}\geq 467$ over 3 years, or at least one Medicare AMI patient every 2.34 days. \
In a conventional way (Rosenbaum 2010; Stuart 2010), the matching
combined some exact matching, a caliper on the propensity score, and optimal
matching based on a Mahalanobis distance.  { Here, the propensity score predicted the 
low or high volume categories using a logit model and the covariates in Table  \ref{match}.} 
\ The training sample's estimate
of risk of death was used as an out-of-sample risk or prognostic score in
the validation sample, as suggested by Hansen (2008). \ Hansen's (2007) \texttt{%
optmatch} package in \texttt{R} was used.

Before discussing the results of this comparison, a few words of caution are needed.
In every observational study, there is reason to be concerned that some important
covariate has not been measured, so that a comparison that corrects for measured
covariates will not correctly estimate the effect under study.  This is a genuine
problem in ranking hospitals, and the best solution is to improve the quality of the
data used in ranking hospitals.  The problem affects both the Bayesian model and
the elementary observational comparison that follows in much the same way -- neither
method addresses potential biases from failure to control an unmeasured covariate.
This issue, though both real and important, is less relevant when the focus is
on calibration.  Calibration asks whether the model's predictions agree with
an examination of the data that does not rely upon the model.  The model is
judged calibrated if its predictions are in reasonable agreement with the
elementary observational comparison.  The two answers may agree yet both be
mistaken estimates of the effects of going to low versus high volume hospitals;
that is an important question, but not a question about calibration with the
observed data.

\ Table \ref{match} gives covariate means before and
after matching, together with differences in means as a fraction of the
standard deviation before matching. \ Notably, the patients at low and high
volume hospitals differed substantially prior to matching, but were similar
in matched samples. \ Patients at low volume hospitals were older on average
(84 versus 78 years old), with a higher estimated mean probability of death
based on patient risk factors (.22 versus .13), a higher proportion of
dementia (22\% versus 12\%), a higher proportion with a history of pneumonia
(24\% versus 12\%), and a somewhat higher history of congestive heart failure
(21\% versus 14\%), all factors that generally increase mortality risk. 
Patients at low volume hospitals also had a lower history
of prior percutaneous transluminal coronary angiography (prior PTCA) or stenting
procedures involving the heart (6\% versus 16\%), the history of which generally lowers risk; lower rates of documented artherosclerosis, a cardiac risk factor; and lower rates of
anterior infarction, a factor also generally associated with worse prognosis.
\ Part of the difference in mortality between low
and high volume hospitals reflects the sicker patient population at low
volume hospitals; however, the matching has made an effort to remove this
pattern to the extent that it is visible in measured covariates.

%Patients at low volume hospitals however also had a lower history
%of prior percutaneous coronary intervention (6\% versus 16\%) (the history of which generally lowers risk), lower rates of documented artherosclerosis (a factor that increases risk), and lower rates
%of anterior infarction, a factor that also generally is associated with worse prognosis.

The final two columns of Table \ref{match} give standardized measures of
covariate imbalance before and after matching.  The standardized difference
is the difference in means, low volume minus high volume, divided by the
standard deviation of the covariate prior to matching.  The standard deviation
prior to matching is based on the 1,302 patients at low volume hospitals and
the 50,278 patients at high volume hospitals, pooling the within group
variances with equal weights; see Rosenbaum and Rubin (1983) for
discussion of this measure of covariate imbalance.  For example, the difference in mean ages before
matching, 84.3 versus 77.7, is 80\% of the standard deviation, but after matching
this drops to 1\% of the same standard deviation.  All of the standardized
differences after matching are less than 10\% of the standard deviation, whereas
many were much larger before matching.  In short, the groups look comparable in
terms of measured covariates after matching.

\begin{table}[!htbp]
\centering
\tiny
\begin{tabular}{|l|ccc|cc|}
  \hline
 &\multicolumn{3}{c}{Covariate Means} &  \multicolumn{2}{|c|}{Standardized Differences} \\ \hline
  Patient  Covariates & Low Volume & \multicolumn{2}{c|}{\, High Volume} & Before & After \\
&  & Matched & All & Matching & Matching \\
%Patient & Low & \multicolumn{2}{c|}{\, High \quad\quad\;  High} & Before & After \\
 %Covariate         & Volume & Matched & All & Matching & Matching \\
  \hline
Number of Patients & 1,302 & 6,510 & 50,278 & 1,302 vs 50,278 & 1,302 vs 6,510 \\ \hline
  Prior PTCA & 0.06 & 0.06 & 0.16 & -0.34 & -0.02 \\
  Prior CABG & 0.08 & 0.09 & 0.10 & -0.07 & -0.03 \\
  Heart Failure & 0.21 & 0.23 & 0.14 & 0.19 & -0.04 \\
  Prior MI & 0.12 & 0.12 & 0.09 & 0.10 & -0.01 \\
  Anterolateral MI & 0.05 & 0.05 & 0.10 & -0.20 & 0.00 \\
  Inferolateral MI & 0.07 & 0.07 & 0.15 & -0.27 & 0.00 \\
  Unstable Angina & 0.03 & 0.03 & 0.02 & 0.03 & 0.01 \\
  Chronic Athero. & 0.46 & 0.47 & 0.82 & -0.80 & -0.01 \\
  CPR Failure Shock & 0.05 & 0.06 & 0.06 & -0.01 & -0.02 \\
  Valvular Heart Dis. & 0.12 & 0.13 & 0.20 & -0.21 & -0.02 \\
  Hypertension & 0.66 & 0.67 & 0.71 & -0.13 & -0.04 \\
  Stroke & 0.02 & 0.02 & 0.01 & 0.04 & 0.00 \\
  Cerebrovasc. & 0.06 & 0.06 & 0.04 & 0.07 & -0.01 \\
  Renal Failure & 0.14 & 0.16 & 0.12 & 0.08 & -0.05 \\
  COPD & 0.24 & 0.24 & 0.21 & 0.08 & -0.01 \\
  Pneumonia & 0.24 & 0.23 & 0.12 & 0.32 & 0.03 \\
  Diabetes & 0.34 & 0.35 & 0.36 & -0.04 & -0.02 \\
  Malnutrition & 0.04 & 0.04 & 0.05 & -0.03 & 0.00 \\
  Dementia & 0.22 & 0.23 & 0.12 & 0.29 & -0.02 \\
  Paraplegia & 0.03 & 0.03 & 0.02 & 0.06 & -0.00 \\
  Peripheral Vas. Dis. & 0.06 & 0.07 & 0.06 & 0.01 & -0.01 \\
  Cancer & 0.03 & 0.03 & 0.02 & 0.03 & 0.00 \\
  Trauma & 0.12 & 0.11 & 0.10 & 0.04 & 0.01 \\
  Psych. & 0.04 & 0.03 & 0.02 & 0.11 & 0.07 \\
  Chronic Liver Dis. & 0.01 & 0.01 & 0.01 & -0.03 & -0.00 \\
  Male & 0.41 & 0.40 & 0.54 & -0.27 & 0.02 \\
  Age (years) & 84.3 & 84.2 & 77.7 & 0.80 & 0.01 \\
  logit(Propensity Score) & -2.94 & -2.97 & -4.34 & 1.26 & 0.02 \\
  logit(Risk Score) & -1.47 & -1.48 & -2.14 & 0.85 & 0.01 \\
   \hline
\end{tabular}
\caption{\small Covariate Balance Before and After Matching.  The table compares all
1,302 patients at low volume hospitals to all 50,278 patients at high volume
hospitals (All) and to 6,510 high-volume controls matched 5-to-1 (Matched).  The matching
controlled the listed covariates that described the condition of the patient
prior to admission.  Standardized differences are differences in means in units of a
pooled standard deviation prior to matching.}
\label{match}
\end{table}

\begin{table}[!htbp]
\centering
\footnotesize
\begin{tabular}{|l |c |c |c |}
\hline \hline
  &Low Volume& \specialcell{High Volume\\Matched}& \specialcell{High Volume\\All}\\ \hline
  Observed Mortality & 0.2834 & 0.1982 & 0.1236 \\ \hline
(C,C) & 0.2311 & 0.2158 & 0.1273 \\
  (L,C) & 0.2842 & 0.2069 & 0.1233 \\
%  (Q,C) & 0.2930 & 0.2084 & 0.1242 \\
%  (S,C) & 0.2965 & 0.2079 & 0.1240 \\
%  (L,L) & 0.2816 & 0.2068 & 0.1231 \\
  (S,L) & 0.2965 & 0.2081 & 0.1239 \\
%  (S+H1,C) & 0.2963 & 0.2080 & 0.1240 \\
%  (S+H3,C) & 0.2965 & 0.2078 & 0.1240 \\
%  (S+H1,L) & 0.2964 & 0.2079 & 0.1239 \\
%  (S,L) & 0.2966 & 0.2080 & 0.1239 \\
%  (C+I,C) & 0.2831 & 0.2069 & 0.1235 \\
%  (S+I,C) & 0.2952 & 0.2098 & 0.1239 \\
% (S+H3+I,C) & 0.2952 & 0.2096 & 0.1239 \\
%  (S+I,L) & 0.2956 & 0.2098 & 0.1239 \\
  (SLI,L) & 0.2961 & 0.2103 & 0.1240 \\
   \hline
\end{tabular}
\caption{\small Out-of-sample predicted mortality compared against observed mortality in the
matched observational study of low and high volume hospitals.}
\label{matchResults}
\end{table}

If we did not have the Bayes model for individualized prediction, we might
rely on a matched observational study to provide general advice. \ As seen
in Table \ref{matchResults}, an out-of-sample matched observational study making no use of
the Bayes model records a 30-day mortality rate of 28.3\% at low volume
hospitals, and a mortality rate of 19.8\% among similar matched patients at
high volume hospitals, or an excess mortality of about 8.5\% at low volume
hospitals, which is consistent with what the literature says. \ { If one had the option}, good general
advice would be to seek care for an AMI at a large volume hospital because the
mortality rates are lower for patients who look similar in measured covariates
describing patients prior to admission.

The remainder of Table \ref{matchResults} sets aside the actual out-of-sample mortality,
and instead uses the Bayes models to { predict the mortality of the
very same patients used in the matched observational study.   Let us consider which Bayes models come close to correct predictions, making individual predictions that aggregate to agree with empirically based general advice.}

The (C,C) model is { very inaccurate} in its predictions. \ That model
assumed hospital mortality is independent of volume, and its predictions
agree with its assumptions rather than with the out-of-sample data. \ It
says, incorrectly, that mortality is only slightly elevated at low volume
hospitals, and it also overstates the mortality at high volume hospitals.
\ In sharp contrast, every one of the other models agree with the general advice that risk is elevated at low
volume hospitals. \ Compared to the (C,C) model, their aggregate predictions are much closer to the actual mortality levels of both the low volume hospitals and their matched counterparts at the high volume hospitals.  It is interesting to note that although the overall out-of-sample performance of the (L,C) model was the weakest of the non-(C,C) models in Table \ref{predictiveBF}, its aggregate predictions for the low volume hospitals were better than the rest here.
A second illustration of our general out-of sample calibration approach is presented in Appendix \ref{USNews}.

The lessons of Table \ref{matchResults} are summarized below.

\begin{itemize}
\item It is important to check models against data in a manner that is
capable of judging { their inadequacies}. \ In the current context,
it is difficult to judge that a model is inadequate by predicting the
mortality experience of three patients at a hospital whose total AMI volume
is three patients. \ Something else needs to be done to check such
predictions.

\item It is important to check aspects of models that we actually care
about. \ A spline is an approximation and no one really cares whether it is
true or false; rather, we care whether it is adequate or inadequate as an
approximation for something else that we do really care about. \ In the
current context, we care about model predictions that might both affect
hospital choice and patient mortality. \ In particular, a model that says
low volume hospitals are safe for AMI treatment when they are { not, is} a
model that is failing in a way that we actually care about.

\item A good model for individualized predictions should produce predictions
that are, in aggregate, consistent with sound empirically based general
advice that we would otherwise fall back on in the absence of individualized
predictions. \ The model should correctly predict the results of sound,
out-of-sample studies of general advice that make no use of the model. \ In
Table \ref{matchResults}, { our models do this, and the (C,C) model} does not.

\item It is popular to { associate} Stein's paradox with Bayes inference, but
they actually point in different directions. \ Stein's paradox is a paradox
because it seems to say that shrinkage is never harmful providing at least
three parameters are estimated; however, it actually refers to a very
special situation. \ In contrast, there is nothing in Bayes inference that
suggests one will get the right answer by assuming things that are false or
by fitting the wrong model. \ That the Bayesian, like the classical
frequentist, can be wrong, that both need to look at the data to avoid being
wrong, to look at the data to judge whether their assumptions are reasonable
and their models are adequate --- this need to look at the data --- is a
strength of the Bayesian and classical frequentist perspectives, not a
weakness.
\end{itemize}

\section{Standardized Mortality Rates For Public Reporting}\label{sec:standardization}

%Aside from statistical estimation error, such estimates would then be attributable entirely to hospital qualities rather than patient characteristics, allowing for fairer comparisons.

%For public reporting of hospital performance, it is necessary to standardize hospital mortality rate estimates $\hP_h$ in a way that eliminates the effect of patient case-mix variation.
%For the purpose of a fair comparison of hospital performance measures, all of our estimated hospital mortality rates $P_h$  suffer from the fact that they are determined in part their patient case-mix.  For public reporting, it is desirable to standardize these estimates in a way that eliminates this case-mix variation.
%Two general approaches for this purpose are indirect standardization which is used by Hospital Compare for the (C,C) model estimates, and The first of these is a method of indirect standardization used by Hospital Compare for the (C,C) model, which we ultimately do not recommend.  The second is an alternative method of direct standardization, which we find to be more transparent and more appropriate.

After modeling mortality rates as a function of hospital and patient attributes, the next major step in preparing hospital  rate estimates for public reporting and further analysis, is to remove patient case-mix effects with some form of standardization.   Devoid of differences due to patient risk factors, such estimates allow for much clearer assessments of hospital quality.  In this vein, Hospital Compare employs a form of indirect standardization for their (C,C) model estimates.  As an alternative, we propose a direct standardization approach that more successfully eliminates patient case-mix effects over a wider range of models, and is better calibrated with the actual overall observed mortality rates.   Let us proceed to describe and illustrate these two different approaches in detail.

To begin with, both standardization approaches make use of expected mortality rate estimates for any patient at any hospital.  If the $hj$th patient with covariates $\x_{hj}$ had been treated at hospital $h^*$, under any of our models this rate would be given by
\begin{equation}\label{phji}
p_{h^*}(\x_{hj}) = \logit^{-1}(\alpha_{h^*} + \x_{hj}'\be),
\end{equation}
where $\alpha_{h^*}$ is now the hospital effect and $x_{hj}'\be$ is the usual patient effect.  Note that unless $h^* = h$, this rate is counterfactual, since patient $hj$ was actually treated at hospital $h$.  Note also that for models which include patient-hospital interactions, these interaction covariates would be included as extra columns of $\x_{hj}$ that change as $h^*$ is varied.  For example, in our (SLI,L) model,  $\age_{hj}* \log(\vol_{h^*})$ would be the added interaction covariate for patient $hj$ at hospital $h^*$.  Rather than add cumbersome notation to indicate such dependence of $\x_{hj}$ on $h^*$, for notational simplicity we shall assume that this dependence is implicitly understood from context.

%Both of these approaches make use of the following expression for the expected mortality rate of the $hj$th patient had they been treated at any hospital $h^*$,
%\begin{equation}\label{phji}
%p_{h^*}(\x_{hj,h^*}) = \logit^{-1}(\alpha_{h^*} + \x_{hj,h^*}'\be).
%\end{equation}
%In terms of our previous $p_{hj}$ notation in \eqref{eq:HC1 model} and \eqref{eq:H1 model}, $p_h(\x_{hj,h}) = p_{hj}$.
%The subscript $h^*$ has been appended in $\x_{hj,h^*}$ to allow the covariate values for the $hj$th patient to depend on treatment at  hospital $h^*$ as they would in the case of patient-hospital interaction terms.  For example, in our (SLI,L) model,  $\age_{hj}* \log(\vol_{h^*})$
%$(\age_{hj}-\age_{h^*\cdot})* \log(\vol_{h^*})$
%would be the interaction covariate for patient $hj$ at hospital $h^*$.  Note that unless $h^* = h$, $p_{h^*}(\x_{hj,h^*})$ refers to a hypothetical hospital assignment.   As we will see, averages of such hypothetical rates play a key role in the standardizations.

\subsection{Indirectly Standardized Mortality Rates}\label{sec:indirect}

As discussed by Ash et al.~(2012), as part of its mandate, CMS is charged with quantifying  ``How does this hospital's mortality for a particular procedure compare to that predicted at the national level for the kinds of patients seen for that procedure or condition at this hospital?''
To address this goal, Hospital Compare reports estimates of indirectly standardized hospital mortality rates of the form
\begin{equation}\label{PIS}
P^{IS}_h = (P_h/E_h) \times \bar y,
\end{equation}
where $E_h$ is an average expected 30-day mortality rate for the hospital $h$ patients had they been treated at the``national level'', and
$\bar y$ is the overall average patient-level mortality rate estimate for AMI.  Beyond its intuitive appeal,  strictly speaking,
%the formula \eqref{PIS} for
$P^{IS}_h$ lacks any probabilistic justification as a hospital mortality rate estimate.

For their choice of $E_h$ in conjunction with the (C,C) model, Hospital Compare uses
\begin{equation}\label{eq:EhHC}
E_h^{HC} = \frac{1}{n_h} \sum_j p_{\mu}(\x_{hj}) =  \frac{1}{n_h} \sum_{j=1}^{n_h}\logit^{-1}(\mu_\alpha +\bm{x}_{hj}'\be),
\end{equation}
where for patient $j$ at hospital $h$, $p_{\mu}(\x_{hj})  = \logit^{-1}(\mu_\alpha +\bm{x}_{hj}'\be)$ replaces the hospital effect $\alpha_h$ in $p_{hj}  = \logit^{-1}(\alpha_h +\bm{x}_{hj}'\be)$ with the mean hospital effect $\mu_\alpha$ from \eqref{eq:HC2 model}.  As opposed to $p_{hj}$, $p_{\mu}(\x_{hj})$ treats every patient as if they went to a hospital with the same average mortality effect $\mu_\alpha$.  To estimate $E_h^{HC}$, Hospital Compare uses SAS's GLIMMIX plug-in estimates of $\mu_\alpha$ and $\be$,  (Yale New Haven Health Services Corporation 2014, p.~58, equation (4)).

Although the choice of $E_h^{HC}$ for $E_h$ is reasonable, it is tied directly to the Hospital Compare (C,C) model \eqref{eq:HC2 model} through $\mu_\alpha$.   To extend indirect standardization beyond the (C,C) model, we propose instead a more flexible and general choice of $E_h$ that essentially reduces to $E_h^{HC}$ under the (C,C) model.  Our proposal, which is generally applicable for all the models considered in Section \ref{sec:hrm},  is
\begin{equation}\label{eq:Eh}
E_h = \frac{1}{n_h} \sum_{j=1}^{n_h} \left[\frac{1}{H} \sum_{h^*=1}^H p_{h^*}(\x_{hj})\right],
\end{equation}
where from \eqref{phji}, $p_{h^*}(\x_{hj})$ is the expected mortality rate for the $hj$th patient, had they been treated at hospital $h^*$.   Intuitively, $E_h$ is the average expected mortality rate of all hospital $h$ patients had they hypothetically been treated at all hospitals, $h^*=1,\ldots,H$.   Such averaging over all hospitals removes hospital-to-hospital variation, leaving only the patient attributes to drive the variation of $E_h$.  With this choice of $E_h$, posterior mean Bayes estimates $\hP^{IS}_h$ of $P^{IS}_h$ in \eqref{PIS} are straightforwardly obtained via the MCMC approach in Section \ref{sec:MCMC}.

Making use of the fact that $\logit(\cdot)$ is close to linear in the range of most mortality rates here, insight into how $E_h$ works, as well as its connection with $E_h^{HC}$, is obtained by the approximation	
\begin{equation}\label{PISappprox}
\frac{1}{H} \sum_{h^*} p_{h^*}(\x_{hj}) \approx \logit^{-1}(\bar\alpha + \bar\x_{hj}'\,\be)
%\frac{1}{H} \sum_{h^*} p_{h^*}(\x_{hj,h^*}) \approx \logit^{-1}(\bar\alpha + \bar\x_{hj,\cd}'\,\be)
\end{equation}
where $\bar\alpha = \frac{1}{H} \sum_{h^*} \alpha_{h^*}$ and $\bar \x_{hj} = \frac{1}{H} \sum_{h^*} \x_{hj}$. Recall that $\x_{hj}$ will implicitly vary over $h^*$ when patient-hospital interaction covariates are present.   In models with no patient-hospital interactions, where $\bar \x_{hj} = \x_{hj}$, $E_h$ essentially treats every patient as if they went to a hospital with the same average mortality effect $\bar\alpha$.  In particular, under the (C,C) model where $\bar\alpha \approx \mu_\alpha$ in \eqref{eq:EhHC}, $E_h$ will be nearly identical to the Hospital Compare choice $E_h^{HC}$.  As a computational shortcut, \eqref{PISappprox} also provides a convenient route to obtain fast approximations for general $E_h$.

\begin{figure}[h!]
\centering
\begin{subfigure}{0.24\textwidth}
\includegraphics[width=\textwidth, height = \textwidth]{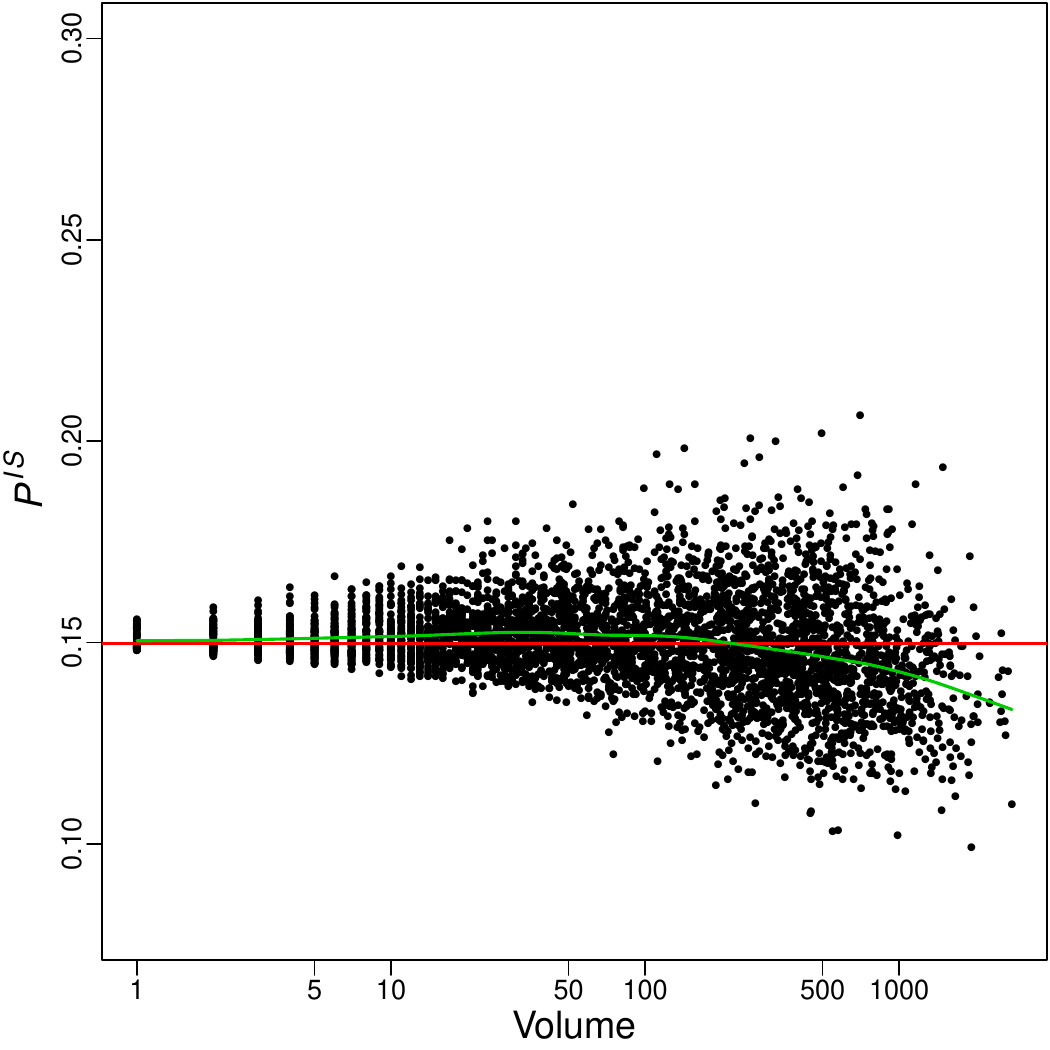}
\caption{\scriptsize $\hP_h^{IS}$ under (C,C)}
\end{subfigure}
\begin{subfigure}{0.24\textwidth}
\includegraphics[width=\textwidth, height = \textwidth]{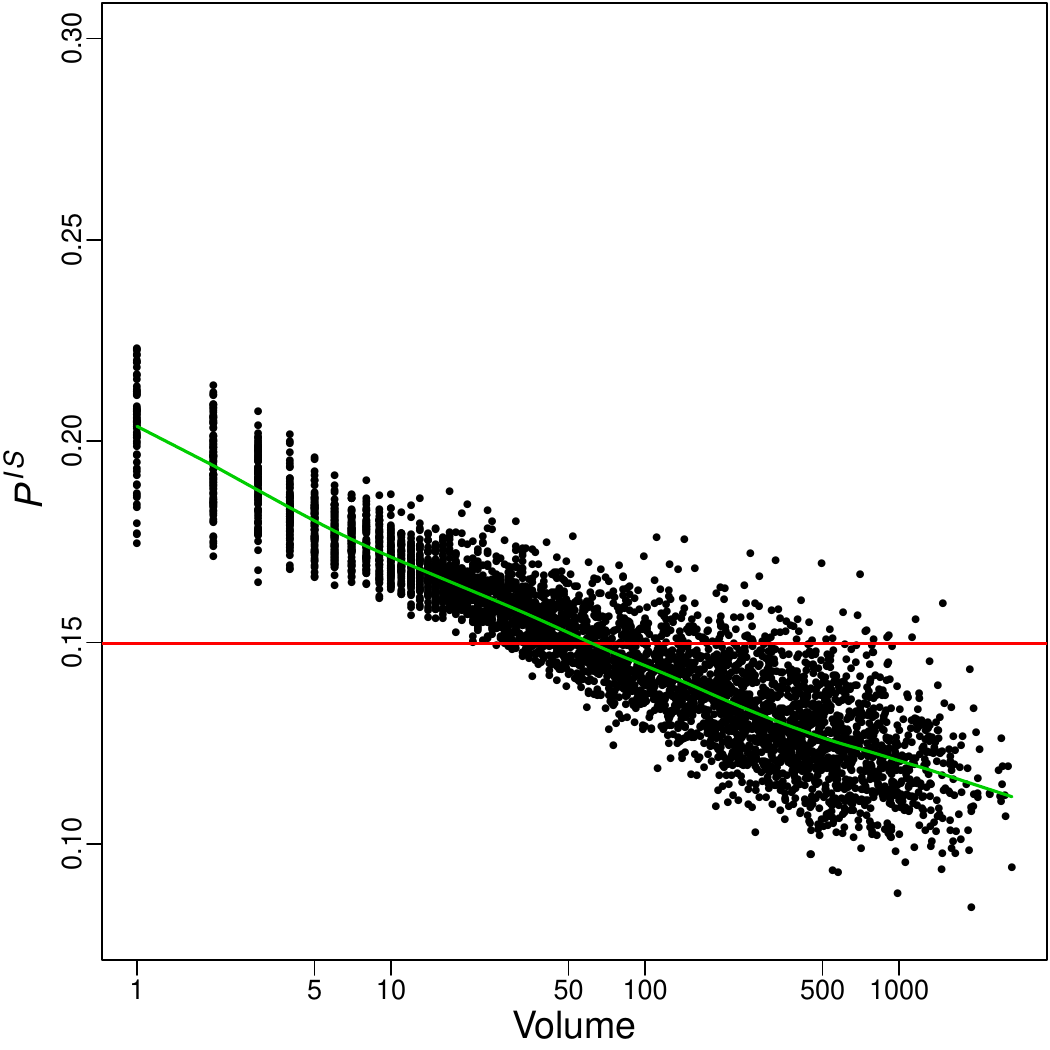}
\caption{\scriptsize $\hP_h^{IS}$ under (L,C)}
\end{subfigure}
\begin{subfigure}{0.24\textwidth}
\includegraphics[width=\textwidth, height = \textwidth]{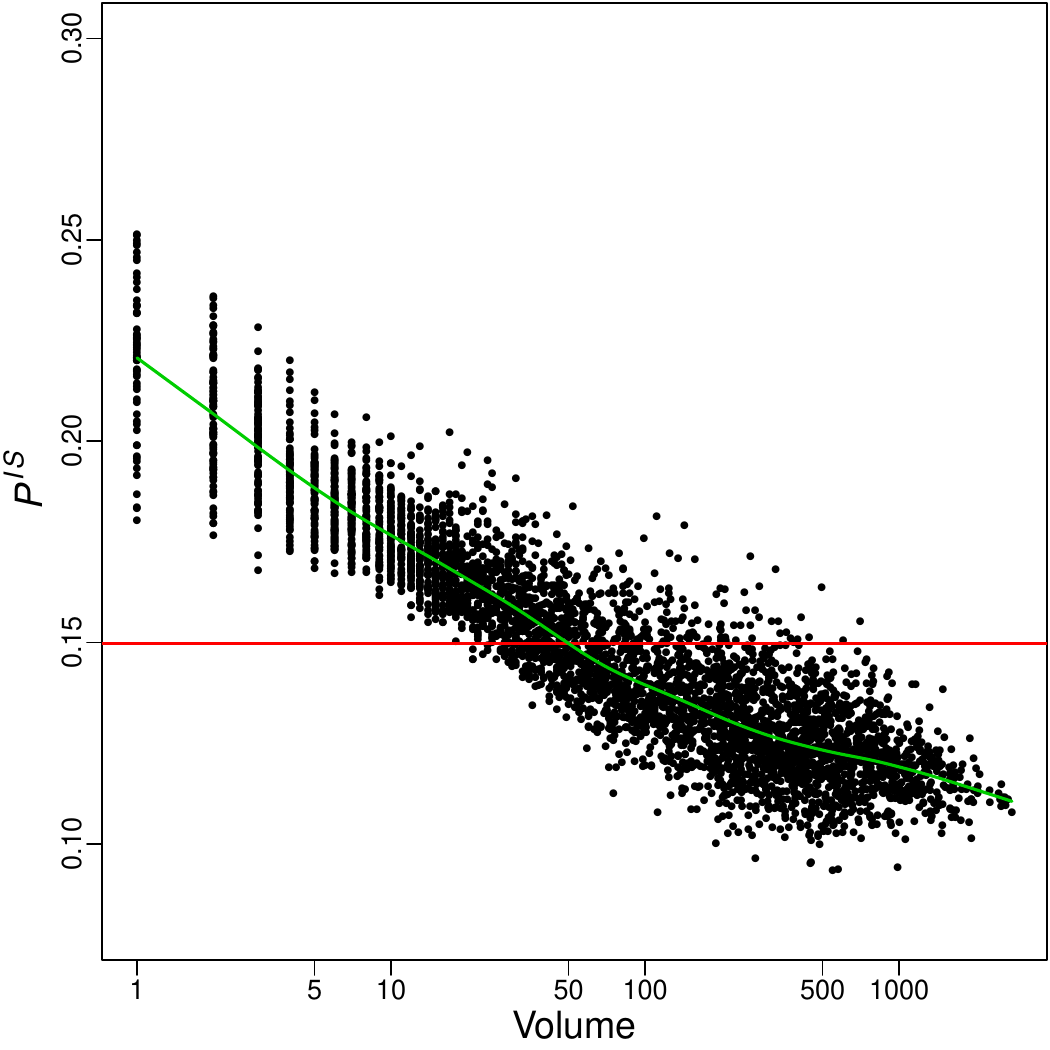}
\caption{\scriptsize $\hP_h^{IS}$ under (S,L)}
\end{subfigure}
\begin{subfigure}{0.24\textwidth}
\includegraphics[width=\textwidth, height = \textwidth]{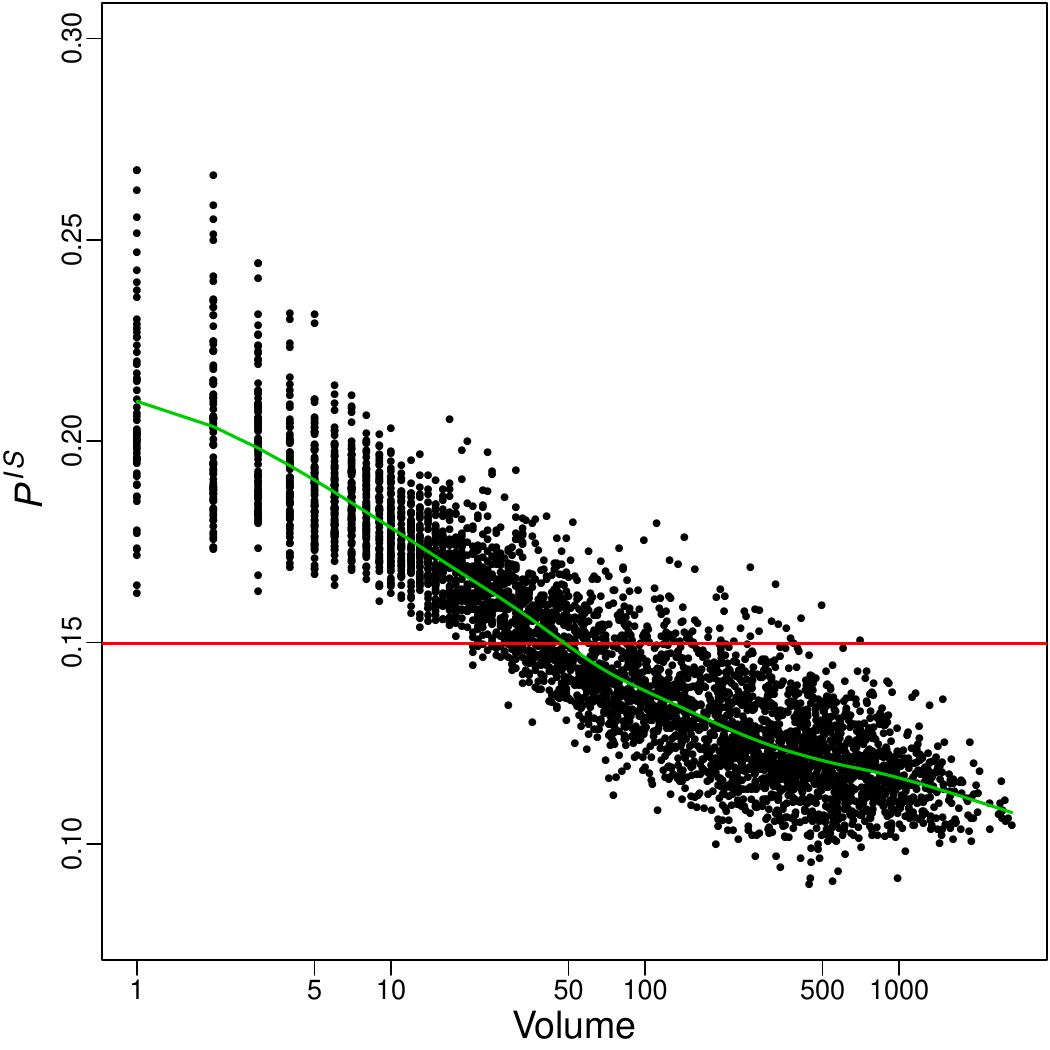}
\caption{\scriptsize $\hP_h^{IS}$ under (SLI,L)}
\end{subfigure}
\centering
\vspace{-.1cm}
\caption{$\hP_h^{IS}$ vs $\vol_h$.}
\label{PISplots}
\end{figure}

To see the effect of the indirect standardization \eqref{PIS} with \eqref{eq:Eh} for the  (C,C), (L,C), (S,L), (SLI,I) models, we apply it to obtain the indirectly standardized mortality rate estimates $\hP^{IS}_h$ in Figures \ref{PISplots}abcd.   In each these plots, $\hP^{IS}_h$ has served to transform the mortality rate estimates $\hP_h$ in Figures \ref{Fig2}abcd into values that much more closely resemble the hospital effect estimates $\ha_h$ in Figures \ref{Fig3}abcd.
%In every case the $\hP_h$
%to the indirectly standardized mortality rate estimates $\hP^{IS}_h$ in Figures \ref{PISplots}abcd.
Beginning with the (C,C) model, the plot of the $\hP^{IS}_h$ in Figure \ref{PISplots}a stands in sharp contrast to the plot of the mortality rate estimates $\hP_h$ in Figure \ref{Fig2}a; note specifically the substantial shrinkage from the two different vertical scales.  The effect of dividing $P_h$ by $E_h$ has left the $\hP^{IS}_h$ estimates looking nearly identical to the hospital effect $\ha_h$ estimates in Figure \ref{Fig3}a.  Evidently, indirect standardization for the (C,C) model has successfully eliminated just about all of the patient case-mix variation from the $\hP_h$ estimates.  Notice also that, unlike the abstract scale of the $\ha_h$'s, standardization has left the $\hP^{IS}_h$'s on a mortality scale which makes them easier to interpret and understand.
%As a result, these $\hP^{IS}_h$ are essentially just a rescaling of the $\alpha_h$  to a mortality rate scale.
%Unfortunately, the inadequacy of the $\alpha_h$'s in terms of capturing the substantial hospital attribute effects as we have seen, renders these $\hP^{IS}_h$  to be unreliable mortality rate estimates.

\begin{figure}[h!]
\centering
\begin{subfigure}{0.24\textwidth}
\includegraphics[width=\textwidth, height = \textwidth]{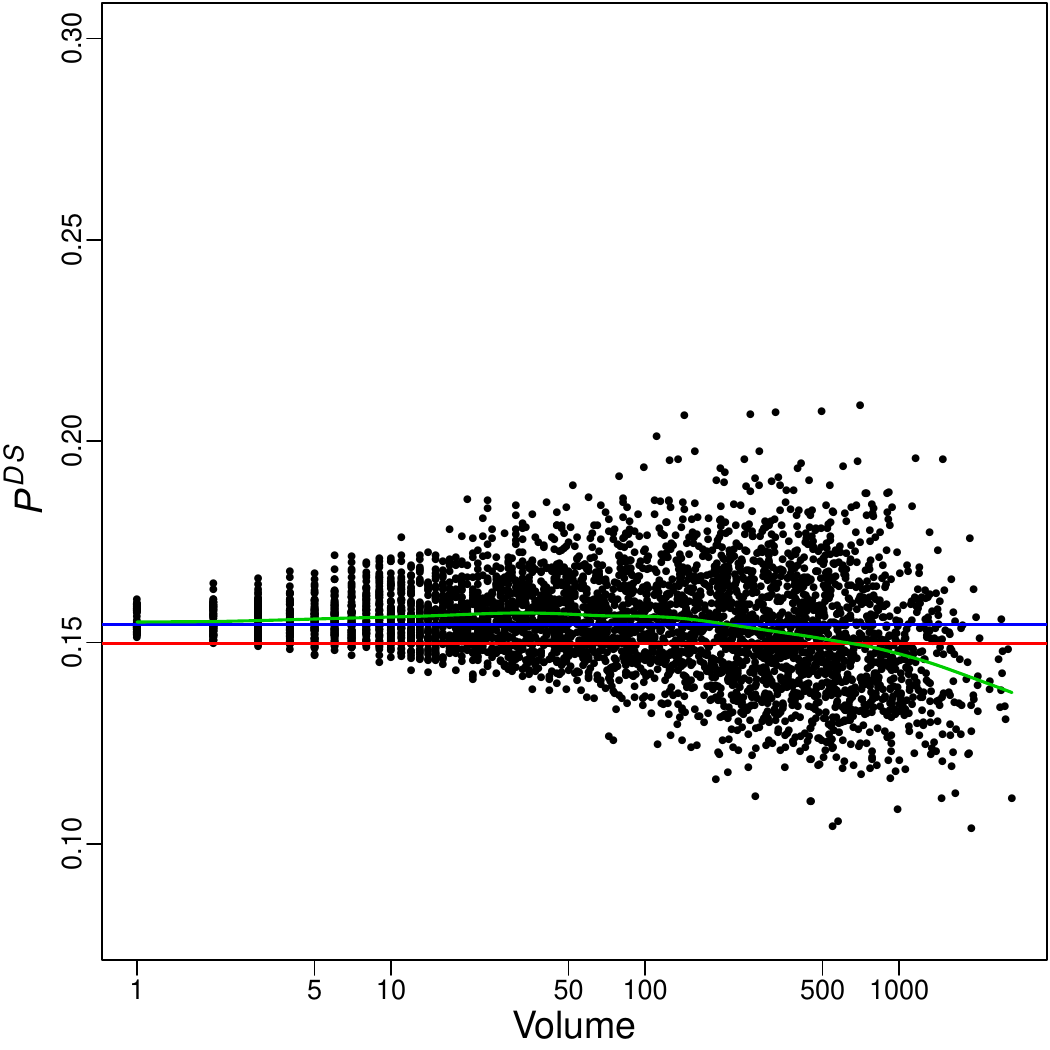}
\caption{\scriptsize $\hP_h^{DS}$ under (C,C)}
\end{subfigure}
\begin{subfigure}{0.24\textwidth}
\includegraphics[width=\textwidth, height = \textwidth]{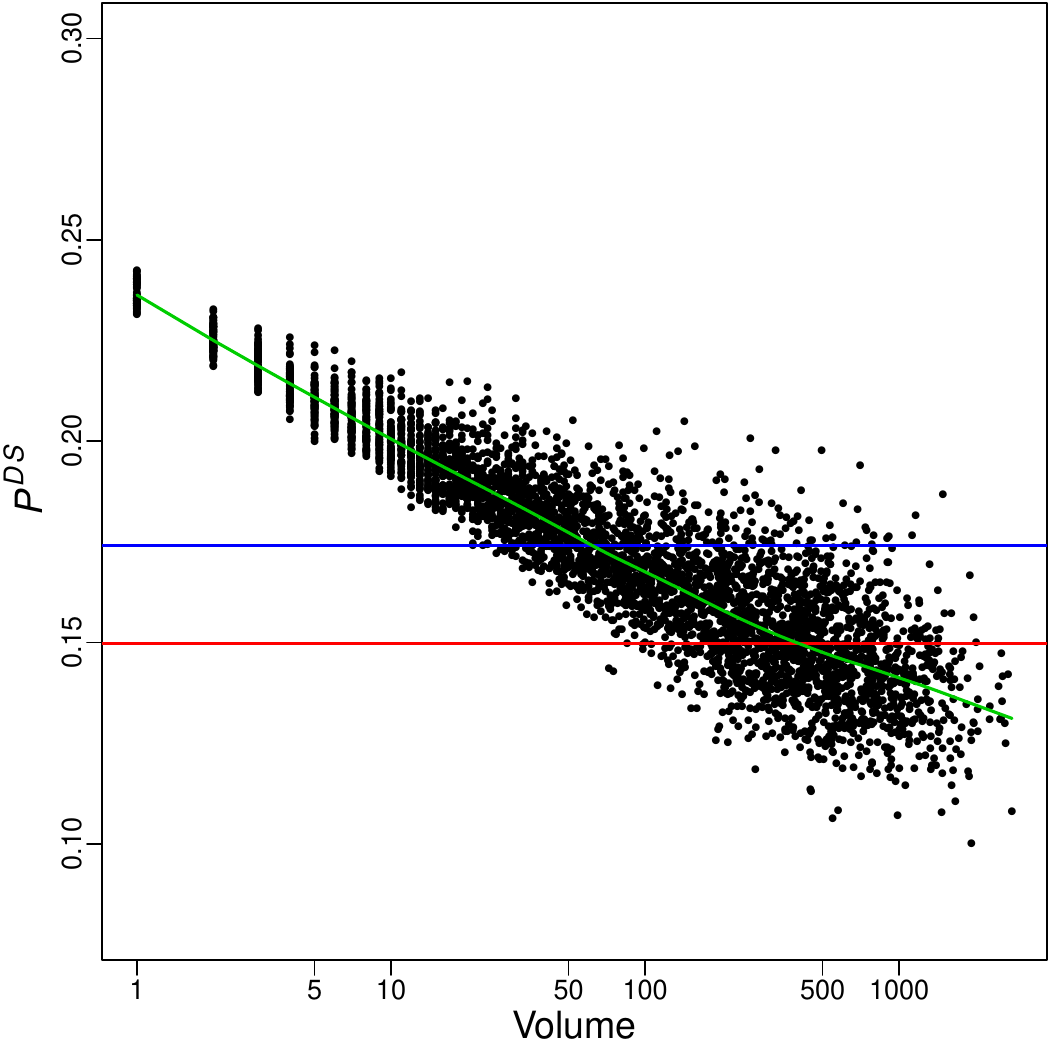}
\caption{\scriptsize $\hP_h^{DS}$ under (L,C)}
\end{subfigure}
\begin{subfigure}{0.24\textwidth}
\includegraphics[width=\textwidth, height = \textwidth]{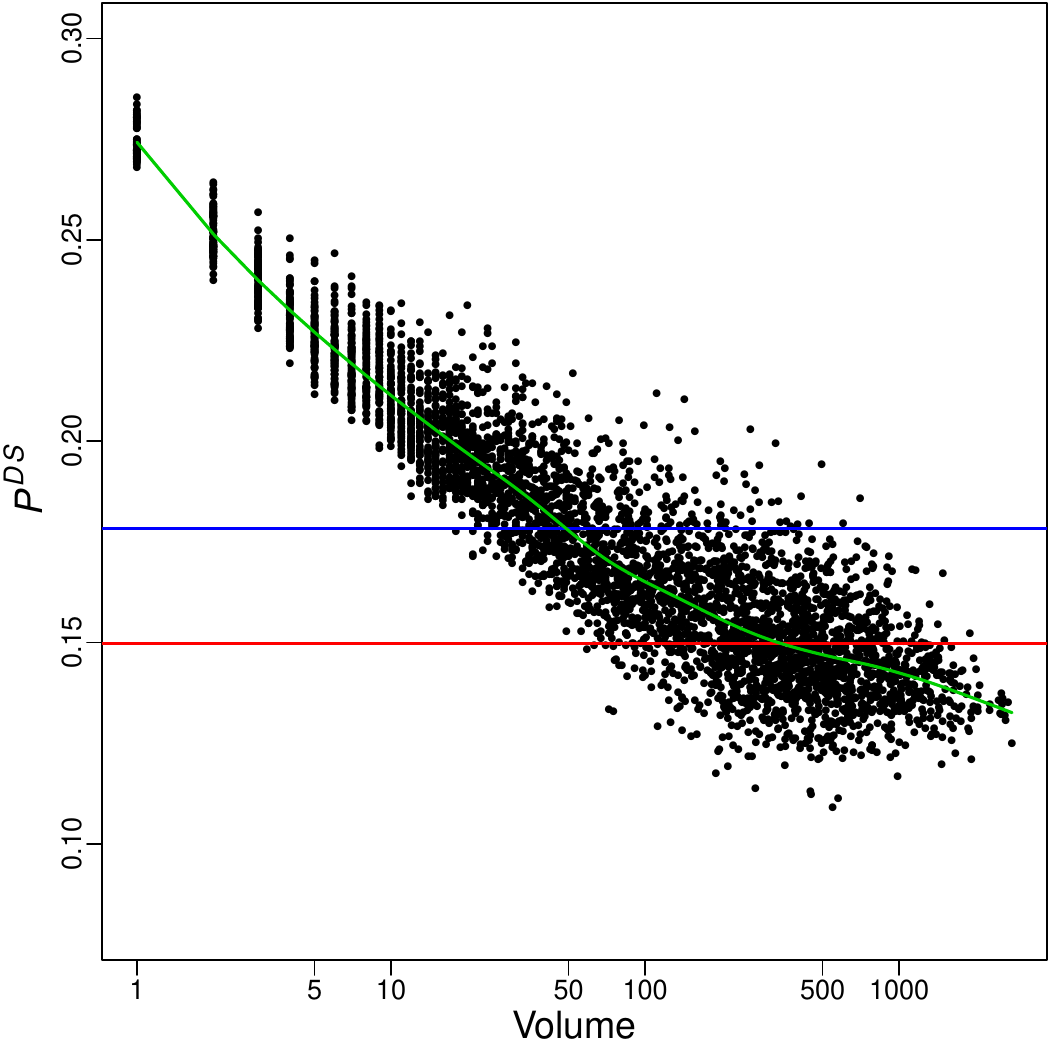}
\caption{\scriptsize $\hP_h^{DS}$ under (S,L)}
\end{subfigure}
\begin{subfigure}{0.24\textwidth}
\includegraphics[width=\textwidth, height = \textwidth]{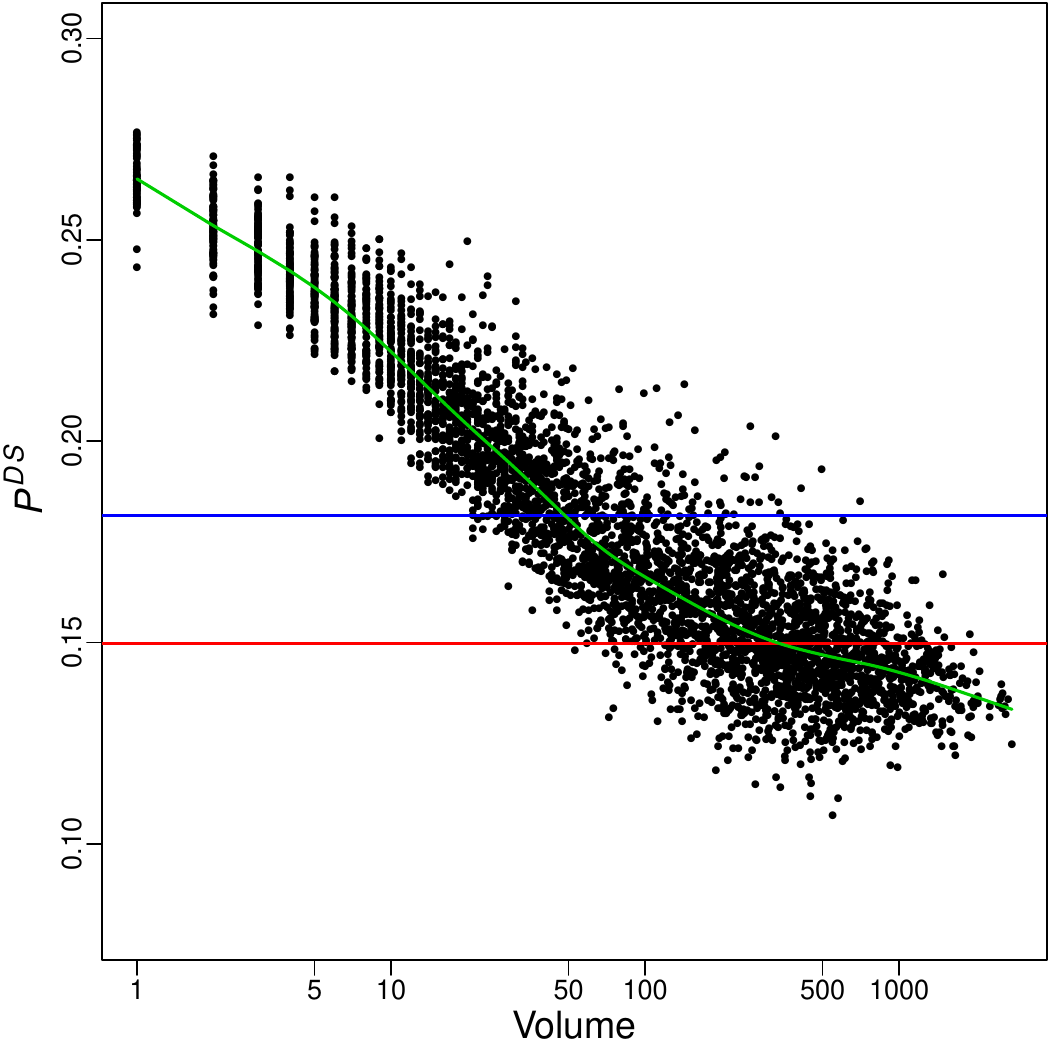}
\caption{\scriptsize $\hP_h^{DS}$ under (SLI,L)}
\end{subfigure}
\centering
\vspace{-.1cm}
\caption{$\hP_h^{DS}$ vs $\vol_h$.}
\label{PDSplots}
\end{figure}

Turning to the $\hP^{IS}_h$ estimates for the (L,C), (S,L) and (SLI,L) models in Figures \ref{PISplots}bcd, we see that these $\hP^{IS}_h$'s have also shrunk the $\hP_h$'s to resemble their corresponding $\ha_h$'s in Figures \ref{Fig3}bcd.  For each model, they are substantially larger at the low volume hospitals, with a clear downward sloping trend.  However, unlike the low volume $\ha_h$'s in Figures \ref{Fig3}bcd which are tightly concentrated around their trend averages, the low volume $\hP^{IS}_h$ values instead fan out dramatically when $\vol_h$ is low.  Insight into this phenomenon is obtained by considering the ratio of $P_h$ to the $E_h$ approximation \eqref{PISappprox},
\begin{equation}\label{eq:PEproblem}
\frac{P_h}{E_h} \approx \frac{\sum_{j=1}^{n_h}\logit^{-1}(\alpha_h +\x_{hj}'\,\be)}{\sum_{j=1}^{n_h}\logit^{-1}(\bar\alpha + \bar \x_{hj}'\,\be)}.
\end{equation}
When the $\alpha_h$ are far from $\bar\alpha$, as occurs for low volume hospitals under the (L,C), (S,L) and (SLI,L) models, the variation of \eqref{eq:PEproblem} will reflect the variation of patient attributes $\x_{hj}'$ across hospitals, especially when $n_h$ is small.   Thus, under these models, the increased variation of the $\hP^{IS}_h$ at the low volume hospitals is an artifact of patient-mix variation rather than of the $\ha_h$ hospital quality variation.   Although this phenomenon does not occur under the (C,C) model where all the low volume $\ha_h$'s are close to $\bar\alpha$, this is precisely where the $\ha_h$'s estimates were seen to be miscalibrated.  For all but the discredited (C,C) model, indirect standardization fails to achieve its goal of eliminating the effect of patient case-mix variation from mortality rates.  Furthermore, as we'll see in the next section, for every one of our models, indirectly standardized mortality rates systemically underestimate actual mortality rates.   Such indirect standardization cannot be recommended for public reporting.

\subsection{Directly Standardized Mortality Rates}\label{sec:direct}

%Direct standardization of $P_h$ is obtained by the average mortality of all patients at hospital $h$, in contrast to indirect standardization which adjusts  $P_h$ by the average mortality of hospital $h$ patients over all hospitals.

Directly standardized mortality rates, an alternative to $P^{IS}_h$, directly eliminate patient-mix effects by averaging the mortality rates of all patients had they (hypothetically) been been treated at hospital $h$. Denoted $P^{DS}_h$, such rates are given by
\begin{equation}\label{PDS}
P^{DS}_h =   \frac{1}{N} \sum_{h^* = 1}^H  \sum_{j = 1}^{n_{h^*}} p_h(\x_{h^*j}),
\end{equation}
where  $p_h(\x_{h^*j})$ in \eqref{phji} is the $h^*j$th patient's mortality rate had they been treated at hospital $h$, and $N =  \sum_{h^* = 1}^H n_{h^*}$ is the total number of patients.  Because every $P^{DS}_h$ is an average over the same set of all patients, there can be no patient-mix differences between them.  Note how $P^{DS}_h$ is complementary to $P^{IS}_h$, which instead adjusts $P_h$ by $E_h$, the average mortality rate of hospital $h$ patients had they (hypothetically) been treated at all hospitals.  Posterior mean Bayes estimates $\hP^{DS}_h$ of $P^{DS}_h$ in \eqref{PDS} are straightforwardly obtained via the MCMC approach in Section \ref{sec:MCMC}.
%Note that $\hP^{DS}_h$ can also be interpreted as the expected hospital $h$ mortality rate had patients been randomly allocated to hospitals.

Further insight into $P^{DS}_h$ is obtained by the approximation
\begin{equation}\label{PDSappprox}
P^{DS}_h \approx \logit^{-1}(\alpha_h +\bar\x_{\cd\cd}'\,\be)
\end{equation}
where $\bar\x_{\cd\cd} = \frac{1}{N} \sum_{h^*} \sum_j \x_{h^*j}$, which follows from the fact that $\logit(\cdot)$ is close to linear in the range of most mortality rates here.  Thus, $P^{DS}_h$ may be regarded as the expected mortality rate at hospital $h$ of a patient with average $\bar\x_{\cd\cd}$ attributes.  In models with no patient-hospital interactions (where $\x_{h^*j}$ does not vary with $h$), $\bar\x_{\cd\cd}$ is simply the mean of $\x_{h^*j}$ over all patients.
%As a computational shortcut, \eqref{PDSappprox} also provides a convenient route to obtain fast approximations for $\hP^{DS}_h$ with large data sets.

Figures \ref{PDSplots}abcd plots directly standardized mortality rate $\hP^{DS}_h$ estimates by $\vol_h$ for each of the (C,C), (L,C), (S,L) and (SLI,L) models.  As with $\hP^{IS}_h$, $\hP^{DS}_h$ has served to transform the $\hP_h$ in Figures \ref{Fig2}abcd into values that much more closely resemble the hospital effect estimates $\ha_h$ in Figures \ref{Fig3}abcd, but that remain on a more meaningful mortality scale.  Up to this rescaling, the $\hP^{DS}_h$ estimates under the (C,C) model are also virtually identical to their $\ha_h$'s, and
%the $\hP^{DS}_h$ estimates
under the (L,C) and (S,L) models now appear much more similar to their $\ha_h$'s.   No longer fanning out at the low volume hospitals as the $\hP^{IS}_h$ rates did, these $\hP^{DS}_h$'s have more successfully eliminated patient case-mix variability.  Indeed, with linear correlations of $0.9967$, 0.9978, 0.9975 under these three models, $\hP^{DS}_h$ serves as meaningfully interpretable, nearly linear rescaling of the $\ha_h$ hospital effect estimates.   For the (SLI,L) model, the $\hP^{DS}_h$'s are more dispersed than the $\ha_h$'s, tracking them less closely with a correlation of 0.9906. This is not surprising because when patient-hospital interactions are present,  as we saw in Section \ref{sec:SLII}, the $\ha_h$'s no longer entirely capture hospital effects.  Evidently, the $\hP^{DS}_h$'s are a much more effective reflection of actual hospital effects, and one that puts them on a natural mortality scale.

It is concerning to see that the overall level of the $\hP^{IS}_h$ rates in Figures \ref{PISplots}abcd is systematically lower than the overall level of the $\hP^{DS}_h$ rates in Figures \ref{PDSplots}abcd.  To understand what is going on, we have put two horizontal lines on each of the $\hP^{DS}_h$ plots.  For each model, the higher line is the simple average of the $\hP^{DS}_h$ rates, while the lower line is the same observed average mortality rate $\bar y =0.1498$ obtained by averaging over all patients in our data. By the indirect standardization construction, the simple average of the $\hP^{IS}_h$ rates will always equal the average patient mortality rate $\bar y$, as is evident in their plots.  In fact, the $\hP^{IS}_h$ rates understate the poor performance of the worst hospitals by overstating the risk faced by the typical patient as the following discussion will show.

As we have seen, low volume hospitals have higher than typical risk, but treat relatively few patients; therefore, the (unweighted) average risk over hospitals is much higher than the average risk faced by patients.  Saying the same thing differently, a random patient likely went to a larger volume hospital -- that's what it means to be a larger volume hospital -- but a random hospital is unlikely to have very high volume -- that's what it means to be a random hospital.  The expected risk $E_h$ used by $\hP^{IS}_h$ in approximation \eqref{PISappprox} is essentially obtained by substituting the average hospital effect $\bar\alpha$ for the specific hospital effect $\alpha_h$ in the various expressions for patient mortality rates.  This average hospital effect $\bar\alpha$ describes the typical hospital, not the hospital that treats the typical patient.  Therefore, $E_h$ is too high: it describes the risk that would be relevant if patients picked hospitals at random with equal probabilities, but they don't; rather, they tend to go to larger volume hospitals with lower risk.  This one problem with $\hP^{IS}_h$ could be fixed with a patch: instead of using the unweighted $\bar\alpha$, one could average over hospitals with weights proportional to their volumes; that would describe the risk faced by the typical patient. However, this would still not resolve the patient-mix variability shortcomings of $\hP^{IS}_h$ discussed in the previous section.  Overall, we recommend $\hP^{DS}_h$ over $\hP^{IS}_h$ as a more reliable standardized mortality rate report, especially for the model elaborations that we have proposed.
%The overall level of the $\hP^{DS}_h$ values compared to the $\hP^{IS}_h$ values also reveals a further shortcoming of the $\hP^{IS}_h$ rates.

%Finally, we note that it is the volume weighted average of the  $\bar P^{DS}$ that will be (very close to) $\bar y$.

\section{Learning from Directly Standardized Mortality Rates}\label{sec:rateinference}

\subsection{Mortality Rate Uncertainty}\label{sec:uncertainty}

For each hospital $h$, each of the Bayesian models induces a posterior distribution $\pi(P^{DS}_h \C \y)$ on its directly standardized mortality rate.  Averages of MCMC simulated samples from each of these posteriors produced the posterior mean estimates $\hP^{DS}_h$ plotted in Figures \ref{PDSplots}abcd.  As is strikingly evident in every plot, the hospital-to-hospital variation of these $\hP^{DS}_h$ values is smallest at the low volume hospitals, gradually increasing as volume increases.  This is a consequence of the stronger shrinkage of the $\ha_h$ estimates to their means by each of the random effects models.  

However, this observed hospital-to-hospital variation of the $\hP^{DS}_h$ values should not be confused with the posterior uncertainty of the accuracy of each estimate, which can be much larger.  This uncertainty is captured by the full posterior distribution of each $\hP^{DS}_h$ value, and can be conveyed with interval estimates based on the simulated posterior samples for each hospital.  For example,  95\% intervals may be obtained from the 2.5\% and 97.5\% sample quantiles.  Such intervals can be used to provide a direct assessment of the reliability of each $\hP^{DS}_h$ estimate, a more informative alternative to the practice of eliminating estimates because of small sample sizes or other reliability adjustments, as advocated for example by Dimick et.~al.~(2010).

\begin{figure}[h!]
\centering
\begin{subfigure}{0.24\textwidth}
\includegraphics[width=\textwidth, height = \textwidth]{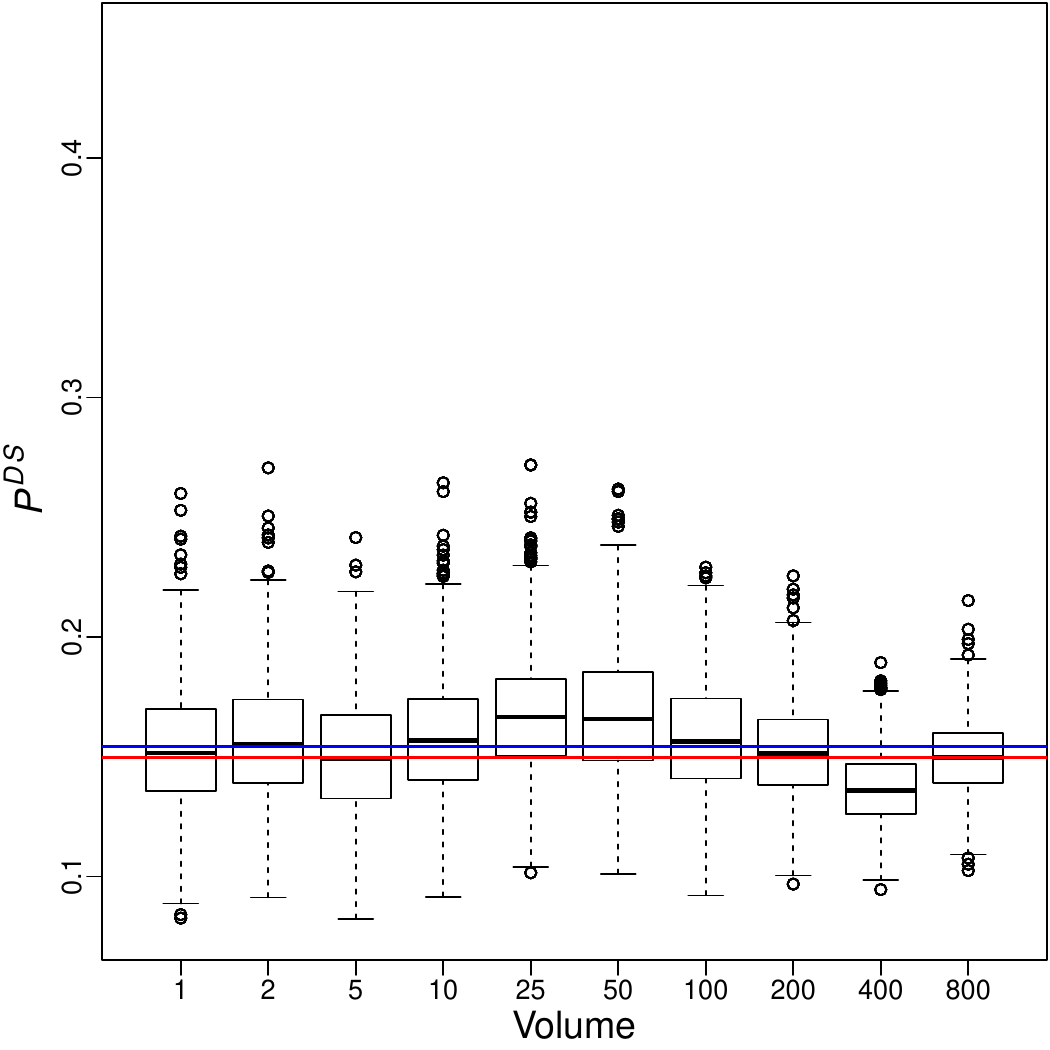}
\caption{(C,C) model}
\end{subfigure}
\begin{subfigure}{0.24\textwidth}
\includegraphics[width=\textwidth, height = \textwidth]{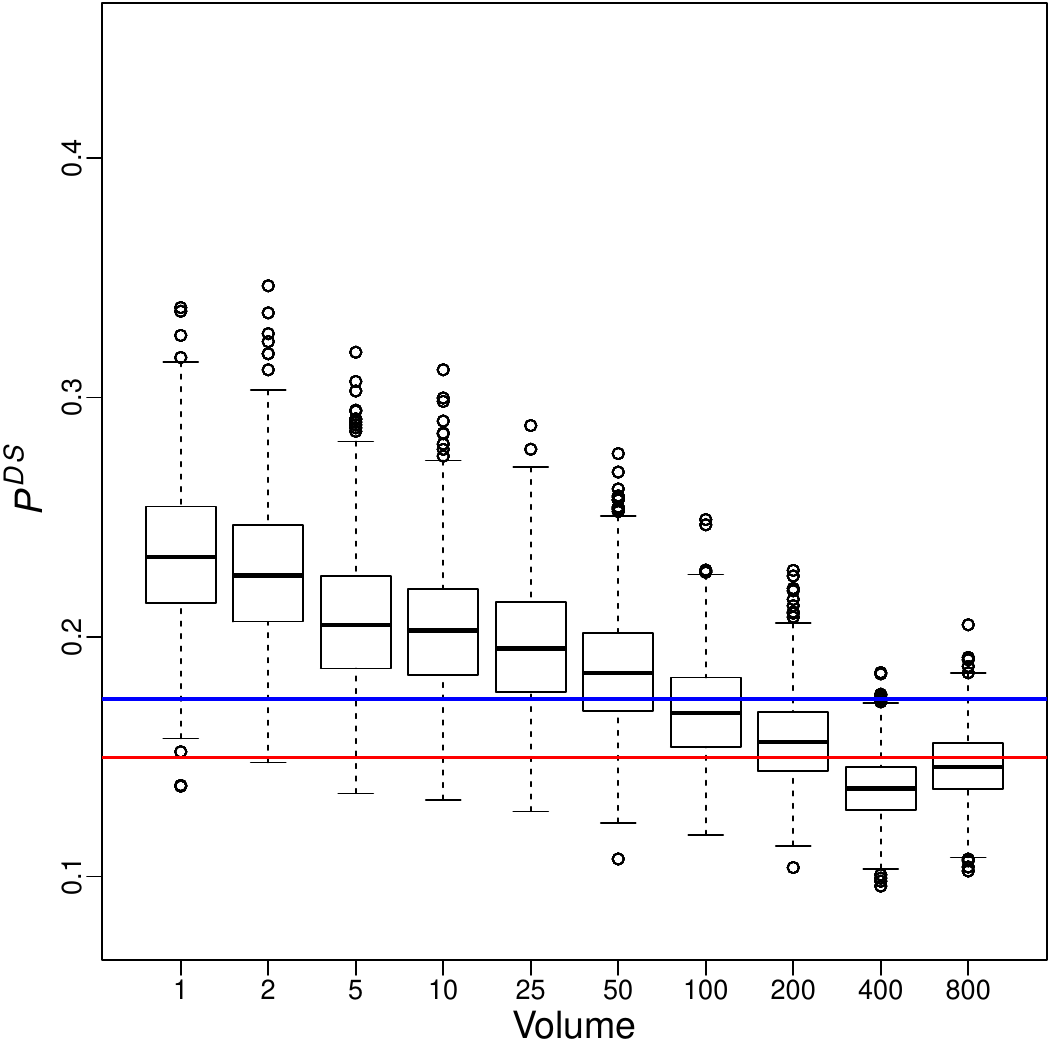}
\caption{(L,C) model}
\end{subfigure}
\begin{subfigure}{0.24\textwidth}
\includegraphics[width=\textwidth, height = \textwidth]{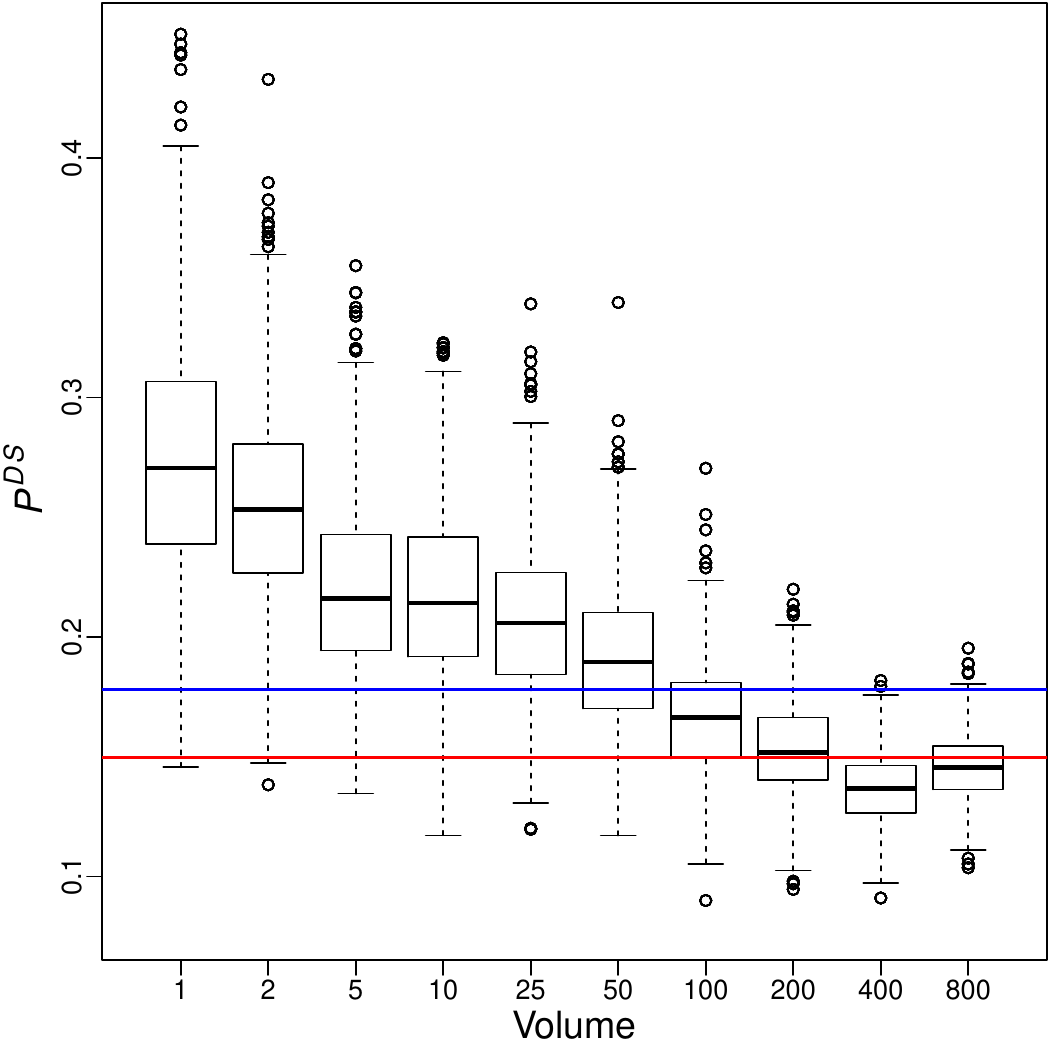}
\caption{(S,L) model}
\end{subfigure}
\begin{subfigure}{0.24\textwidth}
\includegraphics[width=\textwidth, height = \textwidth]{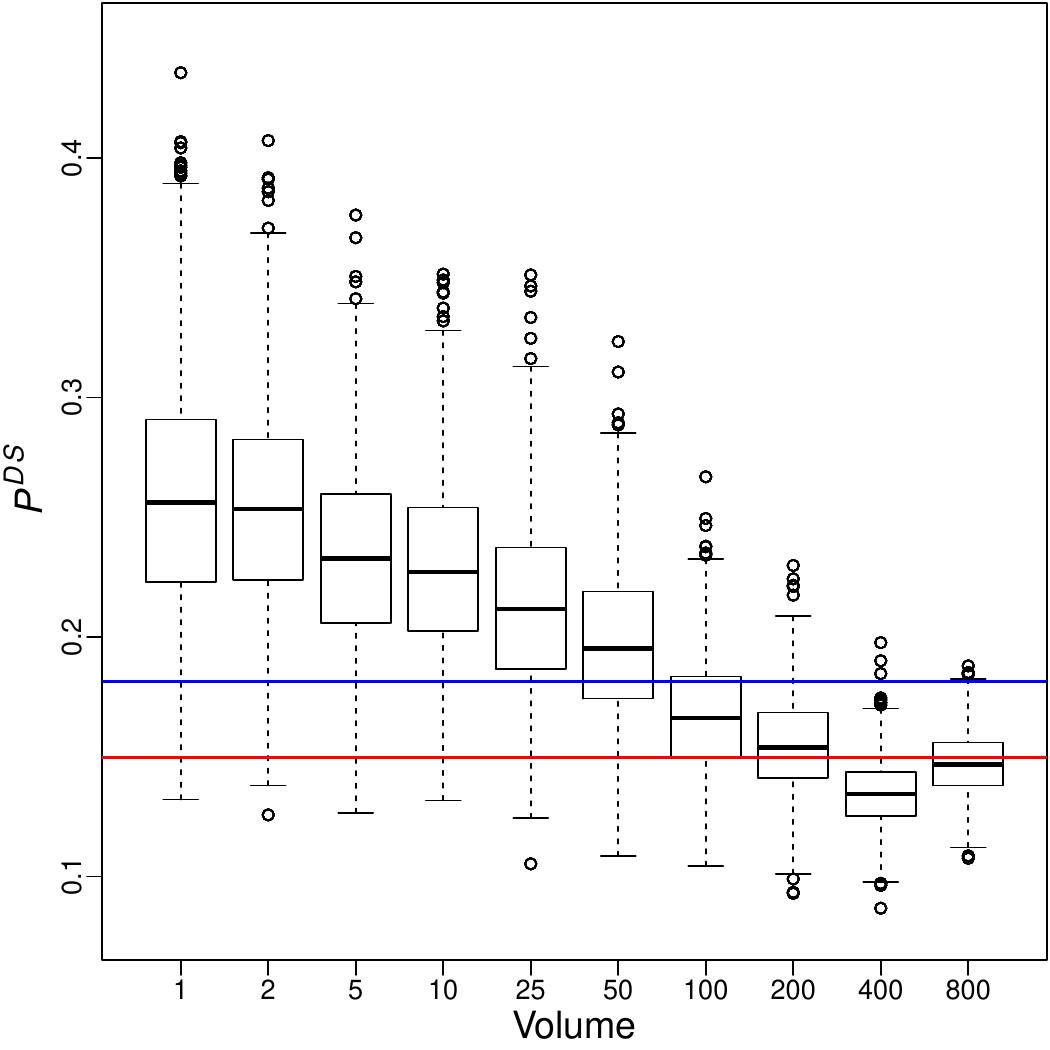}
\caption{(SLI,L) model}
\end{subfigure}
\centering
\caption{$\hP_h^{DS}$ posterior uncertainty at 10 hospitals of varying volume.}
\label{UIplots}
\end{figure}

To illustrate this uncertainty, Figures \ref{UIplots}abcd  display boxplots of simulated posterior samples from the $\pi(P^{DS}_h \C \y)$ distributions for 10 typical hospitals of sizes $\vol_h = 1, 2, 5, 10, 25,50, 100,$ $200, 400, 800$ under the (C,C), (L,C), (S,L) and (SLI,L) models, respectively.  Notice how the profile of mortality rate uncertainty under the (C,C) model stands out from the rest.  Under the (C,C) model, mortality rate uncertainty is hardly related to volume, the level and spread of the posterior distributions being roughly the same at the smaller  $\vol_h$ values.  
%Thus the (C,C) model attaches the roughly the same uncertainly to all mortality rates estimates, regardless of the volume of the hospital. 
But under our models, especially the fully emancipated (S,L) and (SLI,L) models, both the level and spread of the posterior distributions are higher for the low volume hospitals, decreasing steadily as $\vol_h$ increases.   

{ To further illustrate the informative value of mortality rate posterior reporting under our approach,  Silber et.~al.~(2016) apply a hospital attribute enhanced variant of our (L,C) model to estimate mortality rates at five hospitals in Chicago, IL.  Consistent with Figure \ref{UIplots}, mortality rate posteriors for the smaller volume hospitals there are seen to be both higher and more diffuse under the enhanced (L,C) model than under the (C,C) model.}

Thus, the systematically higher mortality rate estimates at low volume hospitals under our models, are also each less precise or reliable in the sense that there is more uncertainty about their accuracy. 
%inversely related to $\vol_h$, the level and spread of the posterior distributions being highest at low volumes and then decreasing as $\vol_h$ increases.  With more uncertainty convey that the low volume estimates are more   less reliable in the sense that a larger  
%Our hierarchical random effects model elaborations in Section \ref{sec:hrm}, have also served to emancipate the spread of these posteriors to better characterize post data uncertainty. 
Figure \ref{UIplots} reminds us that in judging the mortality rate estimate for a given hospital with our models, consideration must given both to the point estimate and its uncertainly as reflected by the posterior distribution.  In particular, if a small hospital was plausibly excellent, an analysis of this form would convey that such excellence is plausible.

\subsection{Hospital Classification by Mortality Rates}

Do many or few hospitals have high mortality rates compared to national rates?  The 95\% credibility intervals for $P^{DS}_h$ described in the previous section can be used to classify hospitals as Low, Average, or High mortality according to whether its 95\% interval is entirely below, intersects or is entirely above the overall average morality rate of 15\%.  

Table \ref{HClass} provides these classifications for the (C,C) and the (SLI,L) models.  Overall, the (C,C) model categorizes most hospitals, 4333, as Average mortality, with only 33 as Low and 30 as High.  Our (SLI,L) model is much more discriminating, classifying 3310 hospitals as Average, with 58 as Low and 1028 as High.  However, much of this discrimination between Average and High mortality hospitals occurs at the lower volume quartile hospitals.  Whereas all 1116 low volume quartile hospitals are classified as Average mortality by the (C,C) model, the (SLI,L)) has recategorized 906 of these as High mortality.  For the higher volume quartile hospitals, the (C,C) model classifies 32 as Low 20 as High, whereas the (SLI,L) model has many more, 57, as Low, and only 4 as High.  The full cross classification leading to Table \ref{HClass} appears as Table \ref{HCClass} in Appendix \ref{CrossClass}. There we see, for example, that of the 57 higher volume hospitals classified as Low mortality by (SLI,L), 28 were classified as Average  by (C,C).  In contrast, of the 32 higher volume hospitals classified as Low mortality by (C,C),  only 4 were classified as Average by (SLI,L).

\iffalse
\begin{table}[!htbp]
\footnotesize
\centering
(a) All Hospitals\\
\begin{tabular}{|l|| c | c| c ||c|}
\hline
 \specialcell{Counts\\(\%)} & Low & Average & High & Total \\ \hline  \hline
(C,C) & 33 & 4333 & 30 & 4396 \\
& (0.752)&(98.57)& (0.68)& \\ \hline
(SLI,L) & 58 & 3310 & 1028 & 4396 \\
& (1.32)&(75.30)& (23.28)& \\ \hline
\end{tabular}\\
\vspace{.5cm}
(b) Lower Quartile Hospitals by Volume\\
\begin{tabular}{|l|| c | c| c ||c|}
\hline
 \specialcell{Counts\\(\%)} & Low & Average & High & Total \\ \hline  \hline
(C,C)  & 0 & 1116 & 0 & 1116\\
& (0.00)&(100.00)& (0.00)& \\ \hline
(SLI,L) & 0 & 210 & 906 &1116 \\
& (0.00)&(18.82)& (81.18)&\\ \hline
\end{tabular}\\
\vspace{.5cm}
(c) Upper Quartile Hospitals by Volume\\
\begin{tabular}{|l|| c | c| c ||c|}
\hline
 \specialcell{Counts\\(\%)} & Low & Average & High & Total \\ \hline  \hline
(C,C) & 32 & 1047 & 20 & 1099\\
& (2.91)&(95.27)& (1.82) &\\ \hline
(SLI,L)  & 57 & 1038 & 4 & 1099 \\
& (5.19)&(94.45)& (.036)& \\ \hline
\end{tabular}
\caption{Hospital Classifications by Mortality Rates.}
\label{HClass}
\end{table}
\fi

\begin{table}[!htbp]
\tiny
\centering
\hspace{2.5cm} All Hospitals \hspace{2.3cm}  Lower Volume Quartile Hospitals \hspace{.8cm} Upper Volume Quartile Hospitals\\
\begin{tabular}{|l|| c | c| c |c| | c | c| c |c|| c | c| c |c||}
\hline
 \specialcell{Counts\\(\%)} & Low & Average & High & Total & Low & Average & High & Total & Low & Average & High & Total\\ \hline  \hline
(C,C) & 33 & 4333 & 30 & 4396 & 0 & 1116 & 0 & 1116 & 32 & 1047 & 20 & 1099\\
& (0.75)&(98.57)& (0.68)& & (0.00) &(100.00)& (0.00)&&(2.91)&(95.27)& (1.82) & \\ \hline
(SLI,L) & 58 & 3310 & 1028 & 4396 & 0 & 210 & 906 &1116 & 57 & 1038 & 4 & 1099 \\
& (1.32)&(75.30)& (23.38)& & (0.00)&(18.82)& (81.18)& & (5.19)&(94.45)& (0.36)&\\ \hline
\end{tabular}\\
\vspace{.5cm}
\caption{Hospital Classifications by Mortality Rates.}
\label{HClass}
\end{table}

\subsection{The Influence of Hospital Attributes}\label{sec:attributes}

\begin{figure}[h!]
\centering
\begin{subfigure}{0.28\textwidth}
\includegraphics[width=\textwidth, height = \textwidth]{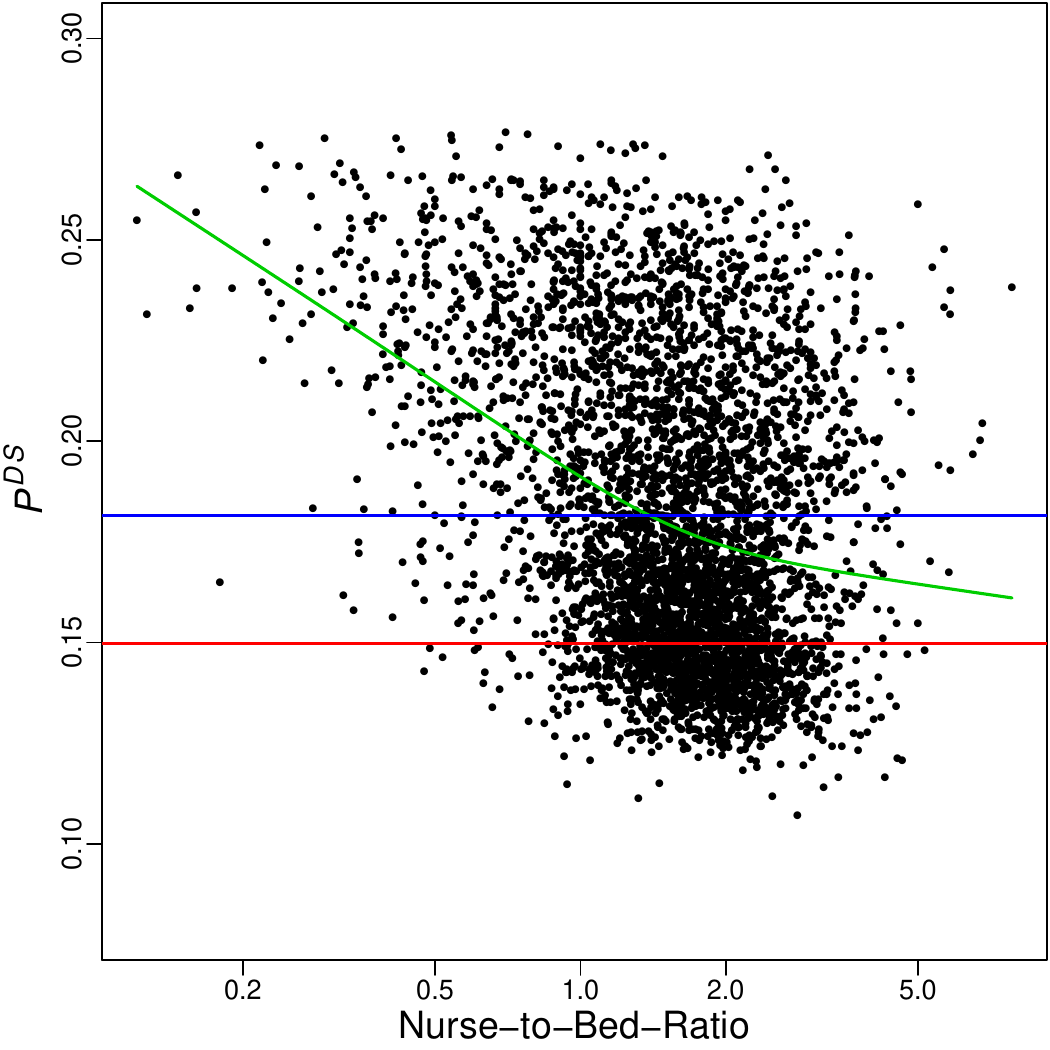}
\caption{$\hP^{DS}_h$ vs NTBR}
\end{subfigure}
\begin{subfigure}{0.28\textwidth}
\includegraphics[width=\textwidth, height = \textwidth]{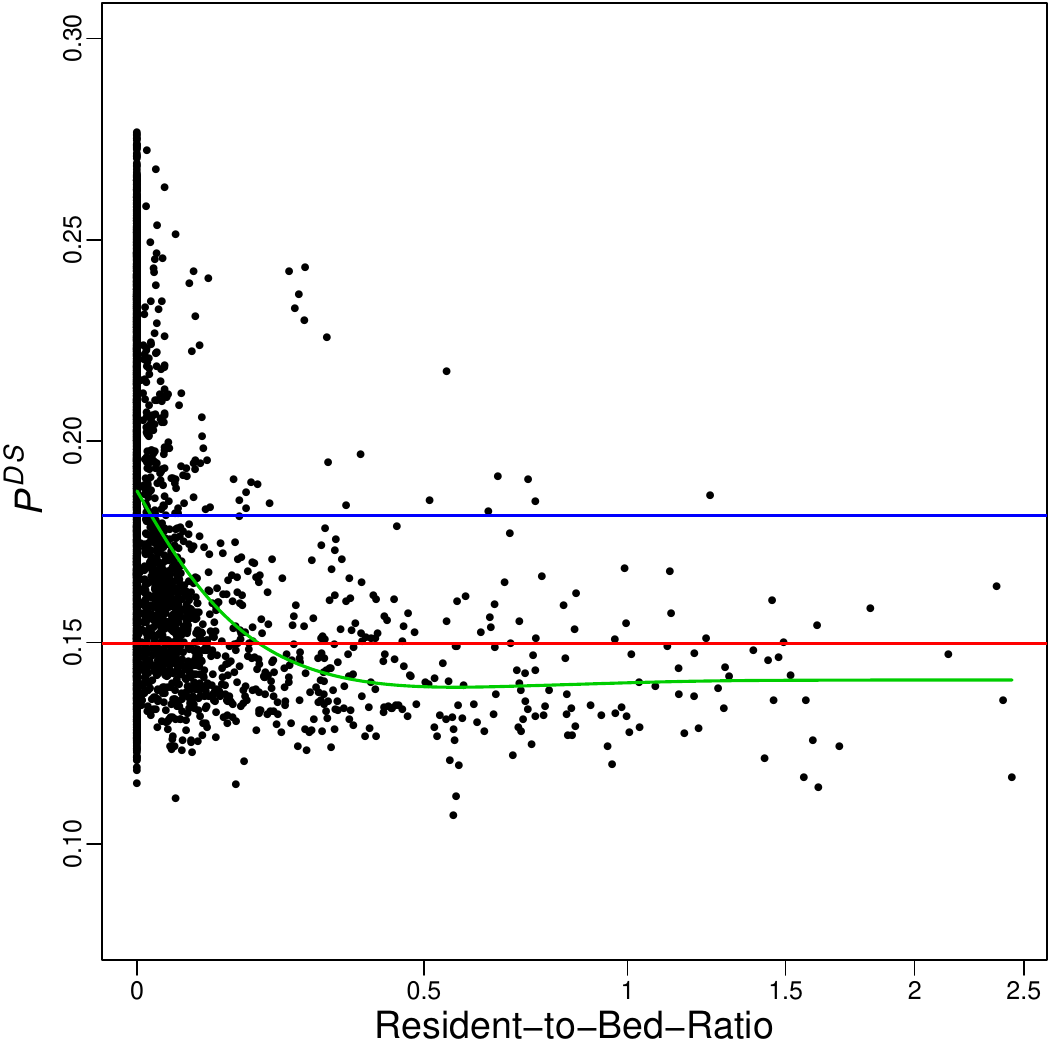}
\caption{$\hP^{DS}_h$ vs RTBR}
\end{subfigure}
\begin{subfigure}{0.28\textwidth}
\includegraphics[width=\textwidth, height = \textwidth]{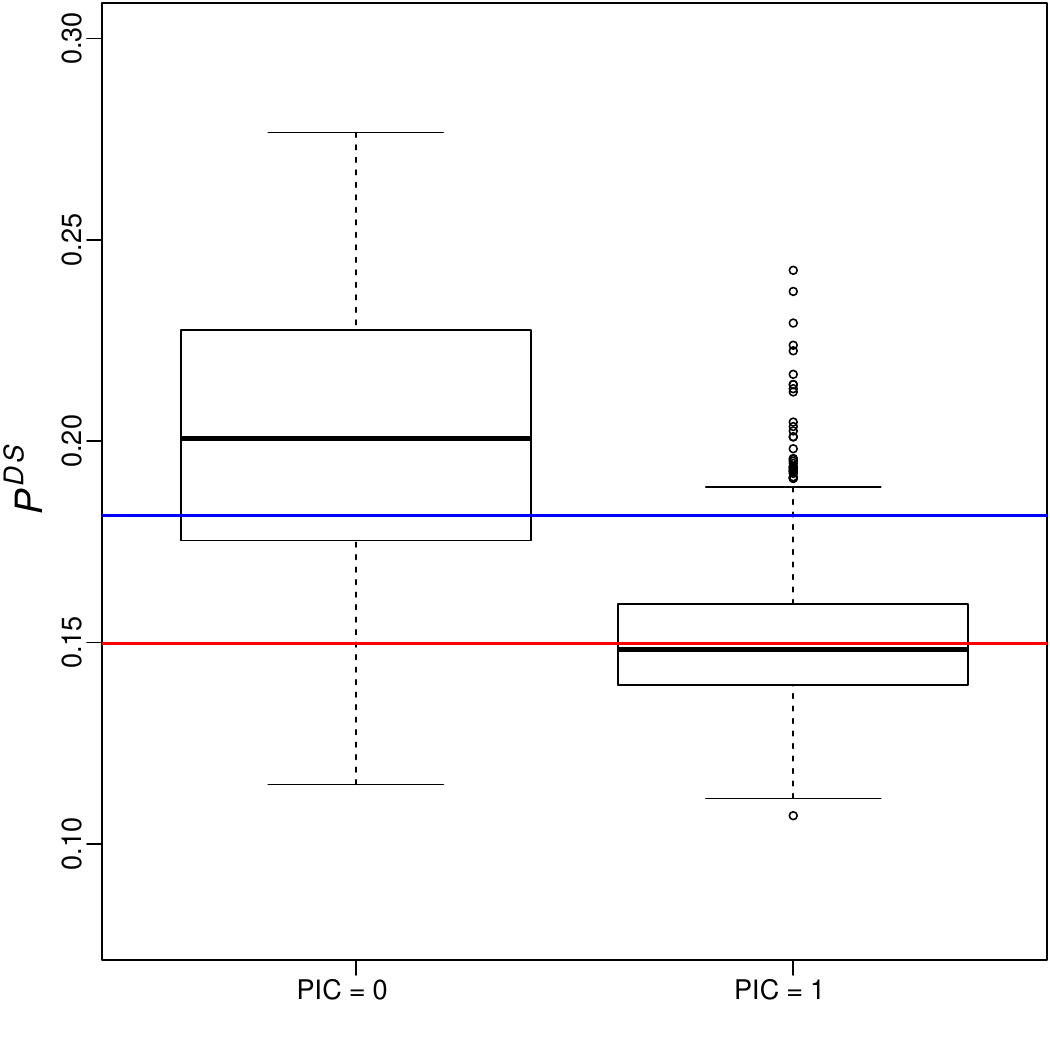}
\caption{$\hP^{DS}_h$ vs PCI}
\end{subfigure}
\centering
\caption{$\hP^{DS}_h$ under the (SLI,L) model.}
\label{Effects}
\end{figure}

The (SLI,L) model included three attributes of hospitals besides volume, namely NTBR, RTBR and PCI.  The 95\% highest posterior density intervals for the coefficients of NTBR and RTBR included only negative values and excluded zero, while the interval for PCI included zero.
However, PCI is highly correlated with hospital volume which is also in the model in the form of a spline.  Figures \ref{Effects}abc plot $\hP^{DS}_h$ versus  NTBR, RTBR and PCI, respectively, for this model.  In Figures \ref{Effects}ab, hospitals with more nurses per bed or more residents per bed are predicted to have lower mortality.  Although not confirmed to be distinct from volume, the ability to perform PCI (PTCA, stents or CABG)  is also associated with lower mortality in Figure \ref{Effects}c.  These patterns are generally consistent with the health services research literature concerning the influence of invasive cardiology (Stukel et al. 2007) and nurse staffing (Person et al. 2004) on AMI survival, though the benefit of treatment at a teaching hospital is more controversial, and studies are inconsistent regarding its influence on mortality (Allison et al., 2000, and Navarthe, et al. 2013).   Lastly, for the age-volume interaction that was also included in the (SLI,L) model, a positive coefficient posterior estimate indicated that large volume hospitals confer a greater survival benefit for younger medicare patients.

\section{Discussion}

As a model for AMI hospital mortality rates, we have found the hierarchical random effect logit model used by Hospital Compare to be inadequate compared to alternatives that model hospital effects as a functions of hospital attributes.  Such models were seen to offer substantial predictive improvements as measured by out-of-sample predictive Bayes factors.  Going further, we have suggested calibrating individualized predictions from Bayesian models against empirically based general advice that would otherwise be used to inform decisions.  This entails conducting an out-of-sample study of general advice without using the model, and then using the model to predict the results of that independent study.  For this purpose, a matched out-of-sample comparison confirmed familiar advice that low volume hospitals tend to have higher mortality rates when treating AMI.  While our models accurately predicted the results of that matched out-of-sample study,  the current Hospital Compare model is not calibrated in this sense, and hence should not be { used. 

One of the main goals of this paper has been to show that the inclusion of hospital attributes in our models leads to better calibrated and more informative mortality rate predictions. In particular, we obtained a vast improvement over the Hospital Compare model with the Medicare data by including hospital volume, staffing by nurses and residents, PCI, and patient-age by hospital-volume interactions.  
While this improvement better serves the needs of patients, its ramifications for public policy must be considered cautiously.  Because low volume hospitals, by definition, have little data regarding mortality, a better small volume hospital may be unable to overcome the poor results of its similarly sized peers to receive the good ranking it deserves.  To some extent, this problem can be mitigated with our models, by including measures of uncertainty along with mortality rate estimates, as described in Section \ref{sec:uncertainty}.  However, further modeling with additional hospital attributes has the clear potential to shed more light on the rankings of such hospitals.   In particular, the addition of hospital attributes which distinguish better low volume hospitals from the rest would be ideal.  Indeed, our models should be considered as the beginning rather than the end of the story. 
While hospital volume is a convenient variable which is strongly associated with mortality, with more information for example about hospital management, it may even turn out not to be the most important predictive variable, thereby diminishing its penalizing effect.  

We have also recommended that the common practice of reporting indirectly standardized rates be avoided.  We found that that for our improved models, indirectly standardized rates fail to eliminate the effect of patient mix differences across hospitals.  Furthermore, such indirectly adjusted rates are also inherently misleading in that they systematically underestimate population hospital mortality rates.}  In contrast, direct adjustment faithfully translates the model's adjustments and mortality predictions into a properly calibrated and easily understood format for public reporting.

The future lies with predictions that are individualized not just to particular hospitals, but to particular patients when treated at particular hospitals.  We scratched the surface of this topic by including patient-by-hospital interactions in (\ref{phji}), finding that younger Medicare patients benefit more that older Medicare patients from treatment at high volume hospitals.  Note that once we have fit one of our models and obtained estimates of the $\alpha_h$ and $\be$, \eqref{phji} can be applied with $p_h(\x)$ for any patient characteristics $\x$ to obtain mortality rate estimates for that patient at any hospital.  Such personalized rate estimates would be more relevant than standardized rates for any particular patient.  

\section*{Supplementary Materials}

As referenced in Sections 2.2, 4.1, 5.1 and 5.2, Appendix A.1 details the MCMC implementation for simulated sampling from our hierarchical logit model posteriors. This entails successive substitution Gibbs sampling from the full conditionals obtained with a suitable Polya-Gamma latent variable posterior augmentation. As referenced in Section 3.1, Appendix A.2 illustrates the relationship between mortality rates and number of hospital beds with a model that excludes volume. This gives further insight into the persistent relationship between hospital mortality and hospital size. As referenced in Section 4.2, Appendix A.3 provides a second out-of-sample calibration example with US News and World Report hospital rankings. As referenced in Section 6.2, Appendix A.4 provides the full cross-classification of low, average and high hospital mortality rates by the (C,C) and the (SLI,L) models.

\spacingset{1.0}

\newpage
\setcounter{page}{1}
\spacingset{1.3}
\begin{center}
{\Large {\bf Mortality Rate Estimation and Standardization \\
for Public Reporting: Medicare's Hospital Compare \\
}}
\vspace{.1in}
{\large {By E.I. George, V. Ro\v{c}kov\'a, P.R. Rosenbaum, V.A. Satop\"a\"a and J.H. Silber
 }}\\
{\large {\it
University of Pennsylvania, University of Chicago and INSEAD}}\\
\vspace{.1in}
{\Large {\bf
Supplemental Material}}\\
\end{center}

\appendix
\section{Appendix}

\subsection{MCMC Posterior Calculation}\label{sec:MCMC}
To fit our fully Bayesian hierarchical models (C,C), (L,C), (S,L) and (SLI,L) we use MCMC simulation sampling from the data induced posterior to calculate quantities of interest. To describe this, it will be convenient to index each of our hierarchical models by $(\al,\be,\ps)$, where $\al = (\alpha_1,\dots,\alpha_H)'$ denotes the $H$ hospital effects, $\be$ denotes the individual fixed effect coefficients, and $\ps$ denotes $\sigma_\beta^2$ and all other hyperparameters associated with $\mu_h(\bm{z})$ and $\sigma^2_h(\bm{z})$.

We use the Gibbs sampler to simulate from $\pi(\al,\be,\ps \C \Y)$.  In principle, this would be obtained by successive substitution sampling from the full conditionals, Casella and George (1992).  However, because $\pi(\al \C \Y,\be,\ps)$ and $\p(\be \C \Y, \al,\ps)$ are not available in closed form, we proceed by Gibbs sampling from an augmented posterior.  Analogous to the augmentation for probit regression with normal latent variables (Albert and Chib 1993), a suitable augmentation for logistic regression is obtained with the introduction of a vector of P\'olya-Gamma latent variables, $\om = \{\omega_{hj}\}$, one for each $hj$, to create a joint posterior  $\p(\om, \al,\be,\ps \C \Y)$, (Polson, Scott and Windle 2013).  The following successive substitution sampling from the full conditionals $\pi( \om \C \Y,\al,\be,\ps)$, $\pi(\al \C \Y,\om,\be,\ps)$, $\pi(\be \C \Y, \om, \al,\ps)$, $\pi(\ps \C \Y, \om, \al,\be)$, is then straightforward.

Simulation from $\pi(\om \C \Y,\al,\be,\ps)$ is obtained by simulating
\begin{equation}
\omega_{hj} \C \al,\be  \sim \mathcal{PG}(1, \alpha_h + \bm{x}_{hj}'\bm{\beta}) \text{ for } h = 1, \dots, H \text{ and } j = 1, \dots, n_h,
\end{equation}
where the P\'olya-Gamma distributions $\mathcal{PG}(b,c)$ are particular infinite convolutions of Gamma distributions.
Polson, Scott and Windle (2013) provide a fast and exact method for simulating from any $\mathcal{PG}(b,c)$ distribution, which is implemented in the R package \text{BayesLogit} (see Windle et al.~(2013) for details).

To describe the simulation of $\al$ and $\be$, let $\bm{X}$ be the complete matrix of patient attributes, $\bm{K}$ be the block diagonal matrix of hospital indicators, $\Omega = \text{diag}(\bm{\omega})$ and $\bm{\kappa} = \Omega^{-1}\left(\bm{Y}-0.5\right)$.  Then, simulation from
$\pi(\al \C \Y,\om,\be,\ps)$ is obtained by simulating
\begin{equation}
\alpha_h \C \om,\be,\ps \sim \mathcal{N}(m_{\alpha_h}, V_{\alpha_h}) \text{ for } h = 1, \dots, H,
\end{equation}
where
$V_{\alpha_h} = \left[ 1/\sigma^2_h(\bm{z}) + \bm{1}_{v_h}' \bm{\omega}_h \right]^{-1}$ and
$m_{\alpha_h} = V_{\alpha_h} \left[ \mu_h(\bm{z})/\sigma^2_h(\bm{z})  + \bm{\omega}_h' (\bm{\kappa}_h - \bm{X}_h \bm{\beta})\right]$.

Simulation from $\pi(\be \C \Y, \om, \al,\ps)$ is obtained by simulating
\begin{equation}
\bm{\beta}\C\om, \al,\ps \sim \mathcal{N}_d(m_{\bm{\beta}}, V_{\bm{\beta}}),
\end{equation}
where
$V_{\bm{\beta}} = \left[\left(1/\sigma^2_{\bm{\beta}}\right)\bm{X}'\bm{X}  + \bm{X}' \Omega \bm{X}\right]^{-1}$ and
$m_{\bm{\beta}} = V_{\bm{\beta}} \left[ \bm{X}'\Omega(\bm{\kappa}-\bm{K}\bm{\alpha}) \right]$.

Finally, simulation from $\pi(\ps \C \Y, \om, \al,\be)$, which does not depend on $\om$, is obtained by well known routine methods and so will not be  further discussed here.

Starting with initial values, successive substitution sampling from these distributions after a suitable burn-in period and with appropriate thinning, yields a sequence
\begin{equation}\label{MCMCsample}
(\al^{(1)},\be^{(1)},\ps^{(1)}), \ldots,(\al^{(S)},\be^{(S)},\ps^{(S)}),
\end{equation}
which may be treated as a sample from $\pi(\al,\be,\ps \C \Y)$.
Letting ${logit} (p_{hj})  = \alpha_h +\bm{x}_{hj}'\bm{\beta}$, the induced sequence $\p^{(1)}, \p^{(2)},\ldots,  \p^{(S)}$
will then be a sample from the induced mortality rate posterior $\pi(\p \C \Y)$.   Posterior estimates of interest are obtained directly from these sequences.  For example, posterior mean estimates of hospital effects are obtained by $\hat{\alpha}_h = \frac{1}{S}\sum_{s=1}^S \alpha_h^{(s)}$.
Posterior mean estimates of individual mortality rates are obtained by $\hat{p}_{hj} = \frac{1}{S}\sum_{s=1}^S p_{hj}^{(s)}$.
Posterior mean estimates of hospital mortality rates are obtained by $\hat{P}_h = \frac{1}{S}\sum_{s=1}^S p_{h\cdot}^{(s)}$, where
$p_{h\cdot}^{(s)} = \frac{1}{n_h}\sum_{j=1}^{n_h} p_{hj}^{(s)}$.  Predictive $(1-\alpha)$\% interval bounds for these rates are obtained by the
corresponding quantiles of the sampled values.

\subsection{Modeling $\alpha_h$ as a Linear Function of the Number of Beds}
All of our models have made use of the strong apparent relationship between hospital mortality and $\vol$.  To get further insight into the relationship between mortality rates and hospital size, we also examined the relationship between mortality rates and the hospital attribute $\beds$, the number of beds in 2008, a variable that is indisputably exogenous to our observed mortality rates.

\begin{figure}[htbp]
   \centering
   \includegraphics[width = 0.5\textwidth]{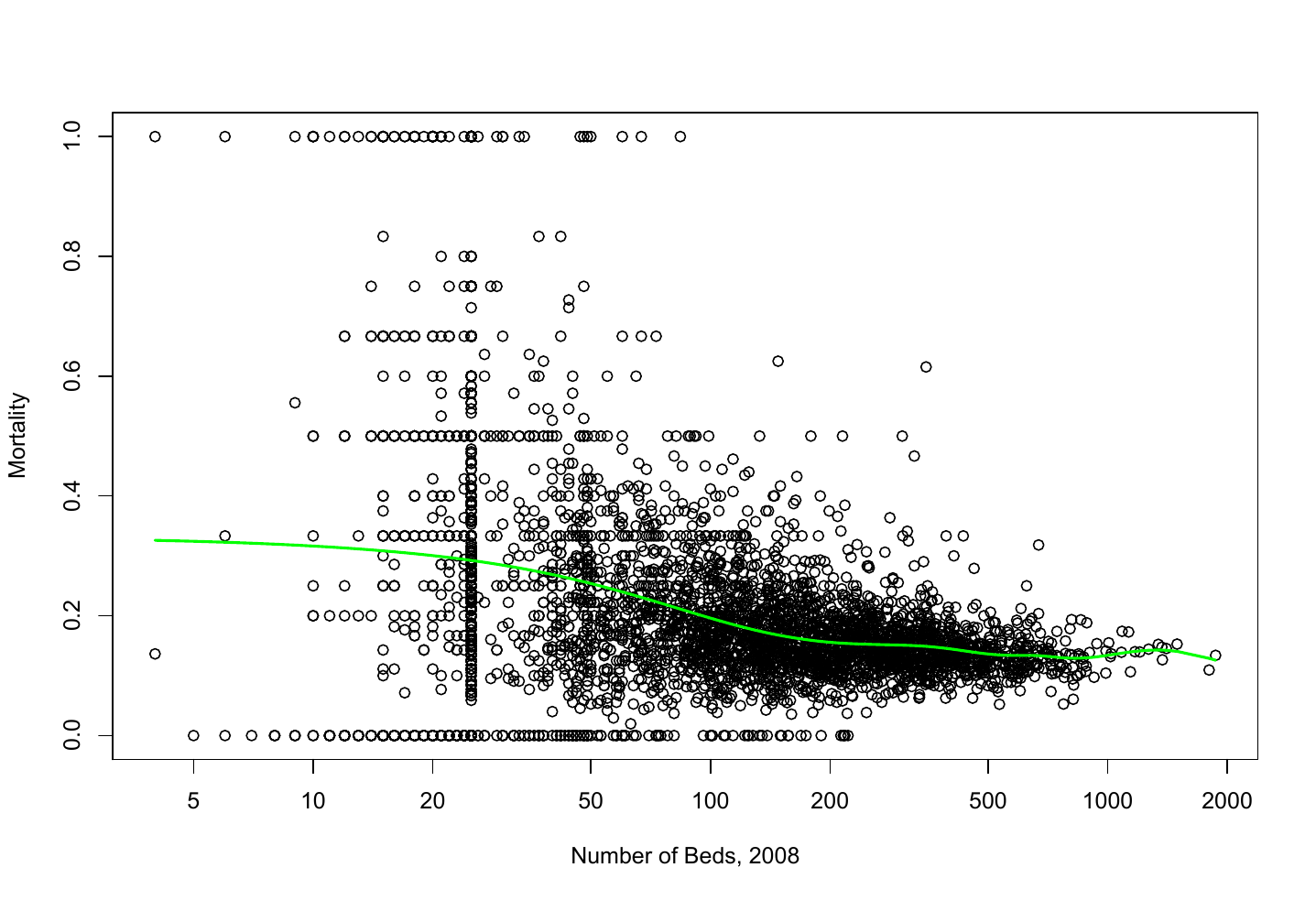} % requires the graphicx package
   \caption{\small Raw observed hospital mortality rates $O_h$ by $\beds_h$.  Average rate by $\beds_h$  summarized by the green superimposed smoothing spline. }
   \label{fig:rawbeds}
\end{figure}

Analogous to Figure \ref{Fig1}, Figure \ref{fig:rawbeds} plots the raw observed mortality rates $O_h$ versus $\beds_h$.  As summarized by the superimposed smoothing spline, the average mortality rate is decreasing as $\beds$ increases.   Note that many hospitals have the same value of $\beds$.  For example, 741 hospitals had $\beds$ = 25, which was the modal number in our data.

\begin{figure}[h!]
\centering
\begin{subfigure}{0.28\textwidth}
\includegraphics[width=\textwidth, height = \textwidth]{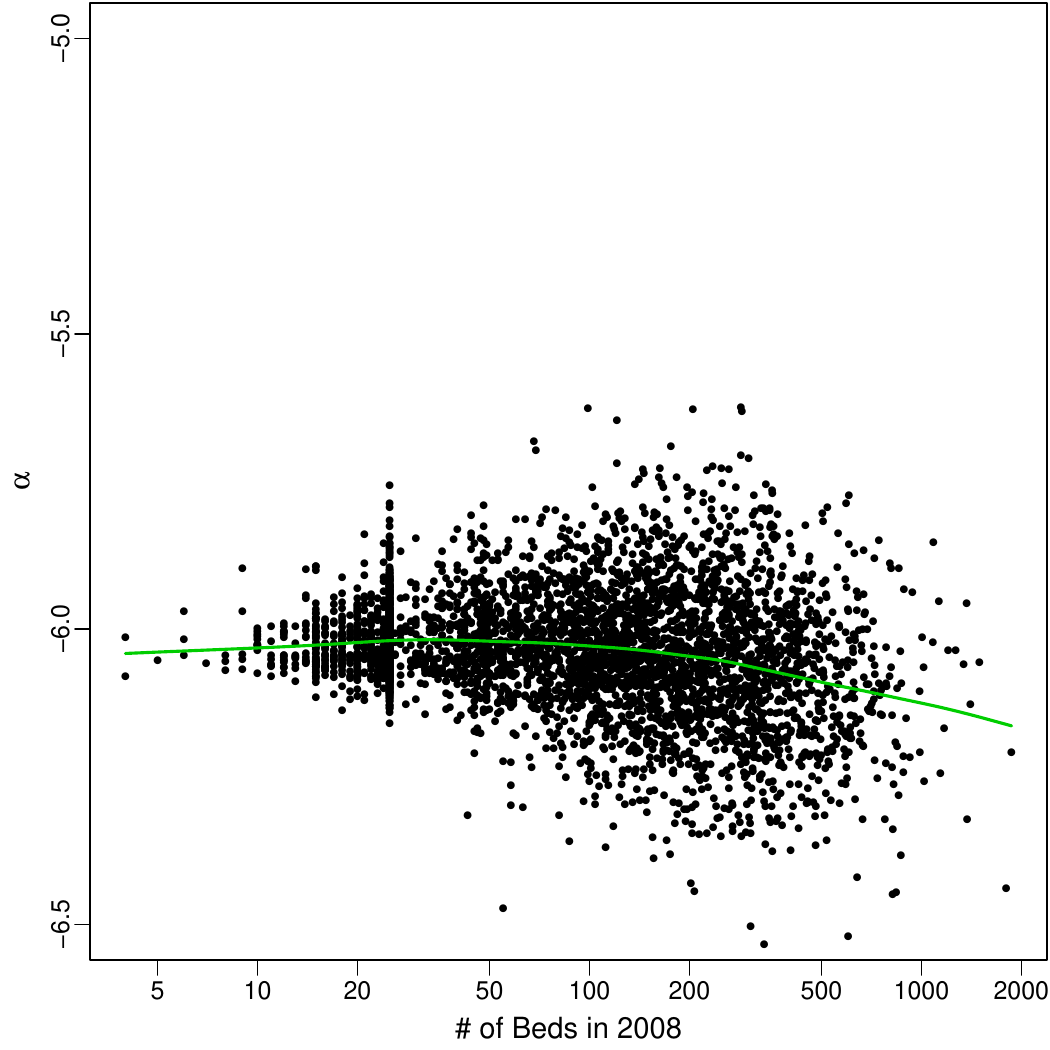}
\caption{\scriptsize $\ha_h$ under (C,C)}
\end{subfigure}
\begin{subfigure}{0.28\textwidth}
\includegraphics[width=\textwidth, height = \textwidth]{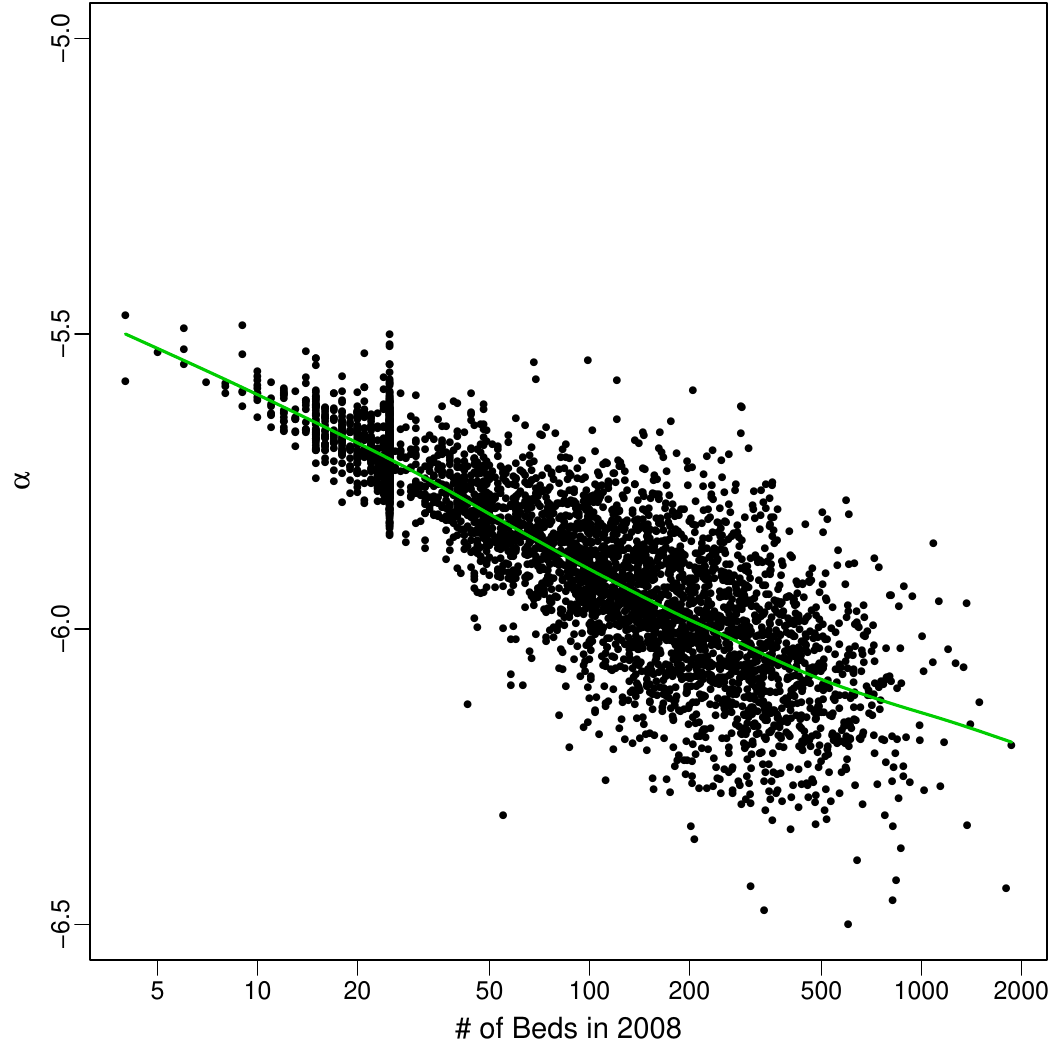}
\caption{\scriptsize $\ha_h$ under (L,C)}
\end{subfigure}
\centering
\vspace{-.1cm}
\caption{$\ha_h$ vs $\beds$. }
\label{FigBeds}
\end{figure}

Figure \ref{FigBeds}a plots the hospital effect estimates $\ha_h$ versus $\beds_h$ for the (C,C) models.  Just as for the plot of $\ha_h$ versus $\vol_h$ in Figure \ref{Fig3}a, the (C,C) model finds no evidence of larger hospital effects at hospitals with smaller $\beds_h$.  However, application of the (L,C) model with $\beds$ as the single hospital attribute, again tells a dramatically different story.  Just as for the plot of $\ha_h$ versus $\vol_h$ in Figure \ref{Fig3}b, Figure \ref{FigBeds}b shows that by emancipating their means as a linear function of $\beds$, the hospitals effects are dramatically higher at the hospitals with smaller $\beds_h$.  Thus, as opposed to the (C,C) model, the (L,C) model here will lead to systematically higher mortality rates at the smaller hospitals.  Confirming that the (L,C) actually leads to improved predictions, the predictive log Bayes factor comparison, as in Section \ref{sec:predcomp}, for (L,C) here vs (C,C) was 19.03, convincing evidence of a strong underlying relationship.

%Only the high tail:
%\begin{verbatim}
%# of Beds     22  24  20  49  15  25
%Count         37  46  49  49  53 741
%\end{verbatim}

\subsection{A Second Calibration with the US News and World Report Rankings}\label{USNews}

To further illustrate our method of matched comparison in Section \ref{sec:calibrations}, we did a second calibration following the same format which we here describe briefly.  The popular magazine, US News and World Report,
ranks hospitals on their ``heart and heart surgery'' program.
We repeated the observational study of low volume
hospitals, replacing low volume hospitals by the top ten hospitals in the
US News and World Report ranking.  In the six-month validation sample, there
were 816 AMI patients in Medicare treated at the top ten hospitals in this
ranking, and we matched them 5-to-1 to patients from other hospitals,
checking covariate balance in parallel with Table \ref{match}.  Unlike
Table \ref{match}, the patients at ``top ten'' hospitals were not very different
from all patients prior to admission.  The matched comparison estimated about
2.0\% lower mortality at ``top ten'' hospitals compared to matched controls without
using the Bayes models.  The Bayes models all predicted lower mortality at
``top ten'' hospitals than at control hospitals, but they slightly underestimated the 2.0\% gain;
for instance, the (SLI,L) model estimated 1.7\% lower mortality at ``top ten'' hospitals.

\subsection{Hospital Cross Classification by Mortality Rates}\label{CrossClass}

\begin{table}[!htbp]
%\footnotesize
\tiny
\centering
(a) All Hospitals\\
\begin{tabular}{|l|| c | c| c ||c|}
\hline
 \specialcell{Counts\\(\%)} & (SLI,L) Low &(SLI,L) Average& (SLI,L) High & Total \\ \hline  \hline
(C,C) Low & 30 & 3 & 0 &33 \\
& (0.68)&(0.07)& (0.00)& (0.75)\\ \hline
(C,C) Average & 28 & 3289 & 1016 &4333 \\
& (0.64)&(78.42)& (23.11)& (98.57)\\ \hline
(C,C) High & 0 & 18 & 12 & 30\\
& (0.00)&(0.41)& (0.27)& (0.68)\\ \hline \hline
Total & 58 & 3310 & 1028 & 4396 \\
& (1.32)&(75.30)& (23.28)& \\ \hline
\end{tabular}\\
\vspace{.5cm}
(b) Lower Volume Quartile Hospitals\\
\begin{tabular}{|l|| c | c| c ||c|}
\hline
 \specialcell{Counts\\(\%)} & (SLI,L) Low &(SLI,L) Average& (SLI,L) High & Total \\ \hline  \hline
(C,C) Low & 0 & 0 & 0 &0 \\
& (0.00)&(0.00)& (0.00)& (0.00)\\ \hline
(C,C) Average & 0 & 210 & 906 &1116 \\
& (0.00)&(18.82)& (81.18)& (100.00)\\ \hline
(C,C) High & 0 & 0 & 0 & 0\\
& (0.00)&(0.00)& (0.00)& (0.00)\\ \hline
Total & 0 & 210 & 906 &1116 \\
& (0.00)&(18.82)& (81.18)&\\ \hline \hline
\end{tabular}\\
\vspace{.5cm}
(c) Upper Quartile Volume Hospitals\\
\begin{tabular}{|l|| c | c| c ||c|}
\hline
 \specialcell{Counts\\(\%)} & (SLI,L) Low &(SLI,L) Average& (SLI,L) High & Total \\ \hline  \hline
(C,C) Low & 29 & 3 & 0 &32 \\
& (2.64)&(0.27)& (0.00)& (2.91)\\ \hline
(C,C) Average & 28 & 1019 & 0 &1047 \\
& (2.55)&(92.72)& (0.00)& (95.27)\\ \hline
(C,C) High & 0 & 16 & 4 & 20\\
& (0.00)&(1.46)& (0.36)& (1.82)\\ \hline \hline
Total & 57 & 1038 & 4 & 1099 \\
& (5.19)&(94.45)& (.036)& \\ \hline
\end{tabular}
\caption{\small Hospital Cross Classifications of Mortality Rates by the (C,C) and (SLI,I) models.}
\label{HCClass}
\end{table}


\begin{thebibliography}{9}
\small
\bibitem{Alber93} Albert, J.H. and S. Chib. (1993), \textquotedblleft Bayesian analysis of binary and polychotomous response data,\textquotedblright\textit{Journal of the American Statistical Association}, 88(422), 669--79.

\bibitem{Allison00} Allison J.J., Kiefe C.I., Weissman N.W., Person S.D., Rousculp M., Canto J.G., Bae S., Williams O.D., Farmer R., Centor R.M. (2000), \textquotedblleft Relationship of hospital teaching status with quality of care and mortality for Medicare patients with acute MI,\textquotedblright\ \textit{Journal of the American Medical Association}, 284, 1256-62.

\bibitem{Ash11}Ash, A., Fienberg, S.E., Louis, T.L., Normand, S.T., Stukel, T.A. and Utts, J. (2012),  \textquotedblleft Statistical Issues in Assessing Hospital Performance. Commissioned by the Committee of Presidents of Statistical Societies.\textquotedblright \;   Available at http://www.cms.gov/Medicare/Quality-Initiatives-Patient-Assessment-Instruments/HospitalQualityInits/Downloads/Statistical-Issues-in-Assessing-Hospital-Performance.pdf.

\bibitem{Berger85} Berger, J. O. (1985), \textit{Statistical Decision Theory and Bayesian Analysis}. SpringerVerlag,
second edition.

\bibitem{Berger06} Berger, J.O. (2006),  \textquotedblleft The Case for Objective Bayesian Analysis,\textquotedblright\ \textit{Bayesian Analysis}, 1(3):385--402.

\bibitem{Box86} Box, G. E. P. and Meyer, R. D. (1986),  \textquotedblleft Dispersion effects from fractional designs,\textquotedblright\ \textit{Technometrics}, 28(1):19-27.

\bibitem{CG92} Casella, G. and George, E.I. (1992), ``Explaining the Gibbs sampler,'' \textit{The American Statistician}, 46 3 167--174.

\bibitem{Cox61} Cox, D. R. (1961), \textquotedblleft Tests of Separate Families of Hypotheses\textquotedblright\ \textit{Proceeding of the Fourth Berkeley Symposium}, 105--123.%

\bibitem{Dawid82} Dawid, A. P. (1982), \textquotedblleft The well-calibrated
Bayesian,\textquotedblright\ \textit{Journal of the American Statistical
Association}, 77, 605-610.%

\bibitem{Dim10} Dimick, J.B., Staiger, D.O. and Birkmeyer, J.D. (2010), \textquotedblleft  Ranking Hospitals on Surgical Mortality: The Importance of Reliability Adjustment,\textquotedblright\ \textit{Health Services Research}, 45,1614--1629.

\bibitem{Gandjour03} Gandjour, A., A. Bannenberg and K. W. Lauterbach. (2003), \textquotedblleft
Threshold Volumes Associated with Higher Survival in Health Care: A Systematic
Review,\textquotedblright\ \textit{Medical Care}, 41, 1129-1141.%


\bibitem{Gelf94} Gelfand, A.E. and Dey, D.K. (1994), Bayesian model choice: asymptotics and exact calculations, \textquotedblright\ \textit{Journal of the Royal Statistical Society, Series B}, 56, 501--514.

\bibitem{Gri12} Grieco, N., Ieva, F. and Paganoni, A.M. (2012), \textquotedblleft Performance assessment using mixed effects models: a case study on coronary patient care,\textquotedblright\ \textit{IMA Journal of Management Mathematics}, 23(2), 117--131.

\bibitem{Gug14} Guglielmi, A., Ieva, F., Paganoni, A.M., Ruggeri, F. and Soriano, J. (2014). \textquotedblleft Semiparametric Bayesian modeling for the classification of patients with high observed survival probabilities,\textquotedblright\ \textit{Journal of the Royal Statistical Society - Series C}, 63 (1): 25--46.


\bibitem{Gu09} Gu, Y., Fiebig, D. G., Cripps, E., Kohn, R. (2009), \textquotedblleft Bayesian estimation of a random effects heteroscedastic probit model,\textquotedblright\ \textit{Econometrics Journal}, 12, 324--339.%

\bibitem{Halm02} Halm, E. A., C. Lee and M. R. Chassin. (2002), \textquotedblleft Is Volume
Related to Outcome in Health Care? A Systematic Review and Methodologic
Critique of the Literature,\textquotedblright\ \textit{Annals of Internal
Medicine}, 137, 511-520.%

\bibitem{Hansen07} Hansen, B. B. (2007), Optmatch: flexible, optimal
matching for observational studies, {\itshape{R News}}, 7, 18-24.

\bibitem{Hansen08} Hansen, B.B. (2008), \textquotedblleft Prognostic
analogue of the propensity score,\textquotedblright\ \textit{Biometrika},%
\textit{\ }95, 481-8.

\bibitem{Luft87} Luft, H.S., S.S. Hunt and S. C. Maerki. (1987) \textquotedblleft The
Volume-Outcome Relationship: Practice-Makes-Perfect or Selective-Referral
Patterns?\textquotedblright\ \textit{Health Services Research}, 22, 157-182.%

\bibitem{Kass95}  Kass, R.E. and Raftery, A.E. (1995) \textquotedblleft Bayes Factors, \textquotedblright\ \textit{Journal of the American Statistical Association}, 90, 773--795.

\bibitem{Krum06} Krumholz, H. M., Y. Wang, J. A. Mattera, Y. Wang, L. F. Han, M. J. Ingber, S. Roman and S.-L. T. Normand. 2006, ``An Administrative Claims Model Suitable for Profiling Hospital Performance Based on 30-Day Mortality Rates among Patients with an Acute Myocardial Infarction,'' \textit{Circulation} 113 (13): 1683--92.

\bibitem{Navathe13} Navathe AS, Silber JH, Zhu J, Volpp KG (2013) \textquotedblleft Does admission to a teaching hospital affect acute myocardial infarction survival?\textquotedblright\ \textit{Academic Medicine}, 88, 475-82.

\bibitem{Person04} Person SD, Allison JJ, Kiefe CI, Weaver MT, Williams OD, Centor RM, Weissman, NW (2004), \textquotedblleft Nurse Staffing and Mortality for Medicare Patients with Acute Myocardial Infarction,\textquotedblright\ \textit{Medical Care}, 42, 4-12

\bibitem{Polson13} Polson, N. G., Scott, J. G. and Windle, J. (2013) \textquotedblleft Bayesian inference for logistic models using Polya-Gamma latent variables,\textquotedblright \textit{Journal of the American Statistical Association (Theory and Methods)}, 108(504), 1339-1349.%

\bibitem{DOS} Rosenbaum, P. R. (2010), \textit{Design of Observational
Studies}, New York: Springer.

\bibitem{match83} Rosenbaum, P. R. and Rubin, D. B. (1983), \textquotedblleft Constructing a control group
by multivariate matched sampling methods that incorporate the propensity score,\textquotedblright\ \textit{American Statistician}, 39, 33-38.

\bibitem{Shahian03} Shahian, D. M. and S. L. Normand. (2003) \textquotedblleft The Volume-Outcome
Relationship: From Luft to Leapfrog,\textquotedblright\ \textit{Annals of
Thoracic Surgery}, 75, 1048-1058.%

\bibitem{Silber10} Silber, J. H., Rosenbaum, P. R., Brachet, T. J., Ross, R. N., Bressler, L. J.,
Even-Shoshan, O., Lorch, S. A. and Volpp, K. G. (2010), \textquotedblleft The
Hospital Compare Mortality Model and the Volume--Outcome
Relationship,\textquotedblright\ \textit{Health Services Research}, 45, 1148-1167.

{
\bibitem{Silber16} Silber, J.H, Satop\"a\"a , V.A., Mukherjee, N., Ro\v{c}kov\'a, V., Wang, W., Hill, A.S, Even-Shoshan, O., Rosenbaum, P.R. and George, E.I. (2016), \textquotedblleft Improving Medicare's Hospital Compare Mortality Model,\textquotedblright\ \textit{Health Services Research}, 51, 1229-1247.
}

\bibitem{Spiegelhalter12} Spiegelhalter, D., Sherlaw-Johnson, C., Bardsley, M., Blunt, I., 
Wood, C. and Grigg, O.  (2012), \textquotedblleft 
Statistical methods for healthcare regulation: rating, screening and 
surveillance,\textquotedblright\ \textit{Journal of the Royal Statistical Society}, A, 175, 
1-47.

\bibitem{Stuart10} Stuart, E. A. (2010), \textquotedblleft Matching methods
for causal inference,\textquotedblright\ \textit{Statistical Science}, 25:
1-21.

\bibitem{Stukel07} Stukel, T. A., Fisher E.S., Wennberg, D. E., Alter, D. A. Gottlieb, D.J. and Vermeulen, M.J., \textquotedblleft Analysis of Observational Studies in the Presence of Treatment Selection Bias: Effects of Invasive Cardiac Management on AMI Survival Using Propensity Score and Instrumental Variable Methods,\textquotedblright\ (2007), \textit{Journal of the American Medical Association}, 297, 278-285.

\bibitem{Wind13} Windle, J., Polson, N.G. and Scott, J.G. (2013),  ``BayesLogit: Bayesian logistic regression,'' 
URL http://cran.r-project.org/web/packages/BayesLogit/index.html. R package version 0.2-4.

\bibitem{Yale14} Yale New Haven Health Services Corporation, (2014) \textit{Measures Updates and Specifications (version 8.0)}, Washington: Centers for Medicare and Medicaid Services.

\end{thebibliography}
\end{document}